\documentclass[11pt,a4paper]{elsarticle}
\usepackage[colorlinks]{hyperref}

\AtBeginDocument{%
\usepackage{color}
\definecolor{mgray}{cmyk}{0,0,0,.8}
\hypersetup{
    bookmarks=true,         
    unicode=false,          
    pdftoolbar=true,        
    pdfmenubar=true,        
    pdffitwindow=true,      
    pdfnewwindow=true,      
    colorlinks=true,       
    linkcolor=mgray,          
    citecolor=blue,        
    filecolor=magenta,      
    urlcolor=blue          
}
}

\usepackage[utf8]{inputenc}
\usepackage{eso-pic}
\usepackage{amsmath,amsfonts,amsthm,graphicx,ulem,tocvsec2}
\usepackage{titlesec}
\usepackage{sectsty}

\allsectionsfont{\normalfont\sffamily\bfseries}

\journal{\hspace{-10em}\colorbox{white}{\phantom{Preprint for submission to Nuclear Physics B}}}

\makeatletter
\def\paragraph{\secdef{\els@aparagraph}{\els@bparagraph}}
\def\els@aparagraph[#1]#2{\elsparagraph[#1]{#2.}}
\def\els@bparagraph#1{\elsparagraph*{#1.}}

\usepackage{wrapfig}

\usepackage{multirow}

\usepackage{mathrsfs}
\usepackage{empheq}

\newcommand\secref[1]{section~\ref{#1}}

\newcommand\appref[1]{\ref{#1}}

\newcommand{\bal}{\begin{align}}
\newcommand{\eal}{\end{align}}
\newcommand{\ba}{\begin{align}}
\newcommand{\ea}{\end{align}}
\newcommand{\beq}{\begin{equation}}
\newcommand{\eeq}{\end{equation}}
\newcommand\beqa{\begin{eqnarray}}
\newcommand\eeqa{\end{eqnarray}}
\newcommand\bea{\begin{array}}
\newcommand\eea{\end{array}}
\newcommand\comment[1]{{}}

\newcommand{\su}{\mathfrak{su}}
\renewcommand{\sl}{\mathfrak{sl}}
\newcommand{\psu}{\mathfrak{psu}}

    \newcommand{\COMMENT}[1]{}
    
    \newcommand{\neqa}{\nonumber\end{eqnarray}}
    \newcommand{\la}[1]{\label{#1}}

\renewcommand{\sl}{\mathfrak{sl}}

\def\a{{\alpha}}

\def\[{\left[}
\def\]{\right]}

\def\s{\sigma}
\def\a{\alpha}
\def\b{\beta}

\def\<{\langle}
\def\>{\rangle}

\def\i2{\frac{i}{2}}

\def\<{\langle}
\def\>{\rangle}

\def\i2{\frac{i}{2}}

\DeclareMathOperator{\Tr}{Tr}
\DeclareMathOperator{\Pf}{Pf}

\def\1h{\hat 1}
\def\2h{{\hat 2}}
\def\3h{{\hat 3}}
\def\4h{{\hat 4}}

\def\be{\begin{eqnarray}}
\def\ee{\end{eqnarray}}
\def\no{\nonumber}

    \def\CD{{\nabla}}

    \def\CN{{\cal N}}
    \def\CO{{\cal O}}
    \def\CP{{\cal P}}

    \def\<{\left\langle\,}
    \def\>{\, \right\rangle}
    \def\[{\left[}
    \def\]{\right]}

\newcommand{\ps}{{\bf p}}

   \def\su{{\mathfrak{su}}}
   \def\sl{{\mathfrak{sl}}}

\def\m{\mu}

\def\i{{\mathsf{i}}}

\renewcommand{\Im}{{\rm Im}\,}

\newcommand{\bP}{{\bf P}}

\usepackage{amssymb}
\usepackage{subfig}
\usepackage{overpic}

\usepackage{varioref}

\begin{document}


 \begin{frontmatter}
\title{\sffamily\LARGE{Quantum spectral curve as a tool for a perturbative quantum field theory}}

\author[1]{Christian Marboe}
\ead{marboec@tcd.ie}
\author[1,2,3]{Dmytro Volin}
\address[1]{School of Mathematics, Trinity College Dublin, College Green, Dublin 2, Ireland}
\address[2]{Nordita,
KTH Royal Institute of Technology and Stockholm University,
Roslagstullsbacken 23,
SE-106 91 Stockholm,
Sweden}
\address[3]{Bogolyubov Institute for Theoretical Physics,\\ 14-b, Metrolohichna str.
Kiev, 03680, Ukraine}
\ead{volind@tcd.ie}
\begin{abstract}
An iterative procedure perturbatively solving the quantum spectral curve of planar $\mathcal{N}=4$ SYM for any operator in the $\sl$(2) sector is presented. A {\texttt{Mathematica}} notebook 
executing this procedure is enclosed. The obtained results include 10-loop computations of the conformal dimensions of more than ten different operators.

We prove that the conformal dimensions are always expressed, at any loop order, in terms of multiple zeta-values with coefficients from an algebraic number field determined by the one-loop Baxter equation. We observe that all the perturbative results that were computed explicitly are given in terms of a smaller algebra: single-valued multiple zeta-values times the algebraic numbers.
\end{abstract}

\begin{keyword}

Quantum spectral curve \sep integrability \sep perturbative quantum
field theory \sep  AdS/CFT correspondence 

\end{keyword}

\end{frontmatter}

\AddToShipoutPictureBG*{%
  \AtPageUpperLeft{%
    \hspace{0.9\paperwidth}%
    \raisebox{-5\baselineskip}{%
      \makebox[0pt][r]{\it TCD-math-2014-07}}
    \raisebox{-6.2\baselineskip}{
      \makebox[-8pt][r]{\it NORDITA-2014-129}
}}}%

\newpage
\thispagestyle{empty}

\begingroup
\hypersetup{linkcolor=black}
\tableofcontents
\endgroup\newpage

\setcounter{tocdepth}{2}

\newpage
\section{Introduction and historical overview}

A distinct benchmark of our understanding is the ability to perform computations, explicitly and efficiently. Computations of conformal dimensions in planar $\mathcal{N}$=4 super-symmetric Yang-Mills theory (SYM) both traced and advanced the progress in the long-standing study of the AdS/CFT spectrum \cite{Beisert:2010jr}. Since the discovery of integrable structures at weak \cite{Minahan:2002ve} and strong \cite{Bena:2003wd} 't Hooft coupling, $\lambda$, about a decade ago, the understanding of the conjectured \cite{Beisert:2003tq} AdS/CFT integrability has been steadily advancing towards arbitrary values of the coupling. The computational capability was not only improving in parallel, but explicit results were a source of insights that promoted the overall progress. 

Study of the conformal dimensions of the so-called twist $L$ spin $S$ operators played a particularly important role. These operators form a closed $\sl(2)$ sector in the spectrum, and they can be represented as linear combinations of the basis states
\be\label{genericoperator}
\Tr \nabla^{s_1}_+Z\nabla_+^{s_2}Z\ldots\nabla_+^{s_L}Z\,
\ee
where $s_1+s_2+\ldots+s_L=S$, $\nabla_+$ is a light-cone covariant derivative, and $Z$ is a complex scalar field of $\mathcal{N}$=4 SYM.

Early approach of factorised scattering \cite{Staudacher:2004tk,Beisert:2005tm} resulted in the Beisert-Staudacher asymptotic Bethe Ansatz equations \cite{Beisert:2005fw} to describe the spectrum of very long operators, with $L\to\infty$. It turned out that the equations were also suited for the large $S$ case as the conformal dimensions of the $\sl(2)$ sector exhibit the universal scaling  $\Delta\to 2\Gamma_{\rm cusp}(\lambda)\log S+\CO(1)$ \cite{Belitsky:2006en} with $S\to\infty$ and $L$ arbitrary. Precisely computation of $\Gamma_{\rm cusp}$ was considered when an integral equation \cite{Beisert:2006ez} was proposed fixing the last missing piece of the Bethe Ansatz -- the dressing phase \cite{Arutyunov:2004vx,Janik:2006dc}. Solving this integral equation had allowed computing $\Gamma_{\rm cusp}$ at any value of $\lambda$ \cite{Beisert:2006ez,Benna:2006nd,Basso:2007wd}. It was the first explicit example of a non-trivial function interpolating between perturbative results of the gauge theory at weak coupling \cite{Bern:2006ew} and the string theory at strong coupling \cite{Roiban:2007dq}  thus strongly supporting the integrability conjecture, as well as the conjecture of the AdS/CFT correspondence.
 
The Bethe Ansatz equations are insufficient at finite $L$, and this was first explicitly demonstrated in the example of the twist 2 operators. For these, several orders of the weak coupling expansion can be found as an analytic function of $S$. Its analytic continuation to $S=-1$ should have a particular pole structure that can be determined from the Balitsky-Fadin-Kuraev-Lipatov (BFKL) equation; however, starting from the fourth loop order, this structure is not reproduced from the Bethe Ansatz \cite{Kotikov:2007cy}. This clearly indicated the relevance of finite-size effects and the necessity to correct the Bethe Ansatz. It was found in \cite{Bajnok:2008bm} that the L\"uscher wrapping corrections do the job, in the example of the four loop conformal dimension of the Konishi operator compared against an explicit field theory computation \cite{Fiamberti:2007rj,Fiamberti:2008sh}. Later L\"uscher corrections were computed for four \cite{Bajnok:2008qj} and five \cite{Lukowski:2009ce} loops and arbitrary $S$ correctly reproducing the prediction from the BFKL equation. Five \cite{Beccaria:2009eq} and six \cite{Velizhanin:2010cm} loops were also successfully computed for certain twist 3 operators at arbitrary spin, with the results passing several non-trivial checks related to continuation to negative spins, and the five-loop result for $S=2$ was also confirmed by a direct perturbative field theory computation \cite{Fiamberti:2009jw}.

With the realization that finite size effects are important, the focus on what is considered as difficult shifted. Indeed, originally integrability was perceived as a tool to diagonalise the dilatation operator which is a large complicated matrix for the case of long operators. This problem, more precisely the part about finding eigenvalues, was solved by the Bethe Ansatz. Therefore, it were the small matrices which became difficult to treat as no Bethe Ansatz exist for short operators. One of the best cases for study is the Konishi operator, $\Tr Z\nabla_+^2Z$. It is the smallest operator with non-protected anomalous dimension. It actually does not mix with other operators, hence the question is to find its multiplicative renormalization, yet this question was notoriously difficult. For instance, from the point of view of the dual string theory, the Konishi operator is a highly quantum state. Application of quasi-classical approaches is questionable, it even gave rise to contradicting results originally \cite{Gromov:2009zb,Roiban:2009aa}, though currently there is agreement about the two leading terms \cite{Gromov:2011de,Roiban:2011fe,Vallilo:2011fj} and two more terms were suggested \cite{Gromov:2011bz,Gromov:2014bva}.

\begin{figure}
\begin{center}
\includegraphics[width=0.6\textwidth]{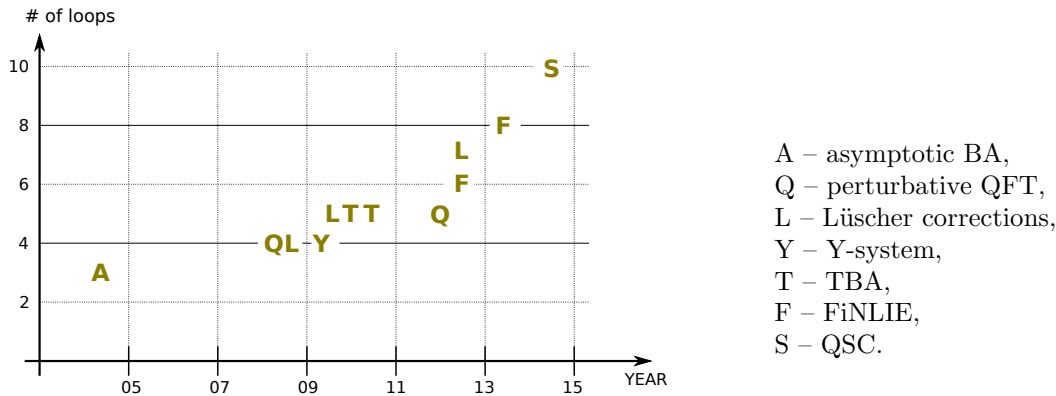}
\hspace{3em}
\parbox[b]{0.3\textwidth}
{
\small{
\begin{tabbing}
A -- asymptotic BA,\hspace{3em} \\  Q -- perturbative QFT,\\
L -- L\"uscher corrections, \\ Y -- Y-system, \\T -- TBA, \\
F -- FiNLIE, \\ S -- QSC.
\end{tabbing}
}
}
\caption{
\label{fig:history}
History of perturbative computations of the Konishi anomalous dimension in planar $\mathcal{N}$=4 SYM.
}
\end{center}
\end{figure}

On the weak coupling side, computing the Konishi anomalous dimension is a very good indicator of the available ideas and computation techniques, as shown in Fig.~\ref{fig:history}. Asymptotic Bethe Ansatz is applicable only up to three loops, after that the wrapping corrections start to be important. The single-wrapping orders can be captured by an adoption of L\"uscher formulae to the AdS/CFT case and it was done up to the maximal possible order - seven loops \cite{Bajnok:2012bz}. The double L\"uscher formulae that cover higher loops were suggested in \cite{Bombardelli:2013yka}, however they seem to be very complicated technically and no explicit computation has been achieved so far. 

Instead of computing corrections in wrappings, the thermodynamic Bethe Ansatz (TBA) approach \cite{Zamolodchikov:1989cf} was adopted to the AdS/CFT case and the TBA equations which are expected to be exact at any $\lambda$ were derived \cite{Gromov:2009tv,Bombardelli:2009ns,Gromov:2009bc,Arutyunov:2009ur}. Alongside the original success of TBA that allowed numerical computation of the conformal dimension of the Konishi operator \cite{Gromov:2009zb,Frolov:2010wt} and a handful of other states from the $\sl(2)$ sector \cite{Frolov:2012zv} at reasonable values of $\lambda$, the attempt to analytically solve these equations was much less encouraging: Konishi anomalous dimension was computed only up to five loops \cite{Arutyunov:2010gb,Balog:2010xa}, after a considerable effort. In comparison, the recent quantum field theory based computation reached the same order \cite{Eden:2012fe}. We therefore faced a situation where the advantage of the integrability techniques was questionable.

Fortunately, the TBA equations appeared to be not the simplest way to encode the spectrum. Using integrability of the underlying Hirota dynamics \cite{Gromov:2010km} in conjunction with analytic properties of the system, the TBA equations were reduced to FiNLIE - a finite set of non-linear integral equations \cite{Gromov:2011cx}, see also \cite{Balog:2012zt}. These equations allowed for a 6-loop \cite{Leurent:2012ab} and then for an 8-loop \cite{Leurent:2013mr} computation. The latter one  is interesting in several ways. First,  the double wrapping order was reached for the first time. Second, this was a first example of a "physical" QFT quantity (not a separate Feynman diagram) which contains a special combination of non-reducible multiple zeta-values (MZV's), predicted in 1995 \cite{Broadhurst:1995km}, based on general analysis of possible graphs. Finally, this computation allowed the identification of the full basis of functions of the spectral parameter which appear in the perturbative expansion and hence allowed to automatize perturbative computations in terms of algebraic manipulations in this basis. We rely on this algebraic structure and further develop it in the current work.

Finally, the FiNLIE was further significantly simplified, after a deep analysis of the interplay between algebra and analyticity, to a concise set of Riemann-Hilbert equations -- the quantum spectral curve (QSC) \cite{Gromov:2013pga,Gromov:2014caa}. QSC already demonstrated its power in computations of near-BPS quantities \cite{Gromov:2013pga,Gromov:2013qga,Gromov:2014bva} (which for the case of $\sl(2)$ operators correspond to analytic continuation to $S=0$ and various expansions around this point) and deriving the Pomeron pole contribution exactly \cite{Alfimov:2014bwa}, in contrast to only few perturbative orders reproduced previously. 

In this work we present another application of QSC -- an efficient perturbative weak coupling expansion of conformal dimensions in the $\sl(2)$ sector of the theory. In comparison, in the previous developments, almost any new loop order was a subject of a new publication. At highest loops, the computations were specially tailored for one chosen operator (Konishi), and although the conceptual possibility existed to repeat analogous computations for other states, this was unthinkable without applying a significant human effort.  The approach we propose works for any operator from the sector, and the recursive algorithm implemented in {\texttt{Mathematica}} is universal and can be run to any order of the perturbative expansion provided the computer memory is sufficient. For instance, we computed the conformal dimension of the Konishi operator and more than ten other operators up to ten loops\footnote{To go beyond ten loops one should update the table of relations between multiple zeta-values, which is straightforward though not included in the published code. To increase the efficiency at these high loops, one might benefit from the MZV datamine \cite{Blumlein:2009cf}.}.

The computation times for the Konishi case, achieved on a {\it single} 3.2 GHZ core of an iMac desktop, are given in the table:
\be
\begin{tabular}{c|l}
$\#$ of loops & time
\\
5 & 4 sec
\\
6 & 15 sec
\\
7 & 1 min
\\
8 & 5 min
\\
9 & 27 min
\\
10 & 3.1 hours 
\end{tabular}
\ee
About 3 GB of memory was used for the 10-loop computation.

In general, an operator is specified by $L$, $S$ and a Baxter polynomial $Q(u)$ -- a degree $S$ polynomial solution of the Baxter equation
\be
\left(u+\frac{i}{2}\right)^L Q(u+i)+\left(u-\frac{i}{2}\right)^L Q(u-i) = T(u) Q(u)\,,
\ee
where $T$ is requested to be a polynomial as well. To account for the cyclicity of the trace \eqref{genericoperator}, one should consider only solutions which satisfy the "zero-momentum" condition $Q(i/2)=Q(-i/2)$. Only a discrete set of solutions is possible. For not too large $L$ and $S$, it is easy to produce all of them on a computer, see  \appref{Qap}. There is a one-to-one correspondence between the solutions and the states in the spectrum. For instance, the Konishi operator is associated to the Baxter polynomial $Q(u)=u^2-\frac{1}{12}$.

The {\texttt{Mathematica}} program receives $L$, $S$ and $Q(u)$ as an input and is then able to compute the corresponding anomalous dimension. The limitations for the computation efficiency are purely of combinatorial nature: one would not want $Q$ to be a polynomial of too high degree, neither should the coefficients in this polynomial be too complicated algebraic numbers. We successfully performed the computation for more than 100 different states with $L+S\lesssim 20$, computing at least seven and up to ten loops, depending on the complexity of the input. Therefore one can state that finally a part of the finite-volume AdS/CFT spectral problem is perturbatively solved in the practical sense: there is a working black box machine which explicitly computes anomalous dimensions.

The presented results should not be perceived only as a technical report. In fact, an early version of this computation was being developed alongside the development of QSC itself, and it was one of the sources for  cross-checks and conceptual ideas that helped to formulate QSC. This perturbative computation give an explicit demonstration of how QSC is used to encode the spectrum. It also has an interesting interconnection with perturbative QFT since we can prove the following generic statement: at any order of perturbation theory, the result is given solely in terms of MZV's, and algebraic numbers from a field where the coefficients of $Q(u)$ live. This is only a subclass of what is expected from the generic perturbative QFT analysis \cite{Broadhurst:2014fga}. In fact, one should be able to build up a Hopf algebra structure behind the presented perturbative computation and question how it can be compared with the algebra of Feynman diagrams.

The article is organised as follows. In \secref{sec:preliminaries} we define the algebra of functions encountered in the computations and formulate the necessary properties of the quantum spectral curve equations. In \secref{sec:procedure} we explain the perturbative algorithm which has two important parts: the leading order solution, which is somewhat specific and requires an ansatz that singles out the $\sl(2)$ sector, and the iterative cycle. Each period of the cycle increases the precision by one loop. We end this section by explaining the possible verifications of the algorithm's implementation. Finally, \secref{sec:results} is devoted to a summary of the obtained results and discussion, whereas appendices are devoted to technical clarifications.

\section{\label{sec:preliminaries}Preliminaries}
\subsection{Basic objects and their properties}
\subsubsection{Analytic structure at finite coupling}
The quantum spectral curve is a set of equations for the functions of the spectral parameter $u$. The functions may have branch points at $u=\pm 2g+i\,{\mathbb Z}$, where $g=\frac{\sqrt{\lambda}}{4\pi}$. Interestingly, this is the only place where the coupling constant enters the whole construction. 
\begin{wrapfigure}{r}{0.45\textwidth}
\captionsetup{width=0.4\textwidth}
\begin{center}
\includegraphics[width=0.3\textwidth]{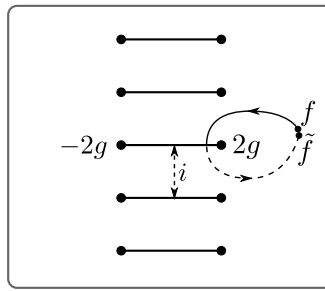}
{\caption{\label{fig:physical}Physical kinematics and continuation to the next sheet.}}
\end{center}
\vspace{-1em}
\end{wrapfigure}
Keeping in mind a weak coupling expansion, we always consider the functions in the so-called physical kinematics, that is we choose a Riemann sheet with short cuts as in Fig.~\ref{fig:physical}. 

One focuses on the branch points $\pm 2g$ only as all other analytic properties follow, see \cite{Gromov:2013pga} and discussion below.  For any function $f(u)$ appearing in AdS/CFT integrability, it is always assumed that an analytic continuation around either $2g$ or $-2g$ gives the same result which is denoted as $\tilde f(u)$. Another standard assumption is that the branch points are of second order:  $\tilde{\tilde f}=f$.

\subsubsection{Shift operators}
The quantum spectral curve leads in particular to finite-difference equations. To account for these, a short-hand notation for shifted functions of the spectral parameter is introduced:  $f^{[n]}(u)\equiv f\left(u+\frac{in}{2}\right)$ and $f^\pm\equiv f^{[\pm 1]}$. The shifts are performed along a contour which avoids short cuts.

Furthermore, we introduce the operator $\CD$,
\be
\CD(f)\equiv f-f^{[2]} \,,
\ee
and the operator $\Psi$ satisfying
\be\label{defPsi}
\CD\cdot\Psi(f)=f\,.
\ee
In principle, \eqref{defPsi} does not define $\Psi$ uniquely. Indeed, one can have $\Psi\cdot\CD
(f)=f+\CP$, where $\CP$ is an arbitrary $i$-periodic function, i.e. $\CD
(\CP)=0$. However, it will be convenient for us to choose  some unambiguous prescription for $\Psi$. This choice is summarized in \appref{sec:Psiresum}. In particular, $\Psi(f) \equiv \sum\limits_{n=0}^\infty f^{[2n]}$ when the sum is convergent. Hence, we will not view the periodic functions as an ambiguity in $\Psi$. Instead, they will reappear in solutions of homogeneous parts of the encountered finite-difference equations.

\subsubsection{\label{sec:algebra}Algebra of functions for perturbative expansion}
In the weak coupling limit, the branch points collide at $u=i\,\mathbb{Z}$. Hence they are not encountered in the perturbative expansion, and instead poles arise at the collision points. Straightforwardly from the below-presented procedure, we can prove that at any order of weak coupling expansion only the following functions are encountered
\begin{itemize}
\item polynomials, $u^a$, and shifted inverse powers, $\frac{1}{(u+in)^a}$,
\item $\eta$-functions, 
$\eta_{a_1,a_2,\hdots,a_k}(u)\equiv \sum\limits_{0\le n_1<\hdots< n_k<\infty}\frac{1}{(u+in_1)^{a_1}\cdots(u+in_k)^{a_k}}$,
\item the $i$-periodic functions $\mathcal{P}_a\equiv \eta_a+\bar{\eta}_a^{[-2]}$,
\end{itemize}
and products thereof.

Note that $\eta$-functions form a ring, and the same is true for the $i$-periodic functions, and for polynomials and inverse powers. Hence, any function appearing in QSC can be represented as at most a trilinear expression in this basis, which is practically used in the computer implementation. Linearity in $\eta_A$ is particularly important because it allows a systematic definition of the action of $\Psi$ on this algebra of functions, with the result still belonging to the same algebra, see \appref{sec:Psiresum} and \cite{Leurent:2013mr}. This is the essential property allowing the computer algorithm to run to arbitrary loop orders.

At $u=i$, $\eta$-functions evaluate to multiple zeta-values (MZV's):
\be
\eta_{a_1, \hdots, a_k}(i)=i^{-\sum a_j} \zeta_{a_1,\hdots,a_k}\,,\ \ \ \ {\rm where}\ \ \ \zeta_{a_1,a_2,\hdots,a_k}\equiv \sum\limits_{1\le n_1<\hdots< n_k<\infty}\frac{1}{n_1^{a_1}\cdots n_k^{a_k}}\,.
\ee
This is how MZV's enter in the computations and ultimately in the perturbative corrections to the conformal dimensions.

\subsection{Quantum Spectral Curve}
\subsubsection{Riemann-Hilbert equations}
For left/right-symmetric states, which is the case for the considered $\sl(2)$ sector, QSC can be defined by the following relations of the Riemann-Hilbert type \cite{Gromov:2013pga,Gromov:2014caa}:
\begin{subequations}
\label{qsc}
\begin{align}
\label{firstRH}
\mu_{ab}-\tilde\mu_{ab} &= \tilde \bP_a\bP_b-\tilde \bP_b\bP_a\,,\\
\label{secondRH}\tilde \bP_a &= (\mu\chi)_{a}{}^{b}\,\bP_b\,,\\ \tilde\mu_{ab} &=\mu_{ab}^{[2]}\,,\label{permu}
\end{align}
\end{subequations}
referred to as the $\bP\mu$-system. Here $\mu_{ab}=-\mu_{ba}$ and $\bP_a$ are functions of the spectral parameter, whereas $\chi$ denotes the constant matrix
\begin{align}\label{chimat}
\chi^{ab}\equiv \left(
\begin{tabular}{cccc}
  0 & 0 & 0 & $-1$\\
  0 & 0 & $+1$ & 0 \\
  0 & $-1$ & 0 & 0 \\
  $+1$ & 0 & 0 & 0 \\
\end{tabular}\right) \,,\ \ (\mu\chi)_a{}^b&\equiv \mu_{ac}\chi^{cb}  \,.
\end{align}
In the left/right-symmetric case, one can always identify
\be\label{mu14eqmu23}
\mu_{14}=\mu_{23}\,.
\ee
For the sake of simplicity, the following notation is interchangeably used:
\be
\mu_{1}\equiv \mu_{12}\,,\ \ \mu_2\equiv\mu_{13}\,,\ \ \mu_3\equiv\mu_{14}=\mu_{23}\,,\ \ \mu_{4}\equiv \mu_{24}\,,\ \ \m_5\equiv\mu_{34}\,.
\ee

The equations \eqref{qsc} are 
defined in the strip $0<\Im u<1$, and elsewhere by their analytic continuation. If this analytic continuation never crosses short cuts, $\bP_a$ and $\mu_{ab}$ are said to be on the physical Riemann sheet.

$\bP_a$ have only one Zhukovsky cut ($u\in[-2g,2g]$) on the physical sheet\footnote{However, infinitely many branch points are present on the other sheets.}. $\mu_{ab}$ have infinitely many branch points, at positions $u=\pm 2g+i\,\mathbb{Z}$, but the monodromies around these points are under control. Indeed, one can derive the following functional relation from \eqref{qsc} \cite{Gromov:2013pga}:
\be\label{fun1}
\mu^{[2]}_{ab}=\mu_{ab}-\big((\mu\chi)_{a}{}^{c}\,\bP_c\bP_b-(\mu\chi)_{b}{}^{c}\,\bP_c\bP_a\big).
\ee  
Furthermore, since \eqref{qsc} also implies
\be\label{ortomu}
(\mu\,\chi-\mu^{[2]}\chi)_{a}{}^{b}\,\bP_b=0\,,
\ee
equation \eqref{fun1} is equivalent to
\be\label{fun2}
\mu^{}_{ab}=\mu_{ab}^{[2]}+\big((\mu^{[2]}\chi)_{a}{}^{c}\,\bP_c\bP_b-(\mu^{[2]}\chi)_{b}{}^{c}\,\bP_c\bP_a\big).
\ee 
Hence $\mu_{ab}^{[2n]}$ can be expressed as a linear combination of $\mu_{ab}^{[0]}$  for any $n\in\mathbb Z$, while the analytical continuation of $\mu_{ab}$ around $u=\pm2g$ is known due to \eqref{permu}.

Equations \eqref{fun1} and \eqref{ortomu}  will be heavily used alongside \eqref{qsc} during the computations. As an illustration of their possible usage, the relation
\be\label{Pf1}
\Pf(\mu)\equiv\mu_{12}\mu_{34}-\mu_{13}\mu_{24}+\mu_{14}\mu_{23}=1\,
\ee
can be derived by noticing that the relation $\bP_a=(\mu\chi)_{a}{}^{b}(\mu\chi)_{b}{}^c\bP_c$ is yet another consequence of \eqref{ortomu}, \eqref{secondRH}, and \eqref{permu}, while, on the other hand, one has the purely algebraic property $(\mu\chi)_{a}{}^{c}(\mu\chi)_{c}{}^b=\Pf(\mu)\delta_a{}^b$\,.

While all the equations are linear in $\mu_{ab}$, the property $\Pf(\mu)=1$ is bilinear. It will be used as a check of the correctness of the computations.

\subsubsection{Asymptotics}
The large $u$ asymptotic behaviour of $\bP_a$ and $\mu_{ab}$ is fixed to\footnote{Here we adopt the conventions of \cite{Gromov:2014caa} which are different by $1\leftrightarrow 2$, $3\leftrightarrow 4$ from \cite{Gromov:2013pga} and which are compatible with the highest-weight description of $\psu(2,2|4)$ representations.}
\begin{eqnarray}\label{Plarge}
\bP_1\simeq A_1\,u^{-\frac{L+2}{2}}\,,\ \ \bP_2\simeq A_2\, u^{-\frac{L}{2}}\,,\ \bP_3\simeq A_3\,u^{\frac{L-2}{2}}\,,\ \ \bP_4\simeq A_4\,u^{\frac{L}{2}}\;,
\no\\
\mu_{1}\sim u^{\Delta-L}\,,\ \ \mu_{2}\sim u^{\Delta-1}\,,\ \ \mu_{3}\sim u^{\Delta}\,,\ \ \mu_{4}\sim u^{\Delta+1}\,,\ \ \mu_5\sim u^{\Delta+L}\,,
\end{eqnarray}
with
\begin{eqnarray}\la{AB}
A_1A_4&=& \frac{[(L-S+2)^2-\Delta^2] [(L+S)^2-\Delta^2]}{16i L (L+1)}\,,  \nonumber  \\
A_2A_3&=& \frac{[(L+S-2)^2-\Delta^2] [(L-S)^2-\Delta^2]}{16iL (L-1) }\,,
\end{eqnarray}
where $L$ and $S$ are, correspondingly, the twist and the spin of the $\sl(2)$ operator and $\Delta$ is its conformal dimension. $L$ and $S$ are integers while $\Delta$ is not, and the main aim of the presented work is to determine this quantity as a power series in $g^2$.

Though the asymptotics of $\bP_a$ contain half-integer powers for odd $L$, the potential sign ambiguity is non-physical. This can be seen by introducing
\be\label{rescaling}
\ps_a\equiv(g\,x)^{\frac{L}{2}}\,\bP_a\,,
\ee where $x$ is the Zhukovsky variable satisfying
\be
\frac u{g}=x+\frac 1x\,,
\ee
with $|x(u)|>1$ on the physical sheet.

Since $\tilde x=\frac{1}{x}$, the rescaling \eqref{rescaling} modifies \eqref{qsc} to
\be
\mu-\tilde\mu=\frac 1{g^L}\tilde\ps\wedge\ps\,,\ \ \tilde\ps=\frac 1{x^L}(\mu\chi)\cdot\ps\,,\ \tilde\mu=\mu^{[2]}\,,
\ee
where the sign ambiguity is no longer present. Moreover, $\ps_a$ appear to be suitably normalized quantities, and they will be used alongside $\bP_a$ in the computations.

\subsubsection{Regularity}
Finally, we require that $\bP_a$ and $\mu_{ab}$ have no poles and that their absolute value is bounded at the branch points (e.g. $\bP_a-\tilde\bP_a$ should behave as $\sqrt{u-2g}$ near $u=2g$, not as $\frac{1}{\sqrt{u-2g}}$.) As typical for integrable models, the regularity condition is used to single out the discrete spectrum of physically-relevant solutions as it will become clear in \secref{sec:Baxterfirst}. 


\subsubsection{Symmetries}
The $\bP \mu$-system \eqref{qsc} is invariant under the transformations (dubbed H-symmetry {\color{blue}\cite{Gromov:2011cx}}):
\be
\bP_a\to H_{a}{}^{b}\,\bP_b\,,\ \ \mu_{ab}\to H_{a}{}^{c}H_{b}{}^{d}\mu_{cd}\,,\ \ \chi^{ab}\to \chi^{cd}(H^{-1})_{c}{}^{a}(H^{-1})_{d}{}^{b}\,,
\ee
where $H$ is a constant\footnote{Generically, $H$ can be $i$-periodic, but in the case considered it should be constant in order to preserve regularity and the power-like behaviour at large $u$.} matrix with $\det H=1$\,. In principle, an arbitrary linear combination of $\bP_a$ can be chosen, however this freedom is significantly constrained by requiring that no two $\bP_a$ have the same exponent of $u$ as their leading asymptotics at $u\to\infty$. This choice is reflected in \eqref{Plarge}. In addition, by keeping \eqref{mu14eqmu23} and the explicit form of $\chi$ \eqref{chimat}, only six parameters of the original 15 remain unfixed in the H-symmetry. 

Two of them amount to constant rescalings,
\be
\bP_1\to\a\,\bP_1\,,\ \ \bP_2\to\beta\, \bP_2\,,\ \ \bP_3\to\b^{-1}\,\bP_3\,,\ \ \bP_4\to\a^{-1}\bP_4\,.\ \
\ee
$\mu_{ab}$ are rescaled accordingly. This symmetry explains why only the products $A_1A_4$ and $A_2A_3$ are fixed in \eqref{AB}.

The remaining four parameters represent the possibility of adding $\bP_a$ with weaker large $u$ asymptotics to $\bP_b$ with stronger asymptotics\footnote{Infinitesimal transformations are given.}:
\be\label{shiftH}
&&\bP_1\to\bP_1\,,
\no\\
&&\bP_2\to\bP_2+\delta_1\bP_1\,,
\no\\
&&\bP_3\to\bP_3-\gamma\,\bP_1+\delta_2\bP_2\,,\no\\
&&\bP_4\to \bP_4+\gamma\,\bP_2+\delta_1\bP_3+\delta_3\bP_1\,.
\ee
Note that for some operators (e.g. Konishi) the solution has an additional parity symmetry ($u\leftrightarrow-u$). In this case $\delta_i=0$ automatically.

In the following, H-symmetry will be fixed completely by the requirements
\be\label{normalization}
A_1=g^2\,,\ \  A_2=1\,,
\ee
leading to the large $u$ asymptotics
\be\label{psnormalisation}
\ps_1\simeq \frac{g^2}{u}\,,\ \ \ps_2\simeq 1 \,.
\ee
The parameters $\gamma$ and $\delta_i$ are fixed by requiring that $\ps_2$ includes no term proportional to $u^{-1}$, that $\ps_3$ has no term proportional to $u^0$, and that $\ps_4$ has no terms proportional to $u^0$ and $u^{-1}$ in their large $u$ expansions.

Except for the $g^2$ scaling of $A_1$, which is chosen for the  transparency of the perturbative algorithm, our prescription to fix H-symmetry is rather arbitrary.

\subsubsection{Formula for $\Delta$} 
The solution of \eqref{AB} with respect to $\Delta$ and $S$ reads
\be\label{eqEnS}
&&\{(S-1)^2\,,\Delta^2\}=
\\
&&i A_3(L-1)-ig^2A_4(L+1)+\frac 12(L^2+1)\pm(L^2-1)\sqrt{\frac{iA_3}{L-1}-\frac{ig^2A_4}{L+1}+\frac 14}\,,
\no
\ee
i.e. either $(S-1)^2$ or $\Delta^2$ is obtained depending on the choice of square root branch. The proper branch is chosen by requiring that at weak coupling 
\be
\Delta=L+S+\CO(g^2)\,.
\ee

In the below-presented computation scheme, $A_3$ and $A_4$ are fixed as a part of the iterative algorithm. Hence \eqref{eqEnS} can be used as an efficient way to find $\Delta$. 

Additionally, two useful checks are gained. First, a different choice of the square root sign should lead to integer $(S-1)^2$, hence we get an all-loop condition to verify. Second, $\Delta$ can also be read off from the asymptotics of $\mu_{ab}$ \eqref{Plarge} so we can verify the consistency of the solution. 

\section{\label{sec:procedure}Procedure}
The algorithm presented in this section determines all the coupling-dependent quantities (i.e. $\mu_{ab}$, $\bP_a$, $\tilde \bP_a$ and $\Delta$) perturbatively in $g^2$ through an iterative procedure which, in principle, can be repeated an arbitrary number of times, increasing the expansion order by one with each iteration. The leading order of the quantities, which encodes the first correction to $\Delta$ and is thus referred to as the first loop, is handled separately. All higher-order corrections (the higher loops) are found from exactly the same algorithm.

\subsection{Scaling of $\bP_a$ at weak coupling}
A systematic knowledge of the structure of the quantities $\ps_a$ is crucial to the algorithm. Since $\bP_a$ have only one cut on the physical sheet, $\ps_a$ can be represented in terms of a convergent $\frac{1}{x}$ expansion, which for $\ps_1$ and $\ps_2$ has the structure
\be\label{p1p2ansatz}
\ps_1=\frac g x+\sum_{k=2}^{\infty}\frac{c_{1,k}(g)\,g^k}{x^k}\,,\ \ \ps_{2}=1+\sum_{k=2}^{\infty}\frac{c_{2,k}(g)\,g^k}{x^{k}}\,,
\ee
where the above-chosen normalization \eqref{psnormalisation} of the leading terms and the convention of how to fix H-symmetry have been taken into account.

Equivalently, $\ps_3$ and $\ps_4$ can be represented as
\be\label{p3p4}
\ps_3=A_3 u\,M_{L-2}(u)+\sum_{k=1}^{\infty}\frac{c_{3,k}(g)g^k}{x^k}\,,\ \ \ps_{4}=A_4 u\,N_{L-1}(u)+\sum_{k=2}^{\infty}\frac{c_{4,k}(g)g^k}{x^{k}}\,,
\ee
where $M_d$ and $N_d$ are polynomials of degree $d$ normalised to $M_{d}=u^{d}+\CO(u^{d-1})$ and $N_{d}=u^{d}+\CO(u^{d-1})$. $A_3$, $A_4$  and the coefficients of the polynomials are functions of $g$.

The expansions converge at least for $|x|>1$ since $\bP_a$ have no singularities on the physical sheet. As $\bP_a$ have no poles on other sheets and the Zhukovsky variable, $x$, resolves the branch points $u=\pm 2g$, the actual radius of convergence is much larger. Indeed, the first branch point in the $x$-plane appears at $x_0=\pm\frac 1{x(2g\pm i)}$. At weak coupling $x_0\simeq\pm i\, g$ and hence the series above converge for $|x|>\Lambda\,g$ with $\Lambda\to 1$ when $g\to 0$.
Therefore, at weak coupling, $c_{a,k}(g)\sim \CO(g^0)$ for sufficiently large $k$.

We put as an ansatz that 
\be\label{scalingansatz}
c_{a,k}(g)\sim \CO(g^0)
\ee
is satisfied for all $k$, and the same non-singular scaling in $g$ is requested for the coefficients in the polynomials $M$ and $N$ \footnote{Instead of introducing $M$ and $N$, one could equivalently consider the parameterizations $\sum\limits_{k=-L+1}^{\infty}\frac{c_{3,k}(g)g^{|k|}}{x^{k}}$ and $\sum\limits_{k=-L}^{\infty}\frac{c_{4,k}(g)g^{|k|}}{x^{k}}$, and then require $c_{a,k}\sim\CO(g^0)$.}. 
As it will become clear below, the whole $\sl(2)$ spectrum can be reproduced within this ansatz. 

Quite interestingly, our preliminary studies show that relaxing this ansatz leads, after a proper adjustment of \eqref{Plarge}, to states outside the $\sl(2)$ sector, in particular from the $\su(2)$ sector. We postpone further study of this observation to future publications.

\subsection{\label{sec:lpone}Loop one}
One important property which we exploit is the particular scaling of $A_a$ at weak coupling. Using the physical requirement that $\Delta=L+S+\CO(g^2)$, one finds
\be\label{Asleading}
A_1A_4=\CO(g^2)\,,\ \ \  A_2 A_3=-i\,S\,\frac{L+S-1}{L-1} +\CO(g^2)\,.
\ee

In the chosen normalization \eqref{normalization}, $A_{a\neq 1}\sim \CO(g^{0})$ and $A_1\sim \CO(g^2)$. Moreover, according to the scaling ansatz \eqref{scalingansatz}, the leading order of $\ps_1$ in the weak coupling expansion is uniquely fixed by the value of $A_1$: $\ps_1=\frac{A_1}{u}=\frac{g^2}{u}$. Hence $\bP_1$ vanishes at the leading order in contradistinction to the remaining $\bP_a$. This significantly simplifies the
$\bP\mu$-system such that it can be solved explicitly\footnote{We will show below that $\mu_{ab}$ scale in the same way precisely for the chosen normalisation of $A_a$ and hence do not spoil the $\bP_1\to 0$ simplification.}. Subleading orders are computed as a perturbation around this solution.

The effect $A_1A_4=0$ can be properly understood in representation theory language. $A_1A_4=0$ is the point where the value of the conformal dimension hits the unitary bound, see appendix C of \cite{Gromov:2014caa}. As we can deduce from \eqref{Asleading},  all operators from the $\sl(2)$ sector reach this bound, therefore precisely at zero coupling all $\sl(2)$ multiplets are "short" while they recombine with other single-trace operators to form long multiplets at finite coupling, precisely like it happens for the Konishi state.

Outside the $\sl(2)$ sector, the same effect occurs for quite a big class of operators, in particular for those from rank-1 sectors, however there are also states for which neither of the functions $\bP_a$ are small at weak coupling.

\subsubsection{\label{sec:Baxterfirst}Baxter equation \{$\mu_1$, $\mu_2$, $\bP_3$, $\tilde \bP_1$\}}
The combinations $\mu_{ab}+\tilde\mu_{ab}$ and $\frac{\mu_{ab}-\tilde\mu_{ab}}{\sqrt{u^2-4g^2}}$ are analytic in the vicinity of the origin. Therefore they do not have poles at $u=0$ in their perturbative expansion, despite that the individual $\mu_{ab}$ may be singular there due to the collision of branch points. Now recall that $\tilde\mu_{ab}=\mu_{ab}^{[2]}$. Since $\mu_{ab}$ and $\mu_{ab}^{[2]}$ are just shifted copies of the same function, we expect them to scale with $g$ in the same way. Then analyticity of $\mu_{ab}+\mu_{ab}^{[2]}$ and $\frac{\mu_{ab}-\mu_{ab}^{[2]}}{\sqrt{u^2-4g^2}}$ implies that $\mu_{ab}$ does not have poles at $u=0$ and $u=i$ at the leading order of the perturbative expansion. 

By recursively applying \eqref{fun1} or \eqref{fun2} and using that $\bP_a$ are analytic outside the origin, we prove that the leading order of $\mu_{ab}$ is free from poles for any $u\in i\,{\mathbb Z}$. Since $\mu_{ab}$ have no singularities elsewhere, by the regularity property of QSC, we conclude that $\mu_{ab}$, at the leading order, are entire functions with power-like asymptotics, i.e. polynomials.

Consider now the functional equation \eqref{fun1} which is written explicitly as
\begin{align}
\label{mufunexplicit}
{
\left(\begin{smallmatrix}
\mu_1^{[2]} \\[0.5mm]
\mu_2^{[2]} \\[0.5mm]
\mu_3^{[2]} \\[0.5mm]
\mu_4^{[2]} \\[0.5mm]
\mu_5^{[2]}
\end{smallmatrix}
\right)
=\no
\left(\begin{smallmatrix}
1-\bP_2\,\bP_3+\bP_1\,\bP_4 & \bP_2^2 & -2\,\bP_1\,\bP_2 & \bP_1^2 & 0 \\[1mm]
-\bP_3^2 & 1+\bP_{2}\,\bP_{3}+\bP_1\,\bP_4 & -2\,\bP_1\,\bP_3 & 0 & \bP_1^2\\[1mm]
-\bP_3\,\bP_4 & \bP_2\,\bP_4 & 1 & -\bP_1\,\bP_3 & \bP_1\,\bP_2\\[1mm]
-\bP_4^2 & 0 & 2\,\bP_2\,\bP_4 &1-\bP_2\,\bP_3-\bP_1\,\bP_4 & \bP_2^2 \\[1mm]
0 & -\bP_4^2 & 2\,\bP_3\,\bP_4 & -\bP_3^2 &1-\bP_1\,\bP_4+\bP_2\,\bP_3\\
\end{smallmatrix}
\right)
\left(\begin{smallmatrix}
\mu_1 \\[2.5mm]
\mu_2 \\[2.5mm]
\mu_3 \\[2.5mm]
\mu_4 \\[2.5mm]
\mu_5 \\
\end{smallmatrix}
\right)}.\\
\end{align}%
For $\bP_1=0$ the above matrix becomes reducible. More precisely, the equations for $\mu_1$ and $\mu_2$ decouple from the rest:
\be
\mu_1^{[2]}&=&(1-\bP_2\,\bP_3)\,\mu_1+\bP_2^2\,\mu_2\,,\no\\
\mu_2^{[2]}&=&-\bP_3^2\,\mu_1+(1+\bP_2\,\bP_3)\,\mu_2\,. \label{mu1mu4}
\ee
By eliminating $\mu_2$ and denoting
\be\label{mu1Q}
\mu_1^+\equiv\alpha\, Q\,,
\ee
where $\a$ is a normalization constant,
one gets {\color{blue}\cite{Gromov:2013pga}}
\begin{equation}\label{Baxter}
-T\,Q+\frac 1{(\bP_2^-)^2}Q^{[-2]}+\frac 1{(\bP_2^+)^2}Q^{[+2]}=0\,,
\end{equation}
where \(T\equiv \frac{\bP_3^-}{\bP_2^-}- \frac{\bP_3^+}{\bP_2^+}+\frac 1{(\bP_2^-)^2}+\frac 1{(\bP_2^+)^2}\)\,. Recall that at the leading order
\be
\hspace{5em}\bP_2= u^{-\frac{L}{2}}\ , \ \ \ \ \ \ \ \frac{\bP_3}{\bP_2}=\ps_3 = A_3\,u\, M_{L-2}(u)\,, \label{p14struc}
\ee
as follows from \eqref{p1p2ansatz} and \eqref{p3p4}. Hence the prefactors of $Q^{[n]}$ in \eqref{Baxter} are polynomials and the equation has precisely the form of the Baxter equation for the $\sl(2)$ XXX spin chain of length $L$ \footnote{In this paper we face this famous second-order equation only at the leading order of the perturbative expansion. Curiously, it is known how to formulate a second-order equation correctly reproducing the whole asymptotic  Bethe Ansatz approximation of the $\sl(2)$ sector, see \cite{Belitsky:2007kf,Volin:2008kd} and also  \cite{Belitsky:2009mu} for a proper account of the pre-wrapping loop order. This higher-loop Baxter-like equation might be a departing point to initiate expansion in wrappings instead of expansion in loops considered in this paper.}. 

It is well known that physical solutions correspond to polynomial $Q$-functions, in a perfect agreement with our conclusion about polynomiality of $\mu_{ab}$ at the leading order.

Following the standard logic, the zeros of $Q$-functions, $u_k$, are fixed by the Bethe equations,
\be\label{BAEQ}
\left(\frac{u_k+\frac{i}{2}}{u_k-\frac{i}{2}}\right)^L=-\frac{Q(u_k-i)}{Q(u_k+i)}\,,
\ee
and each solution corresponds to a state in the $\sl(2)$ spin chain of length $L$. Equivalently, one can directly work on the level of the Baxter equation \eqref{Baxter} and search for polynomial solutions there. The latter approach is more efficient in many cases. 

Furthermore, QSC imposes \cite{Gromov:2013pga} the zero-momentum condition,
\be\label{zeromomentum}
\frac{Q(+\frac{i}{2})}{Q(-\frac{i}{2})}=1\,.
\ee
Indeed, $\mu_{1}$ should have the following analytic structure:
\be
\mu_{1}(u)=f_1(u)+\sqrt{u^2-4g^2}f_2(u),
\ee
where $f_i$ have no branch points in the vicinity of the real axis. Using \eqref{permu},
\be\label{eq40}
\mu_{1}(u+i)=f_1(u)-\sqrt{u^2-4g^2}f_2(u)\,.
\ee 
Note that $f_2(u)$ cannot be singular, since it would contradict the QSC regularity. Therefore, unless $f_1(0)=0$, one has $\frac{\mu_1(0)}{\mu_1(i)}=1$ at the leading order of the weak coupling expansion, which is nothing but \eqref{zeromomentum}. But $f_1(0)=Q(-\frac{i}{2})$ cannot be zero because all solutions of the $\sl(2)$ Baxter equation are real.

The zero-momentum condition is a strict requirement in AdS/CFT integrability due to the cyclicity of the trace, e.g. in \eqref{genericoperator}. In contradistinction to the Bethe Ansatz equations where the zero-momentum condition is an extra requirement to be imposed, the QSC formalism naturally implies it.

We discuss how to find explicit $Q$'s for various $L$ and $S$ in \appref{Qap}. Many interesting solutions are even functions of $u$, for instance $Q_{\rm Konishi}=u^2-\frac 1{12}$. However, there are examples which do not have a particular $u$-parity, the simplest being the two solutions with $L=4$ and $S=3$: $Q=u^3\pm \frac 32u^2+\frac 14u\mp \frac 18$. Solutions with odd $u$-parity are impossible since they are incompatible with \eqref{zeromomentum}.

Once the $Q$-function is chosen, the polynomial $T(u)$ and then $\ps_3$, which is also a polynomial at the leading order, can be computed. Note that $A_3$ is also fixed by the Baxter equation and it is consistent with \eqref{Asleading}.

We choose to normalise $Q$ to $Q(u)=u^S+\CO(u^{S-1})$. The Baxter equation does not define the normalization $\alpha$ in \eqref{mu1Q}. To fix it, consider \eqref{firstRH} for $\mu_{12}\equiv \mu_1$ with $\bP_1=0$:
\be
\label{firstRH12}
&\mu_{1}-\mu_{1}^{[2]}=\tilde \bP_1\bP_2=g^{-L}\tilde\ps_1\ps_2\,.
\ee
Though determining $\tilde\ps_1$ requires the full resummation of \eqref{p1p2ansatz}, its leading order small $u$ expansion can be fixed, since $\tilde\ps_1= g\,x+\CO\left((gx)^2\right)=u+\CO(u^2)$. Consequently,
\be\label{condalpha}
&&\tilde\bP_1\bP_2 = g^{-L}u+\CO(u^2)\,.
\ee
We therefore perform the small $u$ expansion of the l.h.s. of \eqref{firstRH12}:
\be
&&\mu_1-\mu_1^{[2]}=\a(Q^--Q^+)=\a\, u\left(\partial_uQ^--\partial_u Q^+\right)_{|u=0}+\CO(u^2)\,,
\ee
where the zero-momentum property was used.

Now $\a$ can be determined:
\be\label{alpha}
\alpha=\frac {1}{g^{L}Q(\frac{i}{2})\partial_u\log\frac{Q^-}{Q^+}|_{u=0}}\,.
\ee
With $\mu_1$ known, $\mu_2$ can be found from \eqref{mu1mu4} and $\tilde \bP_1$ from \eqref{firstRH12} as $\ps_2=1$ at the leading order.

\subsubsection{Determining $\Delta$ and \{$\tilde \bP_2$, $\bP_4$, $\mu_3$\}}
Consider \eqref{secondRH} for $\tilde\bP_2$ with $\bP_1=0$:
\be
\tilde\bP_2&=&-\mu_3\bP_2+\mu_1\bP_4\,,\no\\
\tilde\bP_2&=&-\mu_3^{[2]}\bP_2+\mu_1^{[2]}\bP_4\,,\label{mu3pt1}
\ee
where the second equality follows from \eqref{ortomu}. Eliminating $\mu_3$ yields 
\be\label{eqforPt1}
\left(\frac{\tilde\bP_2^+}{\bP_2^+}-\frac{\tilde\bP_2^-}{\bP_2^-}\right)=\alpha\, Q\,\left(\frac{\bP_4^+}{\bP_2^+}-\frac{\bP_4^-}{\bP_2^-}\right)
=\alpha\,A_4\,Q\,(u^+N_{L-1}^+-u^-N_{L-1}^-).
\ee
The r.h.s. is a polynomial, so $\tilde\bP_2/\bP_2$ is a polynomial as well, which means that the infinite sum in \eqref{p1p2ansatz}, $\tilde\ps_2=1+\sum_k c_{1,k}(g)\,(g\,x)^k$, is truncated to a finite number of terms at the leading order. The structure of the polynomial can be precised to
\be\label{eq32}
\frac{\tilde\bP_2}{\bP_2}=\left(\frac{u}{g}\right)^L\left(a_{0}u^{S}+a_1u^{S-1}+\ldots+a_{S-2}u^2+0\cdot u+1\right)\,.
\ee
By plugging \eqref{eq32} into \eqref{eqforPt1}, $N_{L-1}$, $a_j$, and $A_4$ can be fixed by matching all orders of $u$. In fact, using that $Q$ solves the Baxter equation, the solution can be written explicitly:
\be
\frac{\tilde\bP_2}{\bP_2}=\left(\frac{u}{g}\right)^L\frac {(iu+\delta)(Q^+-Q^-)+\frac 12(Q^++Q^-)}{{Q(\frac{i}{2})}}\,,
\ee
where $\delta$ is adjusted such that $a_{S-1}=0$ in \eqref{eq32}. 
With $\tilde{\bP}_2$ and $\bP_4$ being fixed, $\mu_3$ can be found from \eqref{mu3pt1}.

Considering \eqref{eqforPt1} at large $u$, 
\be
\frac{i}{g^L\,Q(\frac{i}{2})}\,(L+S)(1-S)u^{L+S-1} +\CO(u^{L+S-2})= i\, \alpha\,A_4\,L\,u^{L+S-1} +\CO(u^{L+S-2})\,, \ee
an explicit expression for $A_4$ can be written:
\be
A_4=\frac{(L+S)(S-1)}{L}\partial_u\log\frac{Q^+}{Q^-}|_{u=0}.
\ee
In consequence, the one-loop $\Delta$, using \eqref{AB}, is the well-known expression
\be\label{dispersion}
\Delta^{(1)}=2ig^2\, \partial_u\log\frac{Q^+}{Q^-}|_{u=0}=2g^2\,\sum_{k=1}^S\frac{1}{u_k^2+\frac 14}\,.
\ee

\subsubsection{\label{sec:inhoBaxter}Inhomogeneous Baxter equation \{$\mu_4$, $\mu_5$\}}
The next nontrivial step is to fix $\mu_4$. Again, we should consider the functional equations \eqref{mufunexplicit} and notice that with $\bP_1=0$ the equations for $\mu_4$ and $\mu_5$ are
\be\label{mu45}
\mu_4^{[2]}-(1-\bP_2\,\bP_3)\,\mu_4-\bP_2^2\,\mu_5&=&U_1,\no\\
\mu_5^{[2]}+\bP_3^2\,\mu_4-(1+\bP_2\,\bP_3)\,\mu_5&=&U_2,
\ee
where the source terms, $U_i$, are functions of  $\bP_a$, $\mu_1$, $\mu_2$ and $\mu_3$. By eliminating $\mu_5$, an inhomogeneous version of the Baxter equation is obtained for $\mu_4$,
\begin{equation}\label{Baxtermu2}
-T^+\,\mu_{4}^{[2]}+u^L\,\mu_4+(u+i)^L\mu_4^{[4]}=U\,,
\end{equation}
where $U$ is a source term as above. As we will see below, exactly this type of equation will be encountered twice, for $\mu_1$ and $\mu_4$, in each perturbative loop. The source term, $U$, will be of gradually increasing complexity. The following discussion of how to solve \eqref{Baxtermu2} is general and the exact same procedure is applied at higher loops.

As the homogeneous part of the equation is the same as in \eqref{Baxter}, we know that one of the homogeneous solutions is the polynomial $Q^-$. 

To find an inhomogeneous solution, consider the ansatz
\be
\mu_4=Q^-f(u)\,.
\ee
After simple manipulations \eqref{Baxtermu2} reduces to
\be
\nabla\left(u^LQ^-Q^+\nabla(f)\right)=Q^+\,U
\ee
which is solved using the $\Psi$-operator \eqref{defPsi}: 
\be
f_{\rm inhomo}=\Psi\left(\frac{1}{u^L\,Q^-Q^+}\Psi \left( Q^+\,U\right)\right)\,.
\ee
Note that by putting $U=0$ and choosing\footnote{Our standard choice is $\Psi(0)=0$, but here, to obtain \eqref{homosol}, we choose differently.} $\Psi(0)=1$, the second homogeneous solution is obtained:
\be
f_{\rm homo}=\Psi\left(\frac{1}{u^L\,Q^-Q^+}\right)\,.\label{homosol}
\ee
Hence, the full solution of the inhomogeneous Baxter equation reads
\be\label{mu2generic}
\mu_4=\Phi_1\,Q^-+\Phi_2\,Q^-\,\Psi\left(\frac{1}{u^L\,Q^-Q^+}\right)+Q^-\,\Psi\left(\frac{1}{u^L\,Q^-Q^+}\Psi\left( Q^+\,U\right)\right)\,,
\ee
where $\Phi_1$ and $\Phi_2$ are $i$-perodic functions. 

An important feature of \eqref{mu2generic} is that it does not generate poles at the (shifted) positions of the Bethe roots despite the $Q^-$ and $Q^+$ in the denominators. This can be understood explicitly from the action of $\Psi(f)=\sum_{n=0}^{\infty}f^{[2n]}$, by observing that the poles from $\frac 1{Q^-}$ are cancelled by the overall prefactor, and that the poles from $\left(\frac 1{Q^-\,\cdot\,\ldots}\right)^{[2n+2]}$ are cancelled by poles from $\left(\frac 1{\ldots\,\cdot\, Q^+}\right)^{[2n]}$ because $Q$ satisfies the Bethe equations. This feature of Bethe root cancellation is essential in proving that the $\bP\mu$-system is consistent with the requirement of regularity.

For computational purposes we will rewrite \eqref{mu2generic} in a form that is explicitly free of $Q$'s in the denominator. Commence by rewriting \eqref{homosol} as
\begin{align}\label{homoin}
\Psi\left(\frac{1}{u^L\,Q^-Q^+}\right)=\Psi\left(\frac 1{u^L}\left(\frac{A}{Q^-}+\frac{B}{Q^+}\right)\right)=\frac{A}{u^L Q^-}+\Psi\left(\frac{A^+(u^-)^L+B^-(u^+)^L}{(u^-u^+)^L\,Q}\right)^{+}\,,
\end{align}
where $A$ and $B$ are polynomials of degree $S-1$ uniquely defined by 
\be\label{defAB}
{A}\,{Q^+}+{B}\,{Q^-}=1\,.
\ee 
From this definition and the Baxter equation \eqref{Baxter},
\be
(u^+)^L\,Q^{[+2]}+(u^-)^L\,Q^{[-2]}=T\,Q\,,
\ee
it follows that $A^+(u^-)^L+B^-(u^+)^L=Q\,R$, where $R$ is a polynomial (of degree at most $L-2$). Therefore one has
\be\label{ABR}
\frac{A^+(u^-)^L+B^-(u^+)^L}{(u^-u^+)^L\,Q}=\frac{R}{(u^-u^+)^L}\equiv\sum_{k=1}^{L}\left(\frac{r_{k,+}}{(u^+)^{k}}+\frac{r_{k,-}}{(u^-)^{k}}\right)\,,
\ee
where $r_{k,\pm}$ are constants. Finally, notice that
\be\label{defC}
\frac{A}{u^L}=Q^-\sum_{k=1}^L\frac{r_{k,+}}{u^k}+C\,,
\ee 
where $C(u)$ is a polynomial. Therefore one gets
\be
Q^-\,\Psi\left(\frac{1}{u^L\,Q^-Q^+}\right)=C+Q^-\sum_{k=1}^{L}\left({r_{k,+}+r_{k,-}}\right)\eta_{k}\,.\label{homofinal}
\ee
By repeating a similar logic, the inhomogeneous solution can be written as\footnote{The sum of $r_{k,+}$ and $r_{k,-}$ can be computed in a faster way, using $\frac 1{Q^-Q^+}=\sum_{k=1}^{L}u^{L-k}(r_{k,+}+r_{k,-})+\CO(u^L)$. However, one still needs to determine $A$ to compute $C$.}
\be
Q^-f_{\rm inhomo}&=&C\,\Psi(Q^+U)+Q^-\,\Psi\left(\sum_{k=1}^{L}\frac{r_{k,+}+r_{k,-}}{u^k}\Psi(Q^+U)-C^{[2]}U\right)\,.
\ee

Note that it is not necessary to know the Bethe roots explicitly, but only the Baxter polynomial, $Q$, whose coefficients are symmetric combinations of the Bethe roots.

We therefore conclude that solving the (inhomogeneous) Baxter equation can produce poles only at $u=i\,\mathbb{Z}$ where, as discussed, the branch points collide. Hence only these singularities are allowed in the periodic functions $\Phi_i$ which, therefore, can always be written as
\be\label{Phiexpansion}
\Phi_i=\phi_{i,0}+\sum_{k=1}^{\Lambda} \phi_{i,k}\, \mathcal{P}_i, \quad \quad \mathcal{P}_i\equiv \eta_i+\bar\eta_i^{[-2]}\,.
\ee
The cut-off $\Lambda$ linearly depends on the order of iteration. The coefficients $\phi_{i,k}$, except for $\phi_{1,0}$, are fixed by the requirement that $\mu_{ab}+\mu_{ab}^{[2]}$ and $\frac{\mu_{ab}-\mu_{ab}^{[2]}}{\sqrt{u^2-4g^2}}$ have no poles on the real axis. For $\mu_4$, the coefficient $\phi_{1,0}$ is determined from the requirement that at the next order $\ps_4$ has no term of order $u^{-1}$, as prescribed by the way H-symmetry is fixed. As described in section \ref{p3pt1}, the singular part of $\ps_4$ at order $n+1$ can be found immediately when $\mu_4$ is known at order $n$. For $\mu_1$, the coefficient $\phi_{1,0}$ is fixed differently, as it will be explained later.

At the leading order, all poles should cancel producing a polynomial answer for $\mu_4$. $\mu_5$ is then found from e.g. the first equation in \eqref{mu45}, and thus all $\mu_{ab}$ have been fixed at the leading order. They all scale as $g^{-L}$ when $g\to 0$, as it follows from \eqref{alpha}.

\subsection{Higher loops}
Having determined all quantities of the $\bP\mu$-system at the leading order, including the one-loop correction to $\Delta$, the next step is to look at perturbations around this solution. Each order in the perturbative expansion is determined through precisely the same iterative algorithm.

\subsubsection{Double scaling \{$\ps_1$, $\ps_2$; singular part of $\tilde\ps_1/u^2$, $\tilde\ps_2/u^2$\}}
\label{sec:singpt1pt2}
The limit $g\to 0$ with $u$ fixed is referred to as {\it normal scaling}. Unless otherwise specified, this is the scaling considered. In this regime, the quantities of the $\bP\mu$-system can be parametrized as a series in $g^2$ with $u$-dependent coefficients, e.g.
\be
\ps_a=\ps_{a,{\rm ns,0}}(u)+g^2\,\ps_{a,{\rm ns},1}(u)+\CO(g^4)\,.
\ee
Though appropriate in most of the computations, the normal scaling cannot be used when computing the analytic continuation of $\bP_a$ around one of its branch points,  because the branch points collide in this scaling\footnote{The issue is not an obstacle for $\mu_{ab}$ due to the property $\tilde\mu_{ab}=\mu_{ab}^{[2]}$.}. To keep the cut non-vanishing at weak coupling, one considers the {\it double scaling}: $g\to 0$, $u\to 0$ with $x$ held constant. Obviously, the expansion \eqref{p1p2ansatz} is well-suited for the double scaling and, even more, it suggests the introduction of another useful parameter: $y=\frac{g}{x}$. Using a slightly imprecise terminology, we refer to the expansion
\be\label{psds}
\ps_a=\ps_{a,{\rm ds},0}(y)+g^2\,\ps_{a,{\rm ds},1}(y)+\CO(g^4)\,
\ee
as {\it double scaling}. For our purposes, it is enough to consider $a=1,2$.

The leading terms ${\bf p}_{a,{\rm ds},0}$ are universal and can be read off from \eqref{p1p2ansatz}:
\be\label{pdslead}
{\bf p}_{1,{\rm ds},0}=y\,,\ \ \ {\bf p}_{2,{\rm ds},0}=1\,.
\ee
Subleading terms are given by the infinite series 
\be\label{eqinfa}
{\bf p}_{a,ds,n}=\sum_{k=2}^\infty c_{a,k}^{(n)}y^k\,,
\ee
where we used the notation
\be
c_{a,k}(g)=c_{a,k}^{(0)}+g^2\,c_{a,k}^{(1)}+\ldots\,.
\ee

The expansion \eqref{psds} allows linking the quantities on the physical and the next-to-physical Riemann sheets. Indeed, by substituting $y=\frac {g}{x}$ and re-expanding \eqref{psds} at constant $u$, the normal scaling expansion of $\ps_a$ is generated. If one instead substitutes $y=\frac{g}{\tilde x}=gx$ and re-expands \eqref{psds}, one generates the normal-scaling expansion of $\tilde\ps_a$. Note that $y=\frac{g^2}{u}+\CO(g^4)$ on the physical sheet, so only a finite number of the terms in the sum \eqref{eqinfa} is needed for computing a given order of the normal scaling expansion. On the contrary, the whole infinite sum should be known to compute $\tilde \ps_{a,{\rm ns},n}$.

Denote by $\tilde\ps_{a,{\rm ns},n}'$ the $n$-th term in the normal scaling re-expansion of the truncated sum $\sum\limits_{k=0}^{n-1}\ps_{a,{\rm ds},k}(g\,x)g^{2k}$. It is a useful object since on one hand, computing $\tilde\ps_{a,{\rm ns},n}'$ requires one term less in the double scaling series compared to computing $\tilde\ps_{a,{\rm ns},n}$, while on the other hand, $\tilde\ps_{a,{\rm ns},n}'$  and $\tilde\ps_{a,{\rm ns},n}$ have the same singular, constant, and linear parts in their small $u$ expansion.  Indeed, the sum \eqref{eqinfa} starts from the $y^2$-term whereas on the next-to-physical sheet one has $y=g\,x=u-\frac{g}{x}=u-\frac {g^2}{u}(1+\CO(\frac{g^2}{u^2}))$. Hence all singularities in $\ps_a/u^2$ at small $u$ are delayed by at least one loop.

In the algorithm, $\tilde \ps_{a,{\rm ns},n-1}$ is taken as an input known from the previous orders of the recursion. First, we produce $\ps_{a,{\rm ds},n-1}$ by\footnote{Let us emphasise that $u$ is not treated as a function of $y$  on the r.h.s. of \eqref{eq74}, but literally each $u$ is replaced by $y$.
}
\be\label{eq74}
\ps_{a,{\rm ds},n-1}(y)=\tilde\ps_{a,{\rm ns},n-1}(u\to y)-\tilde\ps_{a,{\rm ns},n-1}'(u\to  y)\,.
\ee
The leading order is slightly different: $\ps_{a,\text{ds},0}(y)=\tilde{\ps}_{a,\text{ns},0}(u\to y)$ is given by \eqref{pdslead}. 

Then, from the knowledge of $\ps_{a,{\rm ds},k}$ up to $k=n-1$, we compute $\tilde\ps_{a,{\rm ns},n}'$ and expand it at small $u$, thus determining the singular, linear, and constant parts of $\tilde\ps_{a,{\rm ns},n}$. 

Finally, we substitute $y=\frac gx$ and re-expand the double scaling series \eqref{psds} at fixed $u$ producing $\ps_{a,{\rm ns},k}$ on the physical sheet. Since $y=\frac{g^2}{u}+\CO(g^4)$ in this expansion, and the sum \eqref{eqinfa} starts from the $y^2$-term, we can produce $\ps_{a,{\rm ns},k}$ up to $k=n+1$, though $k=n$ is already sufficient to proceed in the algorithm.

\subsubsection{\{$\ps_3$, $\tilde\ps_1$, $\mu_1$, $\mu_2$\}}
In \eqref{p3p4} $\ps_3$ is explicitly separated into the regular polynomial part $A_3\,u\,M_{L-2}$, and the singular (near $u=0$) part obtained from the normal scaling expansion of $\sum\limits_{k=1}^{\infty}c_{3,k}(g)\,(\frac{g}{x})^k$.

To find the singular part, consider \eqref{secondRH} for $\tilde\bP_1$ in the form
\be\label{eq67}
x^{L}\,\tilde\ps_1={\rm PV}[\mu_3]\,\ps_1-{\rm PV}[\mu_2]\,\ps_2+{\rm PV}[\mu_1]\,\ps_3\,,
\ee 
where \eqref{ortomu} was used to replace $\mu$ by ${\rm PV}[\mu]\equiv \frac 12(\mu+\mu^{[2]})$. ${\rm PV}[\mu]$ are regular functions in the vicinity of the real axis.  ${\rm PV}[\mu_1]^{-1}$ is regular, at least perturbatively at any order, since  ${\rm PV}[\mu_1]=(\alpha Q(\frac{i}{2})+\CO(u))+\CO(g^2)$. Finally all $\ps_a$ are regular at the leading order. Hence, to find the order $n$ singular part of $\ps_3$, only knowledge about $\mu_{ab}$ at order $n-1$ is needed. $\ps_1$, $\ps_2$ and $\tilde \ps_1$, more precisely their singular parts, are required at order $n$, and these quantities were found in the previous step.

In the case of $L=2$, knowing the singular part is in principle enough to fully fix $\ps_3$. Indeed, the regular part is explicitly known: $\ps_{3,{\rm reg}}=A_3\,u$, where $A_3$ is fixed from \eqref{AB} by using $\Delta$ from the previous loop order and requiring that the spin $S$ is integer and hence known at all orders. This simplification reflects the fact that a twist 2 state is fully determined by its spin. For instance, there is only one solution to the one-loop Bethe equation. 

For generic $L$, or to avoid the assumption that $S$ is integer but rather derive this fact perturbatively, one should fix $A_3$ and the polynomial $M_{L-2}$ from the QSC equations. 

First, we will determine $\mu_1$ with $A_3$ and $M_{L-2}$ kept arbitrary, by considering the all-loop version of the Baxter equation \eqref{Baxter}:
\be\label{BaxterFull}
\frac{1}{\bP_2^2}\,\mu_1-\left(\frac{\bP_3}{\bP_2}-\frac{\bP_3^{[2]}}{\bP_2^{[2]}}+\frac 1{\bP_2^2}+\frac 1{\left(\bP_2^{[2]}\right)^2}\right)\mu_1^{[2]}+\frac 1{\left(\bP_2^{[2]}\right)^2}\,\mu_1^{[4]}
\no\\
=\frac{\bP_1^{[2]}\tilde\bP_2^{[2]}}{\left(\bP_2^{[2]}\right)^2}-\frac{\bP_1\tilde\bP_2}{\bP_2^2}
+\left(\frac{\bP_1}{\bP_2}-\frac{\bP_1^{[2]}}{\bP_2^{[2]}}\right)\mu_3^{[2]}\,.
\ee
It is most easily derived from \eqref{secondRH} for $\tilde\bP_1$ written in two forms, with $\mu_{ab}$ and with $\mu_{ab}^{[2]}$, and from \eqref{firstRH} for $\mu_{12}$. One should eliminate $\tilde\bP_1$ and $\mu_2$ from these equations to get \eqref{BaxterFull}.

The normal scaling expansion is considered for $\mu_{ab}$,
\be
\mu_{ab}=\frac 1{g^{L}}\left(\mu_{ab,{\rm ns},0}+g^2\mu_{ab,{\rm ns},1}+\CO(g^4)\right)\,.
\ee
Expand \eqref{BaxterFull} to order $n$ and consider it as an equation for $\mu_{1,{\rm ns},n}$. The result is precisely the inhomogeneous Baxter equation \eqref{Baxtermu2} with a source term, which is solved by \eqref{mu2generic}. 

The r.h.s. of \eqref{BaxterFull} is proportional to $\bP_1$ which is zero at the leading order, so $\tilde\bP_2$ and $\mu_3$ should be known only up to order $n-1$. $\bP_1$, $\bP_2$ and the singular part of $\bP_3$ are already known at order $n$, so only $M_{L-2}$ and $A_3$ are yet unfixed, and thus $\mu_{1,{\rm ns},n}$ can be found in terms of these coefficients. The constants $\phi_{i,k}$ are fixed as explained after \eqref{Phiexpansion}, with the exception of $\phi_{1,0}$ which is fixed differently for $\mu_1$ (compared to the prescription for $\mu_4$).

To fix $\phi_{1,0}$ and $M_{L-2}$, return to \eqref{firstRH} for $\mu_{12}$:
\be
\mu_{1}-\mu_{1}^{[2]}=g^{-L}\left(\tilde\ps_1\ps_2-\tilde\ps_2\ps_1\right)\,.
\ee
On the l.h.s., the singular and constant terms depend on $A_3$ and $M_{L-2}$, while the linear term depends on $\phi_{1,0}$. On the other hand, the necessary information to compute these terms on the r.h.s. was already found in \secref{sec:singpt1pt2}. Hence we can fully determine $\phi_{1,0}$, $A_3$, and 
$M_{L-2}$. In addition, the regular part of $\tilde \ps_1$ is found by matching to the regular part of the l.h.s.

Having fully determined $\mu_1$, $\ps_3$ and $\tilde\ps_1$ at order $n$, it is now straightforward to fix $\mu_2$ from e.g. the following version of \eqref{secondRH}:
\be
x^{L}\,\tilde\ps_1=\mu_3\ps_1-\mu_2\ps_2+\mu_1\ps_3\,.
\ee

\subsubsection{\{$\ps_4$, $\tilde\ps_2$, $\mu_3$, $\Delta$\}} \label{p3pt1}
The way to fix $\ps_4$ and $\tilde\ps_2$ is very similar to how $\ps_3$ and $\tilde\ps_1$ was fixed in the previous section. The departing point is the equation \eqref{secondRH} for $\tilde\ps_2$ written in two ways:
\be\label{eq72}
x^{L}\,\tilde\ps_2&=&\mu_4\ps_1-\mu_3\ps_2+\mu_1\ps_4\,,
\no\\
x^{L}\,\tilde\ps_2&=&\mu_4^{[2]}\ps_1-\mu_3^{[2]}\ps_2+\mu_1^{[2]}\ps_4\,.
\ee
In analogy to \eqref{eq67}, taking the sum of these equations produces a relation that includes ${\rm PV}[\mu]$, which can be used to fix the singular part of $\ps_4$. To fix the regular part, $A_4\,u\,N_{L-1}(u)$, $\mu_3$ is eliminated from \eqref{eq72}:
\be
\frac{\tilde\bP_2}{\bP_2}-\frac{\tilde\bP_2^{[2]}}{\bP_2^{[2]}}= \left(\frac{\bP_4}{\bP_2}-\frac{\bP_4^{[2]}}{\bP_2^{[2]}}\right)\mu_1^{[2]} + \left(\frac{\bP_1}{\bP_2}-\frac{\bP_1^{[2]}}{\bP_2^{[2]}}\right)\mu_4^{[2]}\,.
\ee
Applying the $\Psi$-operator to this equation yields
\begin{eqnarray}
x^L\frac{\tilde{\ps}_2}{\ps_2}&=&\Psi\left(\left(\frac{\ps_4}{\ps_2}-\frac{\ps_4^{[2]}}{\ps_2^{[2]}}\right)\mu_1^{[2]}+\left(\frac{\ps_1}{\ps_2}-\frac{\ps_1^{[2]}}{\ps_2^{[2]}}\right)\mu_4^{[2]}\right)+\phi_{0}+\sum_{k=1}^\Lambda \phi_{k}\,\mathcal{P}_k\,.
\end{eqnarray}
Notice that an $i$-periodic function has to be included. $\Lambda$ is some finite number. The only unknown part of the l.h.s. is the regular part (excluding the constant and linear part) of $\tilde\ps_2$ at $u=0$, while the regular part of $\ps_4$ and the coefficients $\phi_k$ need to be fixed on the r.h.s. Considering the equation at order $n$, $\tilde \ps_{2,\text{ns},n}$ is simply multiplied by a factor of $\left(\frac{u}{g}\right)^L$. 
Dividing by this factor and matching the poles at $u=0$ on both sides fixes $\phi_k$ (from the poles of degree $L$ and higher) and the undetermined coefficients of $\ps_4$ (from the poles of degree less than $L$ and the constant term). Furthermore, the equivalence of the regular terms fixes $\tilde \ps_2$.

Having fixed $\ps_4$, $\tilde \ps_2$ and $\mu_1$, it is straightforward to fix $\mu_3$ from e.g. the first equation in \eqref{eq72}.
Finally, $\Delta$ is fixed by using the found $A_4$ in \eqref{eqEnS}.

\subsubsection{\{$\mu_4$, $\mu_5$\}}
With all $\bP_a$ and $\mu_1$, $\mu_2$, and $\mu_3$ fixed, only $\mu_4$ and $\mu_5$ remain. From the fourth and fifth equation in \eqref{mufunexplicit}, the all-loop version of \eqref{Baxtermu2} is derived by eliminating $\mu_5$ and using \eqref{secondRH} to simplify:
\begin{eqnarray}
\frac{1}{\bP_2^2}\mu_4-\left(\frac{\bP_3}{\bP_2}-\frac{\bP_3^{[2]}}{\bP_2^{[2]}}+\frac{1}{\bP_2^2}+\frac{1}{\left(\bP_2^{[2]}\right)^2}\right)\mu_4^{[2]}+\frac{1}{\left(\bP_2^{[2]}\right)^2} \mu_4^{[4]} \nonumber\\ =\frac{\tilde \bP_2 \bP_4}{\bP_2^2}-\frac{\tilde \bP_2^{[2]} \bP_4^{[2]}}{\left(\bP_2^{[2]}\right)^2}+\left(\frac{\bP_4^{[2]}}{\bP_2^{[2]}}-\frac{\bP_4}{\bP_2}\right)\mu_3^{[2]}\,. \label{mu2all}
\end{eqnarray}
Again, this is exactly the inhomogeneous Baxter equation for $\mu_{4,\text{ns},n}$. Following the procedure described above, the solution is completely fixed, except for the constant $\phi_{1,0}$. As explained, this constant is fixed when the singular part of $\ps_4$ is determined at the next order by requiring that the term of order $u^{-1}$ vanishes, as it should due to the way H-symmetry is imposed.

Finally, $\mu_5$ is straightforwardly derived from one of the equations leading to \eqref{mu2all}.

\subsection{Cross-checks}
There are several robust possibilities to cross-check the presented computation:

\begin{itemize}
\item The following equations from \eqref{secondRH} were never explicitly used:
\be
\tilde\bP_3&=&\bP_4\,\mu_2-\bP_3\,\mu_3+\bP_1\,\mu_5\,,
\no\\
\tilde\bP_4&=&\bP_4\,\mu_3-\bP_3\,\mu_4+\bP_2\,\mu_5\,.
\ee
Using them, $\tilde\bP_3$ and $\tilde\bP_4$ can be computed and compared against $\bP_3$ and $\bP_4$ through the re-expansion in the double scaling regime.

\item The fact that $\mu_{ab}+\mu_{ab}^{[2]}$ and $\frac{\mu_{ab}-\mu_{ab}^{[2]}}{\sqrt{u^2-4g^2}}$ have a regular expansion at $u=0$ was only used to fix $\mu_1$ and $\mu_4$. However, it should also apply for $\mu_2$, $\mu_3$ and $\mu_5$. 

\item The large $u$ behaviour of $\mu_{ab}$ \eqref{Plarge} is governed by $\Delta$, and this provides another way to compute the conformal dimension.

\item In the $u$-symmetric cases all $\mu_{ab}$ have certain parity properties which are preserved at all loop orders. 

\item All $\mu_{ab}$ satisfy the bilinear identity \eqref{Pf1}
\be
\mu_1\,\mu_5-\mu_2\,\mu_4+\mu_3^2=1\,.
\ee
\end{itemize}

Though all the mentioned properties can be derived from the equations that were actually used, their explicit check is very nontrivial. It requires the use of the shuffle and stuffle algebraic relations, and, in the case of parity checks, a possibility to express $\bar\eta_{a_1,\ldots, a_k}^{[-2]}$ through $\eta_{a_1,\ldots,a_r}$ and $\bar\eta_{a}^{[-2]}$ by means of the periodicity relations introduced in \cite{Leurent:2013mr}. Hence, these properties provide a very solid verification for the correctness of the implemented algorithm. In comparison, the cross-checks in the FiNLIE approach were significantly less transparent. Almost no means were available to verify the 8-loop computation in \cite{Leurent:2013mr} beyond highest-transcendentality terms.

Apart from self-consistency, there are physical checks. Indeed, the computation produces $S$ and $\Delta$ as an output, see equation \eqref{eqEnS}. $S$ should be a fixed integer, hence it provides an all-loop check of the computation\footnote{If, instead, we use the value of $S$ to simplify the algorithm, e.g. in the twist 2 case, then certain singularities of $\tilde\ps_1$ should automatically cancel which is also a nontrivial check.}.

The approximation to the conformal dimension can be computed from the solution of the asymptotic Bethe Ansatz \cite{Beisert:2005fw,Beisert:2006ez}
\be\label{asympBA}
\left(\frac{x_k^+}{x_k^-}\right)^L=-\prod_{j=1}^{S}\frac {u_k-u_j-i}{u_k-u_j+i}\left(\frac {1-\frac 1{x_k^+x_j^-}}{1-\frac 1{x_k^-x_j^+}}\,\s_{\rm BES}(u_k,u_j)\right)^2\,,
\ee
where $x_k^{\pm}\equiv x(u_k\pm \frac{i}{2})$, using the formula
\be\label{Deltaas}
\Delta_{\rm as}=L+S+2\,i\,g\sum_{k=1}^{S}\left(\frac 1{x_k^+}-\frac 1{x_k^-}\right)\,.
\ee
Naively, Feynman graphs which invalidate the assumption of infinite length appear at $L$ loops, so the asymptotic Bethe Ansatz gives the correct result for $\Delta$ up to $L-1$ loops only. However, in practice the Bethe Ansatz is still valid up to $L+1$ loops. We discuss this bonus effect in \appref{sec:Konishieffect}.

Note that the asymptotic Bethe Ansatz \eqref{asympBA}, the expression \eqref{Deltaas} for $\Delta$, and the integer constraint on the value of $S$ can be analytically derived from QSC \cite{Gromov:2014caa}. However, the only known way of derivation is to consider the curve in its full generality, i.e. to supplement the  $\bP\mu$-system  with the $\mathbf{Q}\omega$-system (which is a consequence of the former) and with the intertwining $QQ$-relations between $\bP_a$ and $\mathbf{Q}_i$. The fact that this information about $\Delta$ and $S$ follow from the $\bP\mu$-system is definitely a nontrivial check for the explicit computations.

There are also several results within single-wrapping orders \cite{Bajnok:2008qj,Lukowski:2009ce,Velizhanin:2010cm}, which  are consistent by themselves with the reciprocity property, BFKL, and double-logarithmic equations (see e.g. \cite{Velizhanin:2013vla}). We checked our computation against these results.

Apart from perturbative comparison, we can do further checks on the generic structure of the answer. In \cite{Leurent:2013mr}, the leading transcendentality terms were computed to all loops for the anomalous dimension of the Konishi operator. Also, one can consider a plausible suggestion  \cite{Schnetz} that the answer is given, at least to a high enough order, in terms of single-valued MZV's \cite{Brown:2013gia,Schnetz:2013hqa}. When an answer satisfies this conjecture (and so far all answers do), one gets a solid verification of the result since the single-valued MZV's are very special combinations of generic MZV's which we cannot  predict in advance from our algorithm. 

Finally, let us note that the equations in the algorithm are universal at any loop order, hence their implementation is. This universal implementation can be thoroughly verified as described above, which gives us an additional certitude about the correctness of the computation.

\section{\label{sec:results}Summary of results and discussion}\label{sec:summaryofresults}

\subsection{Results}
As an example of the structure of the obtained results, the 10-loop conformal dimension of the Konishi operator is:
{\footnotesize
\be
\Delta &=&
4
+12g^2
-48g^4+336g^6+g^8\big(-2496+576\,\zeta_3-1440\,\zeta_5\big)\no\\&&
+g^{10}\big(15168 + 6912 \,\zeta_3 - 5184\,\zeta_3^2  - 8640 \,\zeta_5 + 30240 \,\zeta_7\big)\no\\&&
+g^{12}\big(-7680 - 262656 \,\zeta_3 - 20736 \,\zeta_3^2  + 112320\,\zeta_5 + 155520 \,\zeta_3 \,\zeta_5 + 75600\,\zeta_7 - 489888 \,\zeta_9\big)\no \\&&
+g^{14}\big(-2135040 + 5230080\,\zeta_3 - 421632 \,\zeta_3^2  + 124416 \,\zeta_3^3  - \
229248 \,\zeta_5 + 411264 \,\zeta_3 \,\zeta_5 \no\\&& \indent - 993600 \,\zeta_5^2 - 1254960 \,\zeta_7 - 1935360 \,\zeta_3 \,\zeta_7 - 835488 \,\zeta_9 + 7318080 \,\zeta_{11}\big) \no\\&&
+g^{16}\bigg(54408192  - 83496960\,\zeta_3 + 7934976 \,\zeta_3^2  + 1990656 \,\zeta_3^3  - 19678464 \,\zeta_5 - 4354560 \,\zeta_3 \,\zeta_5 \no\\&&\indent - 3255552 \,\zeta_3^2  \,\zeta_5 + 2384640 \,\zeta_5^2  + 21868704 \,\zeta_7 - 6229440 \,\zeta_3 \,\zeta_7 + 22256640 \,\zeta_5 \,\zeta_7  \no\\&& \indent + 9327744\,\zeta_9 + 23224320 \,\zeta_3 \,\zeta_9  + \frac{65929248}{5}\,\zeta_{11}-106007616 \,\zeta_{13}-\frac{684288}{5}\,Z_{11}^{(2)}\bigg) \no\\&&
+g^{18}   \bigg(-1014549504 + 1140922368 \,\zeta_3 - 51259392 \,\zeta_3^2  -20155392 \,\zeta_3^3  + 575354880 \,\zeta_5 \no\\&&\indent - 14294016 \,\zeta_3 \,\zeta_5 - 26044416 \,\zeta_3^2  \,\zeta_5 + 55296000 \,\zeta_5^2  + 15759360 \,\zeta_3 \,\zeta_5^2  - 223122816 \,\zeta_7 \no\\&& \indent + 34020864 \,\zeta_3 \,\zeta_7 + 22063104 \,\zeta_3^2 \,\zeta_7  - 92539584 \,\zeta_5 \,\zeta_7 - 113690304 \,\zeta_7^2  - 247093632 \,\zeta_9 \no\\&&\indent + 119470464 \,\zeta_3 \,\zeta_9 - 245099520 \,\zeta_5 \,\zeta_9 -\frac{186204096}{5}\,\zeta_{11}  -278505216 \,\zeta_3 \,\zeta_{11} - 253865664 \,\zeta_{13} \no\\&&\indent + 1517836320\,\zeta_{15}+ \frac{15676416}{5}\,Z_{11}^{(2)}- 1306368 \,Z_{13}^{(2)} + 1306368 \,Z_{13}^{(3)}\bigg) \no\\&&
+g^{20}   \bigg(16445313024 - 13069615104\,\zeta_3  - 1509027840 \,\zeta_3^2 + 578949120\,\zeta_3^3 \no\\&& \indent - 14929920 \,\zeta_3^4  - 11247547392 \,\zeta_5 + 1213581312\,\zeta_3 \,\zeta_5 + 1234206720 \,\zeta_3^2 \,\zeta_5 \no\\&&\indent - 70170624 \,\zeta_3^3\,\zeta_5 - 1390279680 \,\zeta_5^2  - 654842880\,\zeta_3\,\zeta_5^2 + \frac{6966252288}{175}\,\zeta_5^3 \no\\&& \indent+ 377212032 \,\zeta_7 - 1610841600 \,\zeta_3 \,\zeta_7 + 154680192 \,\zeta_3^2 \,\zeta_7  + 222341760 \,\zeta_5 \,\zeta_7\no\\&& \indent + 133788672\,\zeta_3 \,\zeta_5 \,\zeta_7 + 868662144 \,\zeta_7^2 + 4915257984 \,\zeta_9 - 332646912 \,\zeta_3 \,\zeta_9  \no\\&& \indent - 91072512 \,\zeta_3^2 \,\zeta_9 + 1099699200\,\zeta_5 \,\zeta_9 + 2275620480 \,\zeta_7 \,\zeta_9 + \frac{9793211904}{5}\,\zeta_{11} \no\\&&\indent
- 2334572928 \,\zeta_3 \,\zeta_{11}  + 2713772160 \,\zeta_5 \,\zeta_{11} - \frac{787483944}{175}\,\zeta_{13}  + 3372969600 \,\zeta_3 \,\zeta_{13}\no\\&&\indent -\frac{4308536566944}{875}\,\zeta_{15} - 21661960320 \,\zeta_{17} 
+\frac{752219136}{5}\,Z_{11}^{(2)}
-\frac{5070791808}{175}\,Z_{13}^{(2)}\no\\&&\indent
-\frac{7159104}{7}\,Z_{13}^{(3)} 
+\frac{2716063488}{175}\,Z_{15}^{(2)}
-\frac{17895168}{25}\,Z_{15}^{(3)}
+11943936\,\zeta_3 \,Z_{11}^{(2)}\bigg)+\CO(g^{22})\,,
\ee}
where $Z_a^{(n)}$ denote single-valued MZV's written in the basis \cite{SchnetzPrivate}
{\footnotesize
\be\label{SVMZV}
Z_{11}^{(2)}&=&-\zeta_{3,5,3}+\zeta_3\,\zeta_{3,5}\no\,,\\
Z_{13}^{(2)}&=&-\zeta_{5,3,5}+11\,\zeta_5\,\zeta_{3,5}+5\,\zeta_5\,\zeta_8\no\,,\\
Z_{13}^{(3)}&=&-\zeta_{3,7,3}+\zeta_3\,\zeta_{3,7}+12\,\zeta_5\,\zeta_{3,5}+6\,\zeta_5\,\zeta_8\no\,,\\
Z_{15}^{(2)}&=&\zeta_{3,7,5}-\zeta_5\,\zeta_{3,7}-3\,\zeta_5\,\zeta_{10}+21\,\zeta_9\,\zeta_6+\frac{175}{2}\,\zeta_{11}\,\zeta_4+\frac{637}{2}\,\zeta_{13}\,\zeta_2\no\,,\\
Z_{15}^{(3)}&=&-\zeta_{3,9,3}+\zeta_3\,\zeta_{3,9}+12\,\zeta_5\,\zeta_{3,7}+30\,\zeta_7\,\zeta_{3,5}+6\,\zeta_5\,\zeta_{10}+15\,\zeta_7\,\zeta_8\,.
\ee}

Selected examples of 10-loop results are given in table \ref{table:ResRat}. The corresponding operators all have Baxter polynomials with rational coefficients. Operators associated to Baxter polynomials with square roots of a prime in the coefficients have also been calculated analytically, up to nine loop orders for the simplest cases. Examples are given in table \ref{table:ResSqrt}. In the case of more complicated algebraic numbers the analytic solution is also possible. However, our attempts of an implementation are slow and maximally five-loop results have been reached. Therefore, the coefficients of $Q(u)$ have been handled numerically while MZV's were kept analytic. Nine loops were reached in the simplest cases of such operators. Examples are given in table \ref{table:ResNon}.

The {\texttt{Mathematica}} notebook {\texttt{Results.m}} includes results for all 91 operators with $L+S\le 10$ to at least eight loop orders as well as some additional results for $L+S > 10$ where the loop order exceeds $L+1$ such that the asymptotic Bethe Ansatz is no longer valid. The notebook {\texttt{Solution of QSC.nb}} contains our implementation of the algorithm and the reader can use it to attempt calculations of the operators and loop orders that we did not cover. The code works efficiently for operators with $L+S\lesssim 15$ and also beyond this range if $S$ or $L$ is small. The published version works analytically for operators with rational coefficients in the Baxter polynomial and semi-numerically when the coefficients are irrational. The modification which works analytically with some irrational expressions is not published but available upon request. The notebooks and the required files containing relations between MZV's can be downloaded from \href{http://www.maths.tcd.ie/~dvolin/QSC/loop10sl2.zip} {www.maths.tcd.ie/$\sim$dvolin/QSC/loop10sl2.zip}, they are also available as the ancillary files of the electronic preprint of this article at arxiv.org. 

\renewcommand{\arraystretch}{1.5}


\subsection{Observations}

\paragraph{\bf Series converge up to $g=\frac{1}{4}$} There is a clear numerical evidence that the radius of convergency for all the obtained series is $\frac{1}{4}$, cf.
\be
\Delta_{\rm Konishi}&=&4+0.7500(4g)^2-0.1875(4g)^4+0.08203(4g)^6-0.05031(4g)^8
\no\\
&&+0.03578(4g)^{10}-0.02728(4g)^{12}+0.02175(4g)^{14}-0.01791(4g)^{16}
\no\\
&&+0.01511(4g)^{18}-0.01299(4g)^{20}+\CO(g^{22})\,.
\ee
This radius of convergency is the same as for the asymptotic Bethe Ansatz answers and is well expected. Indeed, one can envisage such a bound already from the magnon dispersion relation \cite{Santambrogio:2002sb} which becomes singular at $g=\pm\frac i4$. On the level of functions of the spectral parameter, $g=\pm\frac i4$ are the first values of $g$ where Zhukovksy branch points collide, see e.g. discussion in \cite{Volin:2008kd}.

\paragraph{\bf Pad\'e approximation works at least up to $g\simeq 0.7$} One can attempt to resolve the singularities at $g=\pm \frac i4$ by introducing the new variable $w=(1+16g^2)^\alpha$, where the value of $\alpha$ should account for the type of singularity (we assumed it is of branch point type).  We introduced $w$ and constructed diagonal Pad\'e approximations, around $w=1$, to the perturbative answers. We observe empirically that the obtained Pad\'e approximants converge up to $g\simeq 0.7$ independently of $\alpha$, and for $\alpha=1/4$ we get
\begin{table}[t]
\begin{center}
\begin{tabular}{|c|c|l|c|c|c|}
\hline
$L$ & $S$ & $Q$ & $\Delta$ & $\Delta_{\text{num}}$ & $\Delta_{\text{Pad\'e}}$ \\ \hline
2 & 2 & $u^2-\frac{1}{12}$ & \hyperref[sec:22]{\#} & \hyperref[sec:22]{\#} & \hyperref[sec:22]{\#} \\ \cline{2-6}
& 4 & $u^4-\frac{13}{14}u^2+\frac{27}{560}$ &\hyperref[sec:24]{\#} & \hyperref[sec:24]{\#}& \hyperref[sec:24]{\#}\\ \cline{2-6}
& 6 & $u^6-\frac{155}{44}u^4+\frac{329}{176}u^2-\frac{375}{4928}$ &\hyperref[sec:26]{\#} & \hyperref[sec:26]{\#}&\hyperref[sec:26]{\#} \\ \hline
3 & 2 &$u^2-\frac{1}{4}$ & \hyperref[sec:32]{\#}&\hyperref[sec:32]{\#} &\hyperref[sec:32]{\#} \\ \cline{2-6}
& 4 & $u^4-\frac{3}{2}u^2+\frac{11}{48}$ &\hyperref[sec:34]{\#} &\hyperref[sec:34]{\#} & \hyperref[sec:34]{\#}\\\hline
4& 3 & $u^3\pm\frac{3}{2}u^2+\frac{1}{4}u\mp\frac{1}{8}$ & \hyperref[sec:43]{\#}& \hyperref[sec:43]{\#}& \hyperref[sec:43]{\#}\\ \hline
5& 2 & $u^2-\frac{3}{4}$ & \hyperref[sec:521]{\#}& \hyperref[sec:521]{\#}&\hyperref[sec:521]{\#} \\ \cline{3-6}
&& $u^2-\frac{1}{12}$ & \hyperref[sec:522]{\#}& \hyperref[sec:522]{\#}& \hyperref[sec:522]{\#}\\ \hline
\end{tabular}
\caption{Results for operators with $L+S\le 8$ associated to Baxter polynomials with rational coefficients.}
\label{table:ResRat}
\end{center}
\begin{center}
\begin{tabular}{|c|c|l|c|c|c|}
\hline
$L$ & $S$ & $Q$ & $\Delta$ & $\Delta_{\text{num}}$ & $\Delta_{\text{Pad\'e}}$  \\ \hline
4 & 2 & $u^2-\frac{1}{4}-\frac{1}{2\sqrt{5}}$ & \hyperref[sec:421]{\#}& \hyperref[sec:421]{\#}&\hyperref[sec:421]{\#} \\[0.5mm] \cline{3-6}
&  & $u^2-\frac{1}{4}+\frac{1}{2\sqrt{5}}$ & \hyperref[sec:422]{\#}& \hyperref[sec:422]{\#}& \hyperref[sec:422]{\#} \\[0.5mm] \hline
\end{tabular}
\caption{Examples of results for operators associated to Baxter polynomials with a square root of a prime in the coefficients.}
\label{table:ResSqrt}
\end{center}
\end{table}
\begin{table}[h!]
\begin{center}
\begin{tabular}{|c|c|l|l|c|c|c|}
\hline
$L$ & $S$ & $Q$ & $Q_{\text{num}}$ & $\Delta$ & $\Delta_{\text{num}}$ & $\Delta_{\text{Pad\'e}}$  \\ \hline
6 & 2 & $u^2-\frac{1}{4}\cot^2\left(\frac{\pi}{7}\right)$ & $u^2 - 1.07799$ & \hyperref[sec:621]{\#} & \hyperref[sec:621]{\#} & \hyperref[sec:621]{\#} \\ \cline{3-7}
&  & $u^2-\frac{1}{4}\cot^2\left(\frac{2\pi}{7}\right)$ & $u^2-0.158991$ & \hyperref[sec:622]{\#} & \hyperref[sec:622]{\#} & \hyperref[sec:622]{\#} \\ \cline{3-7}
&  & $u^2-\frac{1}{4}\cot^2\left(\frac{3\pi}{7}\right)$ & $u^2-0.0130238$ & \hyperref[sec:623]{\#} & \hyperref[sec:623]{\#} & \hyperref[sec:623]{\#}\\ \hline
\end{tabular}
\caption{Examples of results for operators associated to Baxter polynomials with more complicated algebraic numbers as coefficients.}
\label{table:ResNon}
\end{center}
\end{table}
\hspace{-1em} the best matching against the known numerical results from TBA \cite{Gromov:2009zb,Frolov:2010wt,Frolov:2012zv}: In the described way, a three-digit accuracy was achieved at $g=0.7$. For $g<0.4$, the accuracy from the Pad\'e approximants is empirically estimated to be more than 5 digits and is hence better than that of the known numerical results.

Although $\alpha=\frac{1}{4}$ is favored by comparison with the numerics, it is premature to conclude that this is the true value of the critical exponent. Other values of $\alpha$ still lead to numerically reasonable answers and attempts to fit the exponent by analyzing solutions near $g=\frac{i}{4}$ were inconclusive.

\ \\
{\bf Bethe Ansatz works up to $L+1$ loops,} although naive expectation would be $L-1$ loops. This is a well-known fact for the Konishi operator. We observe this phenomenon for any state from the $\sl(2)$ sector. It has an explanation, see \appref{sec:Konishieffect}.

\paragraph{\bf Only MZV's at any loop order (theorem)} It is a straightforward consequence of the algorithm that the answer at any loop order is expressible only in terms  algebraic numbers (originating from coefficients of Baxter polynomials) times MZV's. Indeed, the algebra of functions used is closed under all operations performed, see \appref{sec:Psiresum}. MZV's appear when we Taylor-expand $\eta$-functions at zero. Nothing else can be generated by this expansion.

\paragraph{\bf Only single-valued MZV's (observation)} We observe that in all computations that were done, the answer is expressible using only the subclass of possible MZV's -- the so-called single-valued MZV's \cite{Brown:2013gia,Schnetz:2013hqa}. This subclass consists of single-indexed zeta-values of odd argument and particular combinations of multi-indexed zeta-values listed, for transcendentally up to 15, in \eqref{SVMZV}. We do not have an analytic explanation of this fact. 

For twist-two states, the complexity of the answer seems to follow the number of wrappings: single-indexed zeta-values appear for the first time at loop four, and the first multiple-index (but single-valued) zeta-value appears at loop eight\footnote{We do not observe similar correlation for arbitrary states. Twist-two states might be special because zeta-values from the dressing phase appear at the same order as the first wrapping.}. It is therefore reasonable to ask whether "single-valuedness" will be preserved at triple wrapping, i.e. at twelve loops.

\subsection{Outlook}
The results presented in this work seem to put us on the eve of the practical systematic computation of the perturbative conformal spectrum of planar $\CN=4$ SYM. Let us discuss how the proposed approach can be extended beyond the $\sl(2)$ sector. One relies on the ${\bf P_1}\to 0$ property to find the leading order solution, see \secref{sec:lpone}. It is satisfied for all the multiplets that reach the unitary bound at zero coupling. For instance, it is satisfied for the most interesting case of rank-one sectors, including the $\su(2)$ sector. If the ${\bf P_1}\to 0$ property is fulfilled, we expect that relaxing the ansatz \eqref{scalingansatz} would be sufficient to generalise the algorithm, however some practical issues may arise, such as $\mu_1$ being zero at $u=0$ which indeed happens for exceptional operators \cite{Arutyunov:2012tx} and which will require extra care when defining the small $u$ expansions.

For the operators that do not reach the unitary bound, the iterative scheme is not directly applicable. However, we know \cite{Gromov:2014caa} how the quantum spectral curve is related to the asymptotic Bethe Ansatz  when the latter one is applicable, hence one can compute the leading order and certain sub-leading orders of the QSC quantities by solving the Bethe Ansatz and use this information as a departing point. Most importantly, it is very likely that algebraic manipulations would not be more complicated than those described in this work and all the functions will belong to the algebra from \secref{sec:algebra}, hence one has all technical tools ready.

The presented number of loop orders, ten, is not a conceptual limit of the procedure. The algorithm works to any order, and restrictions are of purely combinatorial nature. The basis of $\eta$-functions is of dimension $2^{2\cdot\# loops-3}$, and hence the required memory and time for computation are growing (at least) exponentially. Even with this exponential growth, the computation is very fast, allowing to compute ten loops for simple operators in a matter of hours. The algorithm is implemented in {\texttt{Mathematica}}, with the aim to rather be comprehensible than fast, and it can, without doubt, be made several orders of magnitude faster and less memory-consuming with a proper low-level implementation. Also, a single core was used, however all the time-consuming operations are linear or bi-linear and can hence be efficiently parallelized. Thus, we expect several loops more to be computable by improving the code, using more advanced hardware, and allowing longer runtimes.

One should keep in mind that most of the advancements in the study of the $\mathcal{N}=4$ SYM spectrum, including the presented one, rely on the conjecture of integrability which was not proven and which still appears as a miracle. As a step towards a proof, one should devise a better way of deriving the QSC equations than the historical approach through TBA. Hoping for an analogy with the algebraic Bethe Ansatz \cite{Faddeev:1996iy}, one expects that ${\bf P}_a$ and ${\mu_{ab}}$ arise as commuting operators acting on the basis \eqref{genericoperator} (or a more generic one, if outside the $\sl(2)$ sector). The eigenvalues of these operators should be the ones found in this paper. Furthermore, stronger constraints follow from our results. One should expect that ${\bf P}_a$  and $\mu_{ab}$ as operators arise from QFT renormalization and hence their matrix elements are likely to be from the algebra of MZV's over the field of {\it rational} numbers, whereas any algebraic number appears only as a result of diagonalization. It is not trivial to construct a matrix with rational coefficients so as to reproduce given algebraic numbers. Hence the explicit analytic knowledge of the eigenvalues gained in this paper should help in finding  the operatorial version of QSC and hence in the proper formulation of AdS/CFT integrability.

\ \\
{\bf Acknowledgments.} We thank Nikolay Gromov, Philipp H\"{a}hnel and Oliver Schnetz for useful discussions.

\appendix
\renewcommand{\thesubsection}{\Alph{section}.\arabic{subsection}}

\settocdepth{section}

\section{\label{sec:Psiresum}$\Psi$-operator}
In this appendix we explain  how to compute the action of the $\Psi$-operator \eqref{defPsi} on the algebra of functions described in \secref{defPsi}. See also \cite{Leurent:2013mr}. It will be clear that this algebra is invariant under the action of $\Psi$. Recall that the result of a $\Psi$-operation has an ambiguity of adding $i$-periodic functions. In this appendix we give a prescription that fixes this ambiguity. In particular, we demand that $\Psi$ is a linear operator. In the following, small letters denote single indices, e.g. in $\eta_a$, while capital letters denote multiple indices, e.g. in $\eta_A$.

\subsection{Rational functions}
First note that any rational function of the form $\frac{\sum_a b_a u^{a}}{\prod_{n,m}(u+in)^{m}}$ can be rewritten as a sum of a polynomial and shifted inverse powers, $\sum c_a u^a+\sum_{n,m}\frac{d_{n,m}}{(u+in)^m}$.

Applying $\Psi$ to a polynomial $r=\sum_{a=0}^n c_a u^a$ results in a polynomial of one order higher, found from solving the equation
\be\label{poleq}
\Psi(r)(u)-\Psi(r)(u+i)=r(u)\,.
\ee  In practice, we compute only $\Psi(u^a)$ and then extend the result by linearity. However we need to assure linearity, and we therefore require that $\Psi(r)(0)=0$ which removes the constant term ambiguity in the solution of \eqref{poleq}. For instance, $\Psi(1)=i\,u\,.$

The action of $\Psi$ on a shifted inverse power is prescribed to be the following sum:
\begin{eqnarray}
\Psi\left(\frac{1}{(u+in)^a}\right)=\sum_{m=0}^\infty \frac 1{(u+in+im)^a}=\eta_a^{[2n]}\,.
\end{eqnarray}
For $a\geq 2$ the sum is convergent and equals an $\eta$-function by definition. For $a=1$ we regularize the logarithmically divergent sum by postulating that it is equal to $\eta_1^{[2n]}$, which is defined as $\eta_1(u)\equiv i \psi(-iu)$ \cite{Leurent:2013mr}, where $\psi$ is the digamma function.

\subsection{Expressions involving $\eta$-functions}
First note that $i$-periodic functions $\mathcal{P}_a$ are treated as constants by $\Psi$:
\be\label{contact}
\Psi(\CP_a\,f)=\CP_a\,\Psi(f)\,.
\ee
In particular, $\Psi(\CP_a)=\CP_a\Psi(1)=i\,u\,\CP_a$. In contrast to the previous work \cite{Leurent:2013mr} where both $\eta_A$ and complex-conjugated $\bar\eta_A$ were used, we systematically remove all the $\bar\eta_A$ from the expressions using the defining property $\eta_a+\bar\eta_a^{[-2]}=\CP_a$ in order to explicitly simplify the computations using \eqref{contact}.

To handle products of $\eta$-functions and shifted inverse powers, we use our general convention $\Psi(f) \equiv \sum\limits_{n=0}^\infty f^{[2n]}$ when the sum is convergent. Therefore
\begin{eqnarray}
\Psi\left(\frac{\eta_A^{[2n+2]}}{(u+in)^a}\right)=\eta_{a,A}^{[2n]}\,.\label{niceform}
\end{eqnarray}
The logarithmically divergent sums are always regularized so as to satisfy \eqref{niceform}.

When an expression of the kind $\Psi\left(\frac{\eta_A^{[2n]}}{(u+im)^a}\right)$ is encountered, the strategy is to shift the $\eta$-function using the relation
\begin{eqnarray}
\eta_{a,A}=\eta_{a,A}^{[2]}+\frac{\eta_A^{[2]}}{u^a}\,, \label{etared}
\end{eqnarray}
until the produced terms have the form \eqref{niceform} or the $\eta$-function runs out of indices.

Products of polynomial powers and $\eta$-functions are handled using the relation
\be
\CD\left(\Psi(u^a)\eta_{b,A}^{[2n]}\right)=u^a\eta_{b,A}^{[2n]}+\Psi(u^a)^{[2]}\frac{1}{(u+in)^b}\eta_A^{[2+2n]}\,,
\ee
which leads to
\be
\Psi(u^a\eta_{b,A}^{[2n]})&=&\Psi(u^a)\eta_{b,A}^{[2n]}-\Psi\left(\Psi(u^a)^{[2]}\frac{1}{(u+in)^b}\eta_A^{[2+2n]}\right)\label{psitrick}\,.
\ee
The last relation is applied repeatedly until the produced terms are products of shifted inverse powers and $\eta$-functions, or the $\eta$-function runs out of indices.

\section{Determination of Baxter polynomials} \label{Qap}
As described, each operator is characterized by $L$, $S$ and its Baxter polynomial, $Q(u)=\prod\limits_{k=1}^S(u-u_k)$. The possible choices of $Q$ are a consequence of $L$ and $S$ and are defined as the polynomials satisfying the Bethe equations \eqref{BAEQ} and the zero-momentum condition \eqref{zeromomentum}. 

As only symmetric combinations of Bethe roots $u_k$ determine $Q(u)$ and hence only these combinations are relevant in the perturbative procedure, one may opt not to solve the Bethe equations, but instead solve the Baxter equation \eqref{Baxter},
\begin{eqnarray}\label{Baxter33}
0=\left(u+\frac{i}{2}\right)^LQ^{[2]}+\left(u-\frac{i}{2}\right)^LQ^{[-2]}-T\,Q \,,
\end{eqnarray}
from which we can fix $Q(u)$ without knowing $u_k$ explicitly.

In this appendix, it is discussed how to determine the possible Baxter polynomials first in special cases and then in generality. The reader may also consult \cite{Freyhult:2010kc} devoted to a review of the $\sl(2)$ sector of AdS$_5$/CFT$_4$ integrability.

\subsection{Spin 2}
From the zero-momentum condition it follows that solutions with $S=2$ are always $u$-symmetric:  $u_1=-u_2$, so $Q(u)=u^2-u_1^2$. The Bethe equations then reduce to
\be
\left(\frac{u_1+\frac{i}{2}}{u_1-\frac{i}{2}}\right)^{L+1}=1
\ee
with non-singular solutions
\be
u_1=-u_2=\frac 12 \cot\left(\frac{\pi k}{L+1}\right)\,,\ \ k=1,\ldots,\left[\frac{L}{2}\right]\,.
\ee
Though $u_1$ is always an algebraic number, e.g.
\be\label{eqtab}
\raisebox{1.5em}
{
$
\begin{array}{c|c|c|c|c|c|c|c}
 L   & 2 & 3 & 4 & 5 & 6 & 7 &\ldots\\
 k   & 1 & 1 & 1,2 & 1,2 & 1,2,3 & 2;1,3\\
\hline
u_1^2 & \frac 1{12} & \frac 14 & \frac 14 \pm\frac 1{2\sqrt 5}
& \frac 34, \frac 1{12}
& \parbox{10em}{roots of\\ $448x^3-560x^2+84x-1$}
& \frac 14,\frac 34\pm\frac 1{\sqrt{2}}
&\ldots\\
\end{array}
$
}
\,,
\ee
the explicit form of $u_1$ increases in complexity with $S$. Our implementation of the perturbative algorithm is most efficient for Baxter polynomials with rational coefficients. The algebraic number fields generated by square roots of a prime have also been treated analytically, and the $L=4$ case in \eqref{eqtab} is presented in section \ref{sec:summaryofresults}. To get a reasonable speed, we approximate more complicated algebraic numbers numerically while keeping exact expressions for MZV's. As an example of this semi-numerical computation, we considered the $L=6$ case in \eqref{eqtab}, and the results are again given in section \ref{sec:summaryofresults}.

\subsection{Twist 2}
This is the most interesting subclass of operators, in particular because of its relation to $\Gamma_{\rm cusp}$ \cite{Belitsky:2006en} and to the BFKL limit \cite{Kotikov:2007cy}. The one-loop solution is known explicitly \cite{Faddeev:1994zg,Korchemsky:1994um,Eden:2006rx}:
\be\label{F21}
Q(u)=\frac{(S!)^2}{(2i)^S(2S-1)!!}\ {}_3F_2\left(-S,S+1,\frac{1}{2}-i\,u;1,1;1\right)\,,
\ee
which is  a polynomial for integer values of $S$, with rational coefficients. The zero-momentum condition is satisfied only for even $S$. 

As it was recently understood \cite{Janik:2013nqa,Gromov:2014bva}, \eqref{F21} is not the correct solution at non-integer $S$, simply because it has poles. However, only integer $S$ corresponds to the states in the spectrum of single-trace operators and this is the only case considered in this work.

\subsection{Twist 3, rational $Q(u)$}
There is also a known series of solutions for twist-3 operators given by \cite{Beccaria:2007cn,Kotikov:2007cy}
\be\label{twist3}
Q(u)=\frac{(-1)^{\frac{S}{2}}(\frac{S}{2})!^4}{S!} {}_4F_3\left(-\frac S2,\frac S2+1,\frac 12+iu,\frac 12-iu;1,1,1;u\right)\,.
\ee
At even $S$, \eqref{twist3} is a polynomial with rational coefficients that satisfies the zero-momentum constraint. 

Whereas \eqref{F21} exhausts all possible solutions in the twist 2 case, the twist 3 case allows for solutions which are not given by \eqref{twist3}. The first occur at $S=3$: $Q(u)=u^3\pm\frac{3}{2}\sqrt{\frac{5}{7}}u^2+\frac{1}{4}u\mp\frac{1}{\sqrt{35}}$. We checked up to $S=20$ that solutions not given by \eqref{twist3} all have non-rational coefficients. However, for $3\le S\le 8$ and $S=10$ these solutions contain nothing worse than square roots, and they lead to results containing only rational numbers times MZV's in the perturbative corrections to the conformal dimension. 

\subsection{Generic state}
In practice, it is simplest to solve the Baxter equation \eqref{Baxter33}. This is done by looking for polynomial $Q$ and $T$. Hence, we plug in the ansatz $T\equiv 2u^L-\sum_{j=1}^L d_j u^{L-j}$ and $Q\equiv u^S+\sum_{j=1}^S c_j u^{S-j}$, and additionally impose the zero-momentum condition that allows to express the constant $c_1$ by a linear combination of the remaining $c_{j}$ with $j$ odd. Requiring that all powers vanish in \eqref{Baxter33} yields enough conditions on $d_i$ and $c_i$. In particular, one finds that $d_1$ and $d_2$ are universal
\be
d_1=0\,,\ \ d_2=S(S-1)+L\left(S-\frac 14\right)+\frac 14 L^2\,.
\ee
In all cases we have considered, the Baxter polynomial has indeed been obtained in this way.

Finally, let us comment on a different approach, going back to \cite{Yang:1968rm}, that allows to control the completeness of the set of solutions and to prove the reality of the Bethe roots. Consider the logarithmic form of the Bethe equations and introduce an extra real parameter $c$:
\be\label{logBAE}
L\,\arctan(2\,u_k)+\sum_{j=1}^{S}\arctan\left(\frac{u_k-u_j}{c}\right)=\pi\,n_k\,.
\ee
The mode numbers, $n_k$, should all be distinct. They are integers if $L+S-1$ is even and half-integers if $L+S-1$ is odd. It is easy to see that $|n_k|\leq \frac{L+S-3}{2}$.

If $c=+0$, solving \eqref{logBAE} is straightforward and unambiguous for a given set of mode numbers, given that $u_k>u_j$ if $n_k>n_j$. Then one only needs to perform the continuous deformation of the solution up to $c=1$. Reality of the solutions is concluded using the continuity argument. The continuity argument also allows one to compute the total momentum defined by  $\prod_{k}\frac{u_{k}+\frac{i}{2}}{u_{k}-\frac{i}{2}}=e^{ip_{\rm tot}}$ in terms of mode numbers:
\be
p_{\rm tot}=-\frac{2\pi}{L}\sum_{k=1}^{S}n_k-{\pi\, S} \ \ \ \ \  ({\rm mod}\ 2\pi)\,,
\ee
and hence we can easily constrain the possible $n_k$ to zero-momentum states only.

The approach of continuation in $c$ gives us a systematic and well-controlled way to  produce numerical solutions to the Bethe equations and hence to confirm the findings by analytic methods.

\section{\label{sec:Konishieffect}Validity of the Bethe Ansatz up to $L+1$ loops}
Short multiplets which join into a long one at finite coupling can correspond to spin chain states of different length. This happens for instance for the Konishi states. The length-2 operator $\Tr Z\nabla_+^2Z$ is in the same multiplet as the length-4 operator $\Tr (XZXZ-XXZZ)$. Hence these operators share the same anomalous part of the conformal dimension. For length-4 operators the asymptotic Bethe Ansatz is valid up to three loops, hence it should also be valid up to three loops for the length-2 representative, $\Tr Z\nabla_+^2Z$.

It is demonstrated below that this phenomenon is common for all states from the $\sl(2)$ sector. Any state of type \eqref{genericoperator} and length $L$ belongs, at finite coupling, to a multiplet which also contains a state, outside the $\sl(2)$ sector, of length $L+2$. Hence the Bethe Ansatz should be valid up to $L+1$ loops which we indeed observe in practice.

It is not known to us how to explicitly write down the corresponding operator of length $L+2$, except for the Konishi example. Instead, we provide an argument on the level of Bethe Ansatz equations. There, duality transformations affect the length. This phenomenon was one of the first tests to support the Beisert-Staudacher equations \cite{Beisert:2005fw}.

The discussion below is an adaptation of the results of \cite{Beisert:2005fw}.

\ \\
To define the duality transformation, one should not restrict to the $\sl(2)$ sector but consider the full set of asymptotic Bethe equations. This set determines five types of Bethe roots: three types of "bosonic" Bethe roots including the momentum-carrying roots $u_j$ and two auxiliary sets with the elements denoted below as $u_{b\pm}$; and two types of auxiliary "fermionic" Bethe roots parameterised by the Zhukovsky-type variable $y_{\alpha{\pm}}$\footnote{Not related to $y$ from the main text.}, not restricted to the domain $|y|>1$.

The main Bethe equation is written as
\be\label{mBethe}
-\left(\frac{x_k^+}{x_k^-}\right)^L=\prod_{j=1}^S\frac{u_k-u_j-i}{u_k-u_j+i}\left(\frac{1-\frac{1}{x_k^+x_j^-}}{1-\frac{1}{x_k^-x_j^+}}\right)^2\sigma_{\rm BES}(u_k,u_j)^2\prod_{\alpha_{\pm}}\frac{x_k^+-y_{\alpha_\pm}}{x_k^--y_{\alpha_\pm}}\,.
\ee
The Bethe equations for the auxiliary bosonic roots are
\be\label{aBethe}
-1=\prod_{\{ u_{b+}'\}}\frac{u_{b+}-u_{b+}'+i}{u_{b+}-u_{b+}'-i}\prod_{\alpha+}\frac{u_{b+}-w_{\alpha+}-\frac i2}{u_{b+}-w_{\alpha+}+\frac i2}\,,
\no\\
-1=\prod_{\{ u_{b-}'\}}\frac{u_{b-}-u_{b-}'+i}{u_{b-}-u_{b-}'-i}\prod_{\alpha-}\frac{u_{b-}-w_{\alpha-}-\frac i2}{u_{b-}-w_{\alpha-}+\frac i2}\,,
\ee
where $w_{\alpha\pm}\equiv g(y_{\alpha\pm}+\frac 1{y_{\alpha\pm}})\,.$

Instead of writing Bethe equations for the fermionic roots, it will be handy to encode these roots by relations between the following Zhukovsky-Baxter polynomials:
\be
R^{(\pm)}(y)\equiv \prod_{j=1}^S(y-x_j^{\mp})\,,\ \ {\mathbb Q}_{\pm}(u)\equiv \prod_{b\pm}(u-u_{b\pm})\,,\ \ R_{\pm}(y)\equiv \prod_{\alpha\pm}(y-y_{\alpha\pm})\,.
\ee
Introduce the notation $f_1\propto f_2$ to denote that $f_1=\Lambda f_2$ for some constant $\Lambda$.

The relations determining the fermionic roots are
\be\label{fBethe}
R^{(+)}(y){\mathbb Q}_{\pm}(w-i/2)-R^{(-)}(y){\mathbb Q}_{\pm}(w+i/2)\propto R_{\pm}(y)\,\bar R_{\pm}(y)\,,
\ee
with the demand that $\bar R_{\pm}(y)$ is  a polynomial in $y$. We will parameterise it as $\bar R_{\pm}\equiv \prod\limits_{{\bar \alpha}\pm}(y-\bar y_{{\bar \alpha}\pm})$.

Using \eqref{fBethe}, we can make \eqref{mBethe} and \eqref{aBethe} to depend on $\bar y_{\bar \alpha\pm}$ instead of $y_{\alpha\pm}$. This is precisely the above-mentioned duality transformation. We are only interested in its effect on \eqref{mBethe}. The result is:
\be
-\left(\frac{x_k^+}{x_k^-}\right)^L=\prod_{j=1}^S\frac{u_k-u_j+i}{u_k-u_j-i}\sigma_{\rm BES}(u_k,u_j)^2\prod_{\bar\alpha_{\pm}}\frac{x_k^--\bar y_{\bar \alpha_\pm}}{x_k^+-y_{\bar \alpha_\pm}}\,.
\ee
Now note that for the $\sl(2)$ sector there is no auxiliary Bethe roots (prior to the duality transformation). Therefore \eqref{fBethe} is simplified to $R^{(+)}-R^{(-)}\propto \bar R_{\pm}$. But $R^{(+)}(0)-R^{(-)}(0)=0$ due to the zero-momentum constraint $\prod\limits_{k}\frac{x_k^+}{x_k^-}=1$ imposed on solutions of the asymptotic Bethe Ansatz, and thus $y=0$ is one of the zeros of $\bar R_{+}$ and of $\bar R_{-}$. The net effect of two $y=0$ roots is the length change $L\to L+2$:
\be
-\left(\frac{x_k^+}{x_k^-}\right)^{L+2}=\prod_{j=1}^S\frac{u_k-u_j+i}{u_k-u_j-i}\sigma_{\rm BES}(u_k,u_j)^2\prod_{\bar\alpha_{\pm}\neq 0}\frac{x_k^--\bar y_{\bar \alpha_\pm}}{x_k^+-y_{\bar \alpha_\pm}}\,,
\ee
as requested.

\section{\label{sec:resultapp}Results}
In this appendix, we provide selected results in a format that is meant to be easy to parse by a programming language. The notation \texttt{z[a]} and \texttt{Z[a][b]} is used for $\zeta_a$ and $Z_a^{(b)}$, respectively. More results are available online \cite{MathematicaNotebook} and also in the ancillary files at arxiv.org.
\subsection{$L=2$, $S=2$ {\tiny \hyperref[table:ResRat]{Return to table 1}}}\label{sec:22} 
{\footnotesize $\Delta$ \vspace{-0.5em}
{\tiny\begin{verbatim}
4+12g^2-48g^4+336g^6+g^8(-2496+576z[3]-1440z[5])+g^12(-7680-262656z[3]+112320z[5]+155520z[3]z[5]+75600z[7]-489888z[9]-20736
z[3]^2)+g^10(15168+6912z[3]-8640z[5]+30240z[7]-5184z[3]^2)+g^14(-2135040+5230080z[3]-229248z[5]+411264z[3]z[5]-1254960z[7]
-1935360z[3]z[7]-835488z[9]+7318080z[11]-421632z[3]^2+124416z[3]^3-993600z[5]^2)+g^16(54408192-83496960z[3]-19678464z[5]
-4354560z[3]z[5]+21868704z[7]-6229440z[3]z[7]+22256640z[5]z[7]+9327744z[9]+23224320z[3]z[9]-106007616z[13]+7934976z[3]^2
-3255552z[5]z[3]^2+1990656z[3]^3+2384640z[5]^2+65929248z[11]/5-684288Z[11][2]/5)+g^18(-1014549504+1140922368z[3]+575354880z[5]
-14294016z[3]z[5]-223122816z[7]+34020864z[3]z[7]-92539584z[5]z[7]-247093632z[9]+119470464z[3]z[9]-245099520z[5]z[9]-278505216
z[3]z[11]-253865664z[13]+1517836320z[15]-1306368Z[13][2]+1306368Z[13][3]-51259392z[3]^2-26044416z[5]z[3]^2+22063104z[7]z[3]^2
-20155392z[3]^3+55296000z[5]^2+15759360z[3]z[5]^2-113690304z[7]^2-186204096z[11]/5+15676416Z[11][2]/5)+g^20(16445313024
-13069615104z[3]-11247547392z[5]+1213581312z[3]z[5]+377212032z[7]-1610841600z[3]z[7]+222341760z[5]z[7]+133788672z[3]z[5]z[7]
+4915257984z[9]-332646912z[3]z[9]+1099699200z[5]z[9]+2275620480z[7]z[9]-2334572928z[3]z[11]+2713772160z[5]z[11]+3372969600
z[3]z[13]-21661960320z[17]+11943936z[3]Z[11][2]-1509027840z[3]^2+1234206720z[5]z[3]^2+154680192z[7]z[3]^2-91072512z[9]z[3]^2
+578949120z[3]^3-70170624z[5]z[3]^3-14929920z[3]^4-1390279680z[5]^2-654842880z[3]z[5]^2+868662144z[7]^2+9793211904z[11]/5
+752219136Z[11][2]/5-7159104Z[13][3]/7-17895168Z[15][3]/25-787483944z[13]/175-5070791808Z[13][2]/175+2716063488Z[15][2]/175
+6966252288z[5]^3/175-4308536566944z[15]/875)
\end{verbatim}}
\noindent $\Delta_{\text{num}}$\vspace{-0.5em}
{\tiny\begin{verbatim}
4+0.750000000000(4g)^2-0.187500000000(4g)^4+0.0820312500000(4g)^6-0.0503050413694(4g)^8+0.0357813554374(4g)^10-0.0272807716912
(4g)^12+0.0217501134701(4g)^14-0.0179107691403(4g)^16+0.0151113823572(4g)^18-0.0129922111546(4g)^20
\end{verbatim}}

\noindent $\Delta_{\text{Pad\'e}}$\vspace{-0.5em}
{\tiny\begin{verbatim}
(2.42770-5.22440w+5.07017w^2-2.69218w^3+0.75942w^4-0.07548w^5)/(1-2.38873w+2.43525w^2-1.30596w^3+0.36745w^4-0.04170w^5)
/.w->(1+(4g)^2)^(1/4)
\end{verbatim}}
}

\subsection{$L=2$, $S=4$ {\tiny \hyperref[table:ResRat]{Return to table 1}}}\label{sec:24} {\footnotesize
\noindent $\Delta$\vspace{-0.5em}
{\tiny 
\begin{verbatim}
6+50g^2/3-1850g^4/27+241325g^6/486+g^8(-25000z[5]/9+114500z[3]/81-8045275/2187)+g^12(-945000z[9]+1250000z[3]z[5]/3+43098125z[7]
/81-13625000z[3]^2/81+299430575z[5]/729-1918473250z[3]/2187+12344860375/118098)+g^10(175000z[7]/3-125000z[3]^2/9-3357500z[5]/81
+24048500z[3]/729+3007398125/157464)+g^14(42350000z[11]/3-140000000z[3]z[7]/27+12500000z[3]^3/27-71875000z[5]^2/27+975687500z[3]
z[5]/243-1756580750z[9]/243-1808233750z[3]^2/729-17907365875z[7]/2916+11516727625z[5]/4374+290741688625z[3]/19683-25166596925125
/4251528)+g^16(-204490000z[13]+560000000z[3]z[9]/9+1610000000z[5]z[7]/27-13750000Z[11][2]/27+31625000z[5]^2/27+8830456250z[11]/81
-981250000z[5]z[3]^2/81-13079106250z[3]z[7]/243+4030000000z[3]^3/243-52302912500z[3]z[5]/2187+100399413550z[9]/2187-1452895120625
z[5]/8748+330868915000z[3]^2/19683+6593631273125z[7]/52488-33496031056250z[3]/177147+20623221557720125/153055008)+g^18(
2927925000z[15]-1970000000z[5]z[9]/3-2238500000z[3]z[11]/3-43750000Z[13][2]/9+43750000Z[13][3]/9-2741375000z[7]^2/9-21762593750
z[5]z[7]/81-144311199500z[13]/81+1164500000Z[11][2]/81+6650000000z[7]z[3]^2/81+4750000000z[3]z[5]^2/81+204899462500z[3]z[9]/243
-36536133875z[11]/729-174358750000z[5]z[3]^2/729-81702910000z[3]z[7]/2187+1513174862500z[5]^2/2187-340750937500z[3]^3/6561
-15685489033250z[3]z[5]/19683+11765191180625z[3]^2/19683-208130844175375z[9]/118098+1343508816703625z[5]/354294-190566198816875
z[7]/472392+6622119695693125z[3]/4251528-40164947022652931875/16529940864)+g^20(54871250000z[7]z[9]/9+65436250000z[5]z[11]/9
+81331250000z[3]z[13]/9-376075700000z[17]/9-3050000000z[9]z[3]^2/9+40325000000z[3]z[5]z[7]/81+5000000000z[3]Z[11][2]/81-215750000
Z[15][3]/81-29375000000z[5]z[3]^3/81+10915250000Z[15][2]/189-42437500000z[3]^4/243+910079712500z[7]^2/243-153866881250Z[13][2]
/567-10430220087500z[3]z[11]/729+952306718750z[7]z[3]^2/729-3295243750000z[3]z[5]^2/729-11569454483750z[15]/1701+251962250000z[5]
^3/1701-13375653090625z[5]z[7]/2187+8098572962500z[3]z[9]/2187+8629617500000z[5]z[9]/2187+3541152012500Z[11][2]/2187+318419921875
Z[13][3]/5103+70613799125000z[5]z[3]^2/6561-53336803532375z[13]/13608-126850857538750z[3]z[7]/19683+294732037291250z[11]/19683
+63139737837500z[3]^3/19683-167771826267500z[5]^2/19683+974311933258750z[3]z[5]/59049-10115578640260625z[3]^2/354294
+53387963079372875z[9]/1417176-498816377020206875z[7]/17006112-4775163910325555125z[5]/76527504+2083000535488133125z[3]
/344373768+5702048121387295834375/148769467776)
\end{verbatim}}
\noindent{$\Delta_{\text{num}}$}\vspace{-0.5em}
{\tiny
\begin{verbatim}
6+1.04166666667(4g)^2-0.267650462963(4g)^4+0.121228881334(4g)^6-0.0741551486627(4g)^8+0.0519974475681(4g)^10-0.0392261291360
(4g)^12+0.0310399835520(4g)^14-0.0254045270349(4g)^16+0.0213198086377(4g)^18-0.0182430735442(4g)^20
\end{verbatim}}
\noindent{$\Delta_{\text{Pad\'e}}$}\vspace{-0.5em}
{\tiny\begin{verbatim}
(3.83315-11.61394w+14.54178w^2-9.99858w^3+4.08309w^4-0.78480w^5)/(1-3.19099w+4.06001w^2-2.61172w^3+0.88458w^4-0.13177w^5)
/.w->(1+(4g)^2)^(1/4)
\end{verbatim}
}}

\subsection{$L=2$, $S=6$ {\tiny \hyperref[table:ResRat]{Return to table 1}}}\label{sec:26}{\footnotesize
\noindent{$\Delta$}\vspace{-0.5em}
{\tiny
\begin{verbatim}
8+98g^2/5-91238g^4/1125+300642097g^6/506250+g^8(-19208z[5]/5+11736088z[3]/5625-393946504469/91125000)+g^10(403368z[7]/5-2823576
z[3]^2/125-31241812z[5]/375+28848226288z[3]/421875+4156425743851997/205031250000)+g^12(16941456z[3]z[5]/25-32672808z[9]/25
+2171951803z[7]/1875-3939829712z[3]^2/9375+89720524439z[5]/140625-135103809324932z[3]/94921875+63963585215729446667
/369056250000000)+g^14(97615056z[11]/5-21647416z[5]^2/5-210827008z[3]z[7]/25+553420896z[3]^3/625-30518049758z[9]/1875+59185447132
z[3]z[5]/5625-8321262273352z[3]^2/1265625-18988765239829z[7]/1687500+3305589485031253z[5]/284765625+992855735411276149z[3]
/51257812500-1264739078350312951043329/166075312500000000)+g^16(2424510592z[5]z[7]/25+2529924096z[3]z[9]/25-7070119056z[13]/25
-14481180112z[5]z[3]^2/625-12865153448z[5]^2/1875-3043814928Z[11][2]/3125+2299565466814z[11]/9375+437294744656z[3]^3/9375
-3933812906842z[3]z[7]/28125+5520215624672z[9]/84375-5158877019226z[3]z[5]/140625+22909993660202411z[7]/84375000
-14537192107123847z[3]^2/1423828125-3051427713529280386z[5]/7119140625-2278588177560352494067z[3]/15377343750000
+47718535160182440741046032719/298935562500000000000)+g^18(20246250024z[15]/5-26699734656z[5]z[9]/25+784012936Z[11][2]/25
-151693797024z[3]z[11]/125+14019996032z[3]z[5]^2/125-61923845256z[7]^2/125+1321296624564z[3]z[9]/625-2457557677484z[13]/625
-5810919408Z[13][2]/625+5810919408Z[13][3]/625+98139972224z[7]z[3]^2/625-1272664998626z[5]z[7]/3125-20008886376472z[5]z[3]^2
/28125+78750333350143z[11]/140625+2802436953429974z[5]^2/1265625+348946425285724z[3]^3/3515625-4175780825364428z[3]z[7]
/6328125-840371926591144129z[9]/189843750-9353181684276991897z[3]z[5]/2847656250+244731394879850299799z[7]/170859375000
+764512782428890855493z[3]^2/284765625000+249954453809317864731397z[5]/30754687500000-23966753366779005491706347z[3]
/17299511718750000-70658213589762365535612856329047/26904200625000000000000)+g^20(73486388976z[3]z[13]/5-288946481824z[17]/5
+247893031344z[7]z[9]/25+295622760048z[5]z[11]/25+595111936832z[3]z[5]z[7]/625-405104095872z[9]z[3]^2/625+14240036839084z[7]^2
/1875-108224910858036z[3]z[11]/3125+433881982464z[3]Z[11][2]/3125-2549056646976z[5]z[3]^3/3125+39171838795264z[5]z[9]/5625
+1725922124304Z[15][2]/15625-79600372208Z[15][3]/15625-321961276508704z[3]z[5]^2/28125+188944546474754z[7]z[3]^2/46875
+13280111824112z[5]^3/46875+31363843730177Z[13][3]/140625-142168662920704z[3]^4/234375-3983931086282518z[15]/703125
+22079911669943162z[5]z[3]^2/703125-903114566320946Z[13][2]/1171875-56879958713799851z[5]z[7]/2109375+41578152713712916z[3]z[9]
/2109375-567760975469842147z[13]/28125000+273496646465620214Z[11][2]/52734375-7105897779826398217z[5]^2/569531250
-18926613875409552767z[3]z[7]/1423828125+144556615705008184027z[11]/3796875000+51798665840718064814z[3]^3/7119140625
+3374170773264360931729z[9]/34171875000+89923729526501062854319z[3]z[5]/2562890625000-9974605715611757638244477z[3]^2
/115330078125000-39701347306960476836889463z[7]/307546875000000-212774842390295630812877713z[5]/2075941406250000
+2117801074205277617896484802449z[3]/23354340820312500000+8756622316559026747595920980082472627/242137805625000000000000000)
\end{verbatim}}
\noindent{$\Delta_{\text{num}}$}\vspace{-0.5em}
{\tiny
\begin{verbatim}
8+1.22500000000(4g)^2-0.316798611111(4g)^4+0.144985579186(4g)^6-0.0884798495157(4g)^8+0.0617885084251(4g)^10
-0.0464962688166(4g)^12+0.0367236958123(4g)^14-0.0300035188128(4g)^16+0.0251365956037(4g)^18-0.0214737689908(4g)^20
\end{verbatim}}
\noindent{$\Delta_{\text{Pad\'e}}$}\vspace{-0.5em}
{\tiny\begin{verbatim}
(5.46707-16.01265w+20.70198w^2-14.56744w^3+6.16622w^4-1.28494w^5)/(1-3.04508w+3.91782w^2-2.56160w^3+0.88881w^4-0.14117w^5)
/.w->(1+(4g)^2)^(1/4)
\end{verbatim}
}}

\subsection{$L=3$, $S=2$ {\tiny \hyperref[table:ResRat]{Return to table 1}}}\label{sec:32}{\footnotesize
\noindent{$\Delta$}\vspace{-0.5em}
{\tiny 
\begin{verbatim}
5+8g^2-24g^4+136g^6+g^8(-920-128z[3])+g^10(6664+1152z[3]+3840z[5]-2240z[7])+g^12(-49176-17152z[3]-19712z[5]-7680z[3]z[5]-67200
z[7]+64512z[9]+6144z[3]^2)+g^16(-2429336-2828160z[3]-1574400z[5]+488448z[3]z[5]-2380032z[7]+1813504z[3]z[7]-3067904z[5]z[7]
-2795520z[3]z[9]-14755072z[11]+22843392z[13]-6144Z[11][2]+67584z[3]^2+358400z[5]z[3]^2-49152z[3]^3+1085440z[5]^2-15111680z[9]
/3)+g^14(356488+231168z[3]+154112z[5]-161280z[3]z[5]+339136z[7]+172032z[3]z[7]-1300992z[11]-18432z[3]^2-8192z[3]^3+92160z[5]^2
+9164288z[9]/9)+g^18(14231560+32278400z[3]+17756416z[5]+504832z[3]z[5]+18913856z[7]-3526656z[3]z[7]-25174016z[5]z[7]+44820480
z[5]z[9]+40550400z[3]z[11]-374774400z[15]+57344Z[13][3]+559104z[3]^2+1275904z[5]z[3]^2-4243456z[7]z[3]^2-720896z[3]^3+98304z[3]^4
-5478400z[5]^2-4546560z[3]z[5]^2+23106048z[7]^2-63574016z[3]z[9]/3-1357824Z[11][2]/5+314365952z[9]/9+1089140992z[11]/15-1138688
Z[13][2]/25+15755170048z[13]/75)+g^20(-50041880-348687872z[3]-201972480z[5]-31148544z[3]z[5]-162107264z[7]-22644736z[3]z[7]
+113590272z[5]z[7]+98734080z[3]z[5]z[7]-651893760z[7]z[9]-619914240z[5]z[11]-559663104z[3]z[13]+5924638720z[17]+1353728Z[13][3]
-28446720z[3]^2+21235712z[5]z[3]^2-17805312z[7]z[3]^2+47419392z[9]z[3]^2+10797056z[3]^3-3424256z[5]z[3]^3+851968z[3]^4+4864000
z[5]^2-4464640z[3]z[5]^2+157165568z[7]^2+6617088Z[11][2]/5+1425408z[3]Z[11][2]/5-2504385536z[9]/9+186195968z[3]z[9]/9+2774681600
z[5]z[9]/9-7280295424z[11]/15+3864887296z[3]z[11]/15-26732034304z[13]/25-13488128Z[13][2]/25+12292096Z[15][2]/25+474935296z[5]^3
/25-11288576Z[15][3]/75-1227917119744z[15]/375)
\end{verbatim}}
\noindent{$\Delta_{\text{num}}$}\vspace{-0.5em}
{\tiny\begin{verbatim}
5+0.500000000000(4g)^2-0.0937500000000(4g)^4+0.0332031250000(4g)^6-0.0163858533265(4g)^8+0.00931918120466(4g)^10
-0.00560570336609(4g)^12+0.00342921286795(4g)^14-0.00206677281701(4g)^16+0.00117360336233(4g)^18-0.000567983813632(4g)^20
\end{verbatim}}
\noindent{$\Delta_{\text{Pad\'e}}$}\vspace{-0.5em}
{\tiny\begin{verbatim}
(4.70615-11.59234w+7.90565w^2-4.20166w^3+1.67716w^4-0.74998w^5)/(1-2.08037w+0.24092w^2+0.93969w^3-0.63065w^4+0.07940w^5)
/.w->(1+(4g)^2)^(1/4)
\end{verbatim}
}}

\subsection{$L=3$, $S=4$ {\tiny \hyperref[table:ResRat]{Return to table 1}}}\label{sec:34}{\footnotesize
\noindent{$\Delta$}\vspace{-0.5em}
{\tiny 
\begin{verbatim}
7+12g^2-39g^4+957g^6/4+g^8(-216z[3]-28191/16)+g^10(1242z[3]+9360z[5]-5040z[7]+880221/64)+g^12(-50382z[3]+6300z[5]-25920z[3]z[5]
-199080z[7]+145152z[9]+25920z[3]^2-27391071/256)+g^14(-62784z[5]-822960z[3]z[5]+128583z[7]+580608z[3]z[7]+3352608z[9]-2927232
z[11]+80784z[3]^2-41472z[3]^3+311040z[5]^2+3386799z[3]/4+799473405/1024)+g^16(-1022112z[3]z[5]+1125153z[7]+10057824z[3]z[7]
-10354176z[5]z[7]-3692520z[9]-9434880z[3]z[9]-51897456z[11]+51397632z[13]-31104Z[11][2]-216108z[3]^2+1814400z[5]z[3]^2-559872
z[3]^3+6393600z[5]^2-4766031z[5]/2-364105233z[3]/32-19410126015/4096)+g^18(26914734z[3]z[5]+19063296z[3]z[7]-154628352z[5]z[7]
+7094556z[9]-122435712z[3]z[9]+151269120z[5]z[9]+136857600z[3]z[11]-843242400z[15]+290304Z[13][3]+10128321z[3]^2+16363296z[5]
z[3]^2-21482496z[7]z[3]^2-10098432z[3]^3+746496z[3]^4-17874000z[5]^2-23016960z[3]z[5]^2+77982912z[7]^2+365175756z[11]/5-13771296
Z[11][2]/5+1076435523z[5]/16-474911595z[7]/16+19278748656z[13]/25-5764608Z[13][2]/25+17180290539z[3]/128+210275719869/16384
)+g^20(-197035740z[3]z[7]+434780352z[5]z[7]+499841280z[3]z[5]z[7]+219520035z[9]-372272544z[3]z[9]+1903608000z[5]z[9]-2200141440
z[7]z[9]-2092210560z[5]z[11]-1888862976z[3]z[13]+13330437120z[17]+13579488Z[13][3]-366037758z[3]^2+240478632z[5]z[3]^2-196499520
z[7]z[3]^2+240060672z[9]z[3]^2+70703280z[3]^3-26002944z[5]z[3]^3+14805504z[3]^4-288603540z[5]^2-76049280z[3]z[5]^2+976860864
z[7]^2-683484939z[3]z[5]/4-876145788z[11]/5+7472953728z[3]z[11]/5-7546010004z[13]/5-15907752Z[11][2]/5+10824192z[3]Z[11][2]/5
-29585088Z[13][2]/5+62228736Z[15][2]/25-19049472Z[15][3]/25+2404359936z[5]^3/25+6148816641z[7]/32-73280873295z[5]/64
-1594452959568z[15]/125-360974445381z[3]/256+17712213660609/65536)
\end{verbatim}}
\noindent{$\Delta_{\text{num}}$}\vspace{-0.5em}
{\tiny
\begin{verbatim}
7+0.750000000000(4g)^2-0.152343750000(4g)^4+0.0584106445312(4g)^6-0.0308468901227(4g)^8+0.0189494812287(4g)^10
-0.0125872747998(4g)^12+0.00875756540541(4g)^14-0.00627197301169(4g)^16+0.00457165729765(4g)^18-0.00336246471982(4g)^20
\end{verbatim}}
\noindent{$\Delta_{\text{Pad\'e}}$}\vspace{-0.5em}
{\tiny\begin{verbatim}
(5.97027-21.85094w+57.64246w^2-51.60612w^3+26.96925w^4-5.73535w^5)/(1-3.67883w+9.67481w^2-8.30233w^3+3.58313w^4-0.64970w^5)
/.w->(1+(4g)^2)^(1/4)
\end{verbatim}
}}

\subsection{$L=4$, $S=3$ {\tiny \hyperref[table:ResRat]{Return to table 1}}}\label{sec:43}{\footnotesize
\noindent{$\Delta$}\vspace{-0.5em}
{\tiny 
\begin{verbatim}
7+12g^2-42g^4+288g^6+g^8(-2487-144z[3])+g^10(24531+1944z[3]+1440z[5])+g^12(-266229-30348z[3]-2736z[5]-5040z[7]-18144z[9])+g^14(
3109377+307980z[3]+128952z[5]+51840z[3]z[5]-85176z[7]-96768z[3]z[7]-196560z[9]+665280z[11]+65664z[3]^2-8640z[5]^2)+g^16(-2021220
z[3]-2872872z[5]-1603584z[3]z[5]+337500z[7]-1257984z[3]z[7]+1451520z[5]z[7]+2073456z[9]+2903040z[3]z[9]-16061760z[13]-728352
z[3]^2-290304z[5]z[3]^2+82944z[3]^3-673920z[5]^2-153625047/4+33479136z[11]/5-124416Z[11][2]/5)+g^18(-997434z[3]+40419180z[5]
+16977600z[3]z[5]+9085212z[7]+20188224z[3]z[7]+25009344z[5]z[7]-15699888z[9]+21561984z[3]z[9]-38361600z[5]z[9]-58848768z[3]z[11]
+324324000z[15]+145152Z[13][3]+11245824z[3]^2-2985984z[5]z[3]^2+6676992z[7]z[3]^2-1161216z[3]^3-248832z[3]^4+8067600z[5]^2
+6635520z[3]z[5]^2-16027200z[7]^2+1977534657/4-189521424z[11]/5-743370912z[13]/5+1119744Z[11][2]/5-601344Z[13][2]/5)+g^20(
-6554289162+313877349z[3]-423148320z[5]-178974576z[3]z[5]-339657318z[7]-177966720z[3]z[7]-206689536z[5]z[7]-221543424z[3]z[5]z[7]
+86902632z[9]-246881088z[3]z[9]-396299520z[5]z[9]+633225600z[7]z[9]+745303680z[5]z[11]+1017080064z[3]z[13]-5951088000z[17]-
123671664z[3]^2+29507328z[5]z[3]^2+30004992z[7]z[3]^2-107619840z[9]z[3]^2-17895168z[3]^3+14681088z[5]z[3]^3-248832z[3]^4
-103671360z[5]^2+36806400z[3]z[5]^2-202063680z[7]^2+1458139176z[11]/5-1681502976z[3]z[11]/5+1534464Z[11][2]/5+746496z[3]Z[11][2]
/5-850176Z[15][3]/5-15147648Z[13][3]/7+30710016Z[15][2]/35-1310563584z[5]^3/35+114279567336z[13]/175+395442367632z[15]/175
+330106752Z[13][2]/175)
\end{verbatim}}
\noindent{$\Delta_{\text{num}}$}\vspace{-0.5em}
{\tiny
\begin{verbatim}
7+0.750000000000(4g)^2-0.164062500000(4g)^4+0.0703125000000(4g)^6-0.0405898467110(4g)^8+0.0270471330520(4g)^10
-0.0195985184832(4g)^12+0.0150091270785(4g)^14-0.0119534684450(4g)^16+0.00980208645335(4g)^18-0.00822233425140(4g)^20
\end{verbatim}}
\noindent{$\Delta_{\text{Pad\'e}}$}\vspace{-0.5em}
{\tiny\begin{verbatim}
(5.65630+8.03904w-27.28068w^2+27.31725w^3-14.72305w^4+4.08283w^5)/(1+1.27111w-4.41190w^2+3.69745w^3-1.41048w^4+0.29549w^5)
/.w->(1+(4g)^2)^(1/4)
\end{verbatim}
}}

\subsection{$L=5$, $S=2$, $Q=u^2-\frac{3}{4}$ {\tiny \hyperref[table:ResRat]{Return to table 1}}}\label{sec:521}{\footnotesize
\noindent{$\Delta$}\vspace{-0.5em}
{\tiny 
\begin{verbatim}
7+4g^2-6g^4+37g^6/2+g^8(-16z[3]-283/4)+g^10(112z[3]+160z[5]+9597/32)+g^12(-680z[3]-1040z[5]-1680z[7]-86457/64)+g^14(3952z[3]
+6608z[5]+7392z[7]+26208z[9]-7392z[11]+160z[3]^2+1621049/256)+g^16(-37728z[5]-8704z[3]z[5]-55888z[7]+13440z[3]z[7]-1792z[5]z[7]
-43008z[9]-13440z[3]z[9]-496320z[11]+329472z[13]-1392z[3]^2+1600z[5]^2-45765z[3]/2-15625187/512)+g^18(132528z[3]+211341z[5]
+53056z[3]z[5]+334962z[7]+104832z[3]z[7]-214912z[5]z[7]-385344z[3]z[9]+195840z[5]z[9]+506880z[3]z[11]-9266400z[15]-384Z[13][3]
+14636z[3]^2+13696z[5]z[3]^2-12544z[7]z[3]^2-3072z[3]^3+73760z[5]^2-3840z[3]z[5]^2+41664z[7]^2+1358432z[9]/3+4224Z[11][2]/5
+2551648z[11]/15+235746624z[13]/25+10368Z[13][2]/25+1229414557/8192)+g^20(-1185742z[5]-399552z[3]z[5]-639664z[3]z[7]-1762464
z[5]z[7]+440320z[3]z[5]z[7]-3525120z[7]z[9]-6462720z[5]z[11]-12300288z[3]z[13]+211594240z[17]-3840Z[13][3]-130414z[3]^2+59776
z[5]z[3]^2-279552z[7]z[3]^2+380928z[9]z[3]^2-8704z[5]z[3]^3+4096z[3]^4-378840z[5]^2-303360z[3]z[5]^2+2369472z[7]^2-3792575z[7]/2
+45568z[3]^3/3-18159184z[11]/5-26496Z[11][2]/5-4608z[3]Z[11][2]/5-26610272z[9]/9-8977744z[3]z[9]/9+48622720z[5]z[9]/9+114365824
z[3]z[11]/15-12269681z[3]/16+72576Z[13][2]/25-148736Z[15][2]/25+9472Z[15][3]/25+6568304z[13]/75+3560192z[5]^3/75-62872885696
z[15]/375-12274630413/16384)
\end{verbatim}}
\noindent{$\Delta_{\text{num}}$}\vspace{-0.5em}
{\tiny
\begin{verbatim}
7+0.250000000000(4g)^2-0.0234375000000(4g)^4+0.00451660156250(4g)^6-0.00137303024979(4g)^8+0.000572629035928(4g)^10
-0.000294489918381(4g)^12+0.000165718277718(4g)^14-0.0000876743321924(4g)^16+0.0000306539739007(4g)^18+0.0000150151769779(4g)^20
\end{verbatim}}
\noindent{$\Delta_{\text{Pad\'e}}$}\vspace{-0.5em}
{\tiny\begin{verbatim}
(6.78265-16.59004w+22.36820w^2-14.39806w^3+5.33189w^4-0.77603w^5)/(1-2.40746w+3.16729w^2-1.93964w^3+0.66182w^4-0.09363w^5)
/.w->(1+(4g)^2)^(1/4)
\end{verbatim}
}}

\subsection{$L=5$, $S=2$, $Q=u^2-\frac{1}{12}$ {\tiny \hyperref[table:ResRat]{Return to table 1}}}\label{sec:522}{\footnotesize
\noindent{$\Delta$}\vspace{-0.5em}
{\tiny 
\begin{verbatim}
7+12g^2-42g^4+555g^6/2+g^8(-144z[3]-8997/4)+g^10(2160z[3]+1440z[5]+651651/32)+g^12(-27864z[3]-22032z[5]-15120z[7]-12654663/64
)+g^14(325584z[3]+313200z[5]+290304z[7]+187488z[9]-66528z[11]+864z[3]^2+513162183/256)+g^16(-4014432z[5]+387072z[3]z[5]-4156272
z[7]+72576z[3]z[7]-48384z[5]z[7]-4209408z[9]-362880z[3]z[9]-2832192z[11]+2965248z[13]+15984z[3]^2-25920z[5]^2-7991163z[3]/2
-10626282525/512)+g^18(55254960z[3]+48439539z[5]-6094656z[3]z[5]+53001486z[7]-8346240z[3]z[7]+183168z[5]z[7]+59048352z[9]
-2172096z[3]z[9]+5287680z[5]z[9]+63698400z[11]+13685760z[3]z[11]-83397600z[15]+93312Z[11][2]-31104Z[13][3]-361260z[3]^2+155520
z[5]z[3]^2-1016064z[7]z[3]^2+746496z[3]^3-3866400z[5]^2-311040z[3]z[5]^2+1124928z[7]^2+1211324544z[13]/25+839808Z[13][2]/25
+1761731190627/8192)+g^20(-601564050z[5]+58387392z[3]z[5]+111698352z[3]z[7]+119137824z[5]z[7]+35665920z[3]z[5]z[7]-727682400z[9]
+143395920z[3]z[9]+10488960z[5]z[9]-95178240z[7]z[9]-174493440z[5]z[11]-332107776z[3]z[13]+1904348160z[17]-435456Z[13][3]
+28053486z[3]^2-30201984z[5]z[3]^2+30855168z[9]z[3]^2-12234240z[3]^3-2115072z[5]z[3]^3+65280600z[5]^2-2177280z[3]z[5]^2+834624
z[7]^2-1277725185z[7]/2-4312602864z[11]/5+225327744z[3]z[11]/5-7744896Z[11][2]/5-1119744z[3]Z[11][2]/5-13405664655z[3]/16-
24627780432z[13]/25+8366976Z[13][2]/25-12047616Z[15][2]/25+767232Z[15][3]/25+96125184z[5]^3/25-69564393792z[15]/125
-35771191001331/16384)
\end{verbatim}}
\noindent{$\Delta_{\text{num}}$}\vspace{-0.5em}
{\tiny
\begin{verbatim}
7+0.750000000000(4g)^2-0.164062500000(4g)^4+0.0677490234375(4g)^6-0.0369620696114(4g)^8+0.0233208776743(4g)^10
-0.0160524282634(4g)^12+0.0116823463688(4g)^14-0.00882867599217(4g)^16+0.00685415267453(4g)^18-0.00542876307610(4g)^20
\end{verbatim}}
\noindent{$\Delta_{\text{Pad\'e}}$}\vspace{-0.5em}
{\tiny\begin{verbatim}
(5.88287-34.18960w+52.64504w^2-45.06529w^3+21.40017w^4-6.19110w^5)/(1-5.69408w+8.15472w^2-5.78310w^3+1.94918w^4-0.41499w^5)
/.w->(1+(4g)^2)^(1/4)
\end{verbatim}
}}

\subsection{$L=4$, $S=2$, $Q=u^2-\frac{1}{4}-\frac{1}{2\sqrt{5}}$ {\tiny \hyperref[table:ResSqrt]{Return to table 2}}}\label{sec:421}{\footnotesize
\noindent{$\Delta$}\vspace{-0.5em}
{\tiny
\begin{verbatim}
6+g^2(10-2Sqrt[5])+g^4(-34+10Sqrt[5])+g^6(234-414Sqrt[5]/5)+g^8(-2074-80z[3]+16Sqrt[5]z[3]+4078Sqrt[5]/5)+g^10(21050+1104z[3]
-304Sqrt[5]z[3]+800z[5]-160Sqrt[5]z[5]-219586Sqrt[5]/25)+g^12(-227394-4512z[3]-8720z[5]+1360Sqrt[5]z[5]-14000z[7]+6160Sqrt[5]
z[7]-15120z[9]+5040Sqrt[5]z[9]+2656Sqrt[5]z[3]/5+2448714Sqrt[5]/25)+g^18(210441114+(-1096928+1858880Sqrt[5])z[7]+z[5](23114752
-9422720Sqrt[5]+(-24167200+12668000Sqrt[5])z[7]+(-35520000+14208000Sqrt[5])z[9])+(270270000-90090000Sqrt[5])z[15]+(362880
-158592Sqrt[5])Z[11][2]+(-129920+55680Sqrt[5])Z[13][2]+(156800-67200Sqrt[5])Z[13][3]+(-20627296+9266016Sqrt[5]+(9193600-4250240
Sqrt[5])z[5]+(7212800-3091200Sqrt[5])z[7])z[3]^2+(1273600-582400Sqrt[5])z[3]^3+(-320000+140800Sqrt[5])z[3]^4+(-6916000+3122080
Sqrt[5])z[5]^2+(-14840000+5936000Sqrt[5])z[7]^2+z[11](-6402768Sqrt[5]+57544400/3)+z[13](-32520896Sqrt[5]+101293760/3)+z[3]
(161137968+(11823040-5240000Sqrt[5])z[5]+(-591200+375136Sqrt[5])z[7]+(-54489600+21795840Sqrt[5])z[11]+(7168000-3072000
Sqrt[5])z[5]^2+z[9](-263300800/9+132123200Sqrt[5]/9)-1782426032Sqrt[5]/25)+z[9](-28649008/9+157295024Sqrt[5]/45)-11650693614
Sqrt[5]/125)+g^14(2459594+(69600-13408Sqrt[5])z[5]+(114800-24080Sqrt[5])z[7]+z[3](87904-27360Sqrt[5]+(-78400+40000Sqrt[5])z[5]
+(-89600+35840Sqrt[5])z[7])+(215040-130368Sqrt[5])z[9]+(554400-184800Sqrt[5])z[11]+(38400-18112Sqrt[5])z[3]^2+(-8000+3200Sqrt[5])
z[5]^2-134624418Sqrt[5]/125)+g^16(-24950154+(-302000-62416Sqrt[5])z[7]+z[5](-262896-7600Sqrt[5]+(1344000-537600Sqrt[5])z[7])
+(-13384800+4461600Sqrt[5])z[13]+(-26880+11520Sqrt[5])Z[11][2]+(-453824+204160Sqrt[5]+(-313600+134400Sqrt[5])z[5])z[3]^2+
(-192000+89600Sqrt[5])z[3]^3+(784000-408000Sqrt[5])z[5]^2+z[11](2180080Sqrt[5]-8607760/3)+z[3](-5045744+(-168000+80320Sqrt[5])
z[5]+(1724800-873600Sqrt[5])z[7]+(2688000-1075200Sqrt[5])z[9]+10801648Sqrt[5]/5)+z[9](-4782880/3+4080640Sqrt[5]/9)+1376810658
Sqrt[5]/125)
\end{verbatim}
}
\noindent{$\Delta_{\text{num}}$}\vspace{-0.5em}
{\tiny
\begin{verbatim}
6+0.345491502813(4g)^2-0.0454660946289(4g)^4+0.0119271414705(4g)^6-0.00462984336290(4g)^8+0.00226797652141(4g)^10
-0.00133183926997(4g)^12+0.000927620827439(4g)^14-0.000747528404383(4g)^16+0.000667302430772(4g)^18
\end{verbatim}}
\noindent{$\Delta_{\text{Pad\'e}}$}\vspace{-0.5em}
{\tiny\begin{verbatim}
(5.38331-6.17648w+5.02752w^2-0.33146w^3-1.05278w^4+0.89283w^5)/(1-1.32883w+1.47211w^2-0.84162w^3+0.32216w^4)
/.w->(1+(4g)^2)^(1/4)
\end{verbatim}
}}

\subsection{$L=4$, $S=2$, $Q=u^2-\frac{1}{4}+\frac{1}{2\sqrt{5}}$ {\tiny \hyperref[table:ResSqrt]{Return to table 2}}}\label{sec:422}{\footnotesize
\noindent{$\Delta$}\vspace{-0.5em}
{\tiny
\begin{verbatim}
6+g^2(10+2Sqrt[5])+g^4(-34-10Sqrt[5])+g^6(234+414Sqrt[5]/5)+g^8(-2074-80z[3]-16Sqrt[5]z[3]-4078Sqrt[5]/5)+g^10(21050+1104z[3]
+304Sqrt[5]z[3]+800z[5]+160Sqrt[5]z[5]+219586Sqrt[5]/25)+g^12(-227394-4512z[3]-8720z[5]-1360Sqrt[5]z[5]-14000z[7]-6160Sqrt[5]z[7]
-15120z[9]-5040Sqrt[5]z[9]-2656Sqrt[5]z[3]/5-2448714Sqrt[5]/25)+g^14(2459594+(69600+13408Sqrt[5])z[5]+(114800+24080Sqrt[5])z[7]
+z[3](87904+27360Sqrt[5]+(-78400-40000Sqrt[5])z[5]+(-89600-35840Sqrt[5])z[7])+(215040+130368Sqrt[5])z[9]+(554400+184800Sqrt[5])
z[11]+(38400+18112Sqrt[5])z[3]^2+(-8000-3200Sqrt[5])z[5]^2+134624418Sqrt[5]/125)+g^16(-24950154+(-302000+62416Sqrt[5])z[7]+z[5]
(-262896+7600Sqrt[5]+(1344000+537600Sqrt[5])z[7])+(-13384800-4461600Sqrt[5])z[13]+(-26880-11520Sqrt[5])Z[11][2]+(-453824-204160
Sqrt[5]+(-313600-134400Sqrt[5])z[5])z[3]^2+(-192000-89600Sqrt[5])z[3]^3+(784000+408000Sqrt[5])z[5]^2+z[11](-2180080Sqrt[5]
-8607760/3)+z[3](-5045744+(-168000-80320Sqrt[5])z[5]+(1724800+873600Sqrt[5])z[7]+(2688000+1075200Sqrt[5])z[9]-10801648Sqrt[5]/5)
+z[9](-4782880/3-4080640Sqrt[5]/9)-1376810658Sqrt[5]/125)+g^18(210441114+(-1096928-1858880Sqrt[5])z[7]+z[5](23114752+9422720
Sqrt[5]+(-24167200-12668000Sqrt[5])z[7]+(-35520000-14208000Sqrt[5])z[9])+(270270000+90090000Sqrt[5])z[15]+(362880+158592Sqrt[5])
Z[11][2]+(-129920-55680Sqrt[5])Z[13][2]+(156800+67200Sqrt[5])Z[13][3]+(-20627296-9266016Sqrt[5]+(9193600+4250240Sqrt[5])z[5]
+(7212800+3091200Sqrt[5])z[7])z[3]^2+(1273600+582400Sqrt[5])z[3]^3+(-320000-140800Sqrt[5])z[3]^4+(-6916000-3122080Sqrt[5])z[5]^2
+(-14840000-5936000Sqrt[5])z[7]^2+z[11](6402768Sqrt[5]+57544400/3)+z[13](32520896Sqrt[5]+101293760/3)+z[3](161137968+(11823040
+5240000Sqrt[5])z[5]+(-591200-375136Sqrt[5])z[7]+(-54489600-21795840Sqrt[5])z[11]+(7168000+3072000Sqrt[5])z[5]^2+z[9](-263300800
/9-132123200Sqrt[5]/9)+1782426032Sqrt[5]/25)+z[9](-28649008/9-157295024Sqrt[5]/45)+11650693614Sqrt[5]/125)
\end{verbatim}}
\noindent{$\Delta_{\text{num}}$}\vspace{-0.5em}
{\tiny
\begin{verbatim}
6+0.904508497187(4g)^2-0.220158905371(4g)^4+0.102330671030(4g)^6-0.0615983229046(4g)^8+0.0419951251044(4g)^10
-0.0309890115261(4g)^12+0.0241353364599(4g)^14-0.0195275204977(4g)^16+0.0162485659636(4g)^18
\end{verbatim}}
\noindent{$\Delta_{\text{Pad\'e}}$}\vspace{-0.5em}
{\tiny\begin{verbatim}
(4.21948-6.59009w+5.85125w^2-2.17607w^3+0.16282w^4+0.21167w^5)/(1-1.67143w+1.43886w^2-0.62427w^3+0.13669w^4)
/.w->(1+(4g)^2)^(1/4)
\end{verbatim}
}}

\subsection{$L=6$, $S=2$, $Q=u^2-\frac{1}{4}\cot^2\left(\frac{\pi}{7}\right)$ {\tiny \hyperref[table:ResNon]{Return to table 3}}}\label{sec:621}{\footnotesize
\noindent{$\Delta$}\vspace{-0.5em}
{\tiny 
\begin{verbatim}
8+3.01208158513g^2-3.32025395247g^4+7.65808009377g^6+g^8(-22.1489206026-6.33799245425z[3])+g^10(71.4291834408+33.0552045073z[3]
+63.3799245425z[5])+g^12(-245.605950949-150.631494016z[3]-303.152624704z[5]-665.489207696z[7])+g^16(-3258.50080657-2917.91224693
z[3]-5761.75441672z[5]-687.105490212z[3]z[5]-11475.2357328z[7]-36161.4671974z[9]-74784.7008430z[11]-15568.6424759z[13]
-335.928268269z[3]^2)+g^14(881.159987118+666.333998035z[3]+1316.34675267z[5]+2970.47678075z[7]+7453.47912619z[9]+34.3552745106
z[3]^2)+g^18(12325.4572318+12759.9464370z[3]+24944.6053732z[5]+7771.62126345z[3]z[5]+49448.1863872z[7]+642.644054623z[3]z[7]
+2897.29669582z[5]z[7]+116993.785964z[9]+18188.5666941z[3]z[9]-3279.30218932z[5]z[9]+506318.332119z[11]-21643.3944495z[3]z[11]
+486665.949145z[13]+817353.729986z[15]+2210.43450962z[3]^2+2485.89377125z[5]^2-191.292627710z[7]^2)
\end{verbatim}}
\noindent{$\Delta_{\text{num}}$}\vspace{-0.5em}
{\tiny
\begin{verbatim}
8+0.188255099071(4g)^2-0.0129697420018(4g)^4+0.00186964846039(4g)^6-0.000454216738654(4g)^8+0.000168689559060(4g)^10
-0.0000841657227545(4g)^12+0.0000505166349778(4g)^14-0.0000354553823678(4g)^16+0.0000297285531511(4g)^18
\end{verbatim}}
\noindent{$\Delta_{\text{Pad\'e}}$}\vspace{-0.5em}
{\tiny\begin{verbatim}
(7.75418-17.53385w+19.36847w^2-6.81614w^3+1.36801w^4+0.74038w^5)/(1-2.28899w+2.61159w^2-1.09931w^3+0.38685w^4)
/.w->(1+(4g)^2)^(1/4)
\end{verbatim}
}}

\subsection{$L=6$, $S=2$, $Q=u^2-\frac{1}{4}\cot^2\left(\frac{2\pi}{7}\right)$ {\tiny \hyperref[table:ResNon]{Return to table 3}}}\label{sec:622}{\footnotesize
\noindent{$\Delta$}\vspace{-0.5em}
{\tiny
\begin{verbatim}
8+9.78016747165g^2-29.2248600538g^4+167.644254232g^6+g^8(-1171.73225869-103.903347319z[3])+g^10(9072.16629427+1391.08533432z[3]
+1039.03347319z[5])+g^12(-74977.0987975-15758.0286598z[3]-13741.9433097z[5]-10909.8514685z[7])+g^16(-5798156.19976-1809336.66246
z[3]-1734863.40987z[5]-34676.2877436z[3]z[5]-1663045.86523z[7]-1527255.89075z[9]-1306436.06739z[11]-164138.275627z[13]
-45253.6637143z[3]^2)+g^14(648623.139308+170830.753564z[3]+155413.265821z[5]+143503.751558z[7]+122190.336447z[9]+1733.81438718
z[3]^2)+g^18(53021031.7444+19152011.7334z[3]+18963374.6711z[5]+614035.627731z[3]z[5]+18944821.6885z[7]+772720.010164z[3]z[7]
+58868.8671937z[5]z[7]+18484562.8123z[9]+581993.473872z[3]z[9]-112258.728961z[5]z[9]+15783530.6929z[11]-740907.611143z[3]z[11]
+11324398.2566z[13]+8617259.47044z[15]+685111.868083z[3]^2+304245.909732z[5]^2-6548.42585606z[7]^2)
\end{verbatim}}
\noindent{$\Delta_{\text{num}}$}\vspace{-0.5em}
{\tiny
\begin{verbatim}
8+0.611260466978(4g)^2-0.114159609585(4g)^4+0.0409287730059(4g)^6-0.0197850035796(4g)^8+0.0112740828229(4g)^10
-0.00710305507184(4g)^12+0.00478612877698(4g)^14-0.00338980817812(4g)^16+0.00250050013243(4g)^18
\end{verbatim}}
\noindent{$\Delta_{\text{Pad\'e}}$}\vspace{-0.5em}
{\tiny\begin{verbatim}
(7.19318-16.80063w+14.63359w^2-5.60387w^3+0.02318w^4+1.47060w^5)/(1-2.33241w+2.07578w^2-1.04889w^3+0.42003w^4)
/.w->(1+(4g)^2)^(1/4)
\end{verbatim}
}}

\subsection{$L=6$, $S=2$, $Q=u^2-\frac{1}{4}\cot^2\left(\frac{3\pi}{7}\right)$ {\tiny \hyperref[table:ResNon]{Return to table 3}}}\label{sec:623}{\footnotesize
\noindent{$\Delta$}\vspace{-0.5em}
{\tiny 
\begin{verbatim}
8+15.2077509432g^2-59.4548859937g^4+456.697665674g^6+g^8(-4390.11882071-49.7586602268z[3])+g^10(47288.4045223+815.859461177z[3]
+497.586602268z[5])+g^12(-545801.295252-12347.3398462z[3]-8802.90406564z[5]-5224.65932381z[7])+g^14(6599807.70070+183702.912438
z[3]+139238.387426z[5]+98133.7716611z[7]+58516.1844267z[9]-1896.16966169z[3]^2)+g^16(-82352057.2994-2351521.42529z[3]
-1879854.83571z[5]+37923.3932338z[3]z[5]-1513894.89904z[7]-1310726.64205z[9]-1132059.23176z[11]-396869.081897z[13]+50581.5919826
z[3]^2)+g^18(1046299442.80+30436508.3202z[3]+23450656.7236z[5]+9360.75100604z[3]z[5]+19605058.1252z[7]-2073458.65422z[3]z[7]
-889670.163890z[5]z[7]+19219947.4017z[9]-3347318.04057z[3]z[9]-422061.968850z[5]z[9]+22259430.9749z[11]-2785608.99441z[3]z[11]
+26490055.7943z[13]+20835626.7996z[15]+1225061.69741z[3]^2-805611.803503z[5]^2-24620.2815162z[7]^2)
\end{verbatim}}
\noindent{$\Delta_{\text{num}}$}\vspace{-0.5em}
{\tiny
\begin{verbatim}
8+0.950484433951(4g)^2-0.232245648413(4g)^4+0.111498453534(4g)^6-0.0679005670430(4g)^8+0.0465250734120(4g)^10
-0.0342750404688(4g)^12+0.0265235311388(4g)^14-0.0212753813321(4g)^16+0.0175417010481(4g)^18
\end{verbatim}}
\noindent{$\Delta_{\text{Pad\'e}}$}\vspace{-0.5em}
{\tiny\begin{verbatim}
(6.33914-18.56736w+19.13856w^2-11.67028w^3+6.09934w^4-1.50376w^5)/(1-2.83375w+2.37467w^2-0.53720w^3-0.02426w^4)
/.w->(1+(4g)^2)^(1/4)
\end{verbatim}
}}
\bibliographystyle{elsarticle-num}
\bibliography{bibliography}

\begin{thebibliography}{10}
\expandafter\ifx\csname url\endcsname\relax
  \def\url#1{\texttt{#1}}\fi
\expandafter\ifx\csname urlprefix\endcsname\relax\def\urlprefix{URL }\fi
\expandafter\ifx\csname href\endcsname\relax
  \def\href#1#2{#2} \def\path#1{#1}\fi

\bibitem{Beisert:2010jr}
N.~Beisert, C.~Ahn, L.~F. Alday, Z.~Bajnok, J.~M. Drummond, et~al., {Review of
  AdS/CFT Integrability: An Overview}, Lett.Math.Phys. 99 (2012) 3--32.
\newblock \href {http://arxiv.org/abs/1012.3982} {\path{arXiv:1012.3982}},
  \href {http://dx.doi.org/10.1007/s11005-011-0529-2}
  {\path{doi:10.1007/s11005-011-0529-2}}.

\bibitem{Minahan:2002ve}
J.~A. Minahan, K.~Zarembo, {The Bethe-ansatz for N = 4 super Yang-Mills}, JHEP
  03 (2003) 013.
\newblock \href {http://arxiv.org/abs/hep-th/0212208}
  {\path{arXiv:hep-th/0212208}}.

\bibitem{Bena:2003wd}
I.~Bena, J.~Polchinski, R.~Roiban, {Hidden symmetries of the AdS(5) x S**5
  superstring}, Phys. Rev. D69 (2004) 046002.
\newblock \href {http://arxiv.org/abs/hep-th/0305116}
  {\path{arXiv:hep-th/0305116}}, \href
  {http://dx.doi.org/10.1103/PhysRevD.69.046002}
  {\path{doi:10.1103/PhysRevD.69.046002}}.

\bibitem{Beisert:2003tq}
N.~Beisert, C.~Kristjansen, M.~Staudacher, {The dilatation operator of N = 4
  super Yang-Mills theory}, Nucl. Phys. B664 (2003) 131--184.
\newblock \href {http://arxiv.org/abs/hep-th/0303060}
  {\path{arXiv:hep-th/0303060}}, \href
  {http://dx.doi.org/10.1016/S0550-3213(03)00406-1}
  {\path{doi:10.1016/S0550-3213(03)00406-1}}.

\bibitem{Staudacher:2004tk}
M.~Staudacher, {The factorized S-matrix of CFT/AdS}, JHEP 05 (2005) 054.
\newblock \href {http://arxiv.org/abs/hep-th/0412188}
  {\path{arXiv:hep-th/0412188}}.

\bibitem{Beisert:2005tm}
N.~Beisert, {The $su(2|2)$ dynamic S-matrix}, Adv. Theor. Math. Phys. 12 (2008)
  945.
\newblock \href {http://arxiv.org/abs/hep-th/0511082}
  {\path{arXiv:hep-th/0511082}}.

\bibitem{Beisert:2005fw}
N.~Beisert, M.~Staudacher, {Long-range $PSU(2,2|4)$ Bethe Ans\"atze for gauge
  theory and strings}, Nucl. Phys. B727 (2005) 1--62.
\newblock \href {http://arxiv.org/abs/hep-th/0504190}
  {\path{arXiv:hep-th/0504190}}, \href
  {http://dx.doi.org/10.1016/j.nuclphysb.2005.06.038}
  {\path{doi:10.1016/j.nuclphysb.2005.06.038}}.

\bibitem{Belitsky:2006en}
A.~V. Belitsky, A.~S. Gorsky, G.~P. Korchemsky, {Logarithmic scaling in gauge /
  string correspondence}, Nucl. Phys. B748 (2006) 24--59.
\newblock \href {http://arxiv.org/abs/hep-th/0601112}
  {\path{arXiv:hep-th/0601112}}, \href
  {http://dx.doi.org/10.1016/j.nuclphysb.2006.04.030}
  {\path{doi:10.1016/j.nuclphysb.2006.04.030}}.

\bibitem{Beisert:2006ez}
N.~Beisert, B.~Eden, M.~Staudacher, {Transcendentality and Crossing},
  J.Stat.Mech. 0701 (2007) P01021.
\newblock \href {http://arxiv.org/abs/hep-th/0610251}
  {\path{arXiv:hep-th/0610251}}, \href
  {http://dx.doi.org/10.1088/1742-5468/2007/01/P01021}
  {\path{doi:10.1088/1742-5468/2007/01/P01021}}.

\bibitem{Arutyunov:2004vx}
G.~Arutyunov, S.~Frolov, M.~Staudacher, {Bethe ansatz for quantum strings},
  JHEP 10 (2004) 016.
\newblock \href {http://arxiv.org/abs/hep-th/0406256}
  {\path{arXiv:hep-th/0406256}}, \href
  {http://dx.doi.org/10.1088/1126-6708/2004/10/016}
  {\path{doi:10.1088/1126-6708/2004/10/016}}.

\bibitem{Janik:2006dc}
R.~A. Janik, {The AdS$_5$xS$^5$ superstring worldsheet S-matrix and crossing
  symmetry}, Phys. Rev. D73 (2006) 086006.
\newblock \href {http://arxiv.org/abs/hep-th/0603038}
  {\path{arXiv:hep-th/0603038}}, \href
  {http://dx.doi.org/10.1103/PhysRevD.73.086006}
  {\path{doi:10.1103/PhysRevD.73.086006}}.

\bibitem{Benna:2006nd}
M.~K. Benna, S.~Benvenuti, I.~R. Klebanov, A.~Scardicchio, {A test of the
  AdS/CFT correspondence using high-spin operators}, Phys. Rev. Lett. 98 (2007)
  131603.
\newblock \href {http://arxiv.org/abs/hep-th/0611135}
  {\path{arXiv:hep-th/0611135}}, \href
  {http://dx.doi.org/10.1103/PhysRevLett.98.131603}
  {\path{doi:10.1103/PhysRevLett.98.131603}}.

\bibitem{Basso:2007wd}
B.~Basso, G.~P. Korchemsky, J.~Kotanski, {Cusp anomalous dimension in maximally
  supersymmetric Yang- Mills theory at strong coupling}, Phys. Rev. Lett. 100
  (2008) 091601.
\newblock \href {http://arxiv.org/abs/0708.3933} {\path{arXiv:0708.3933}},
  \href {http://dx.doi.org/10.1103/PhysRevLett.100.091601}
  {\path{doi:10.1103/PhysRevLett.100.091601}}.

\bibitem{Bern:2006ew}
Z.~Bern, M.~Czakon, L.~J. Dixon, D.~A. Kosower, V.~A. Smirnov, {The Four-Loop
  Planar Amplitude and Cusp Anomalous Dimension in Maximally Supersymmetric
  Yang-Mills Theory}, Phys. Rev. D75 (2007) 085010.
\newblock \href {http://arxiv.org/abs/hep-th/0610248}
  {\path{arXiv:hep-th/0610248}}, \href
  {http://dx.doi.org/10.1103/PhysRevD.75.085010}
  {\path{doi:10.1103/PhysRevD.75.085010}}.

\bibitem{Roiban:2007dq}
R.~Roiban, A.~A. Tseytlin, {Strong-coupling expansion of cusp anomaly from
  quantum superstring}, JHEP 11 (2007) 016.
\newblock \href {http://arxiv.org/abs/0709.0681} {\path{arXiv:0709.0681}},
  \href {http://dx.doi.org/10.1088/1126-6708/2007/11/016}
  {\path{doi:10.1088/1126-6708/2007/11/016}}.

\bibitem{Kotikov:2007cy}
A.~V. Kotikov, L.~N. Lipatov, A.~Rej, M.~Staudacher, V.~N. Velizhanin,
  {Dressing and Wrapping}, J. Stat. Mech. 0710 (2007) P10003.
\newblock \href {http://arxiv.org/abs/0704.3586} {\path{arXiv:0704.3586}},
  \href {http://dx.doi.org/10.1088/1742-5468/2007/10/P10003}
  {\path{doi:10.1088/1742-5468/2007/10/P10003}}.

\bibitem{Bajnok:2008bm}
Z.~Bajnok, R.~A. Janik, {Four-loop perturbative Konishi from strings and finite
  size effects for multiparticle states}, Nucl. Phys. B807 (2009) 625--650.
\newblock \href {http://arxiv.org/abs/0807.0399} {\path{arXiv:0807.0399}},
  \href {http://dx.doi.org/10.1016/j.nuclphysb.2008.08.020}
  {\path{doi:10.1016/j.nuclphysb.2008.08.020}}.

\bibitem{Fiamberti:2007rj}
F.~Fiamberti, A.~Santambrogio, C.~Sieg, D.~Zanon, {Wrapping at four loops in
  N=4 SYM}, Phys.Lett. B666 (2008) 100--105.
\newblock \href {http://arxiv.org/abs/0712.3522} {\path{arXiv:0712.3522}},
  \href {http://dx.doi.org/10.1016/j.physletb.2008.06.061}
  {\path{doi:10.1016/j.physletb.2008.06.061}}.

\bibitem{Fiamberti:2008sh}
F.~Fiamberti, A.~Santambrogio, C.~Sieg, D.~Zanon, {Anomalous dimension with
  wrapping at four loops in N=4 SYM}, Nucl.Phys. B805 (2008) 231--266.
\newblock \href {http://arxiv.org/abs/0806.2095} {\path{arXiv:0806.2095}},
  \href {http://dx.doi.org/10.1016/j.nuclphysb.2008.07.014}
  {\path{doi:10.1016/j.nuclphysb.2008.07.014}}.

\bibitem{Bajnok:2008qj}
Z.~Bajnok, R.~A. Janik, T.~Lukowski, {Four loop twist two, BFKL, wrapping and
  strings}, Nucl. Phys. B816 (2009) 376--398.
\newblock \href {http://arxiv.org/abs/0811.4448} {\path{arXiv:0811.4448}},
  \href {http://dx.doi.org/10.1016/j.nuclphysb.2009.02.005}
  {\path{doi:10.1016/j.nuclphysb.2009.02.005}}.

\bibitem{Lukowski:2009ce}
T.~Lukowski, A.~Rej, V.~N. Velizhanin, {Five-Loop Anomalous Dimension of
  Twist-Two Operators}, Nucl. Phys. B831 (2010) 105--132.
\newblock \href {http://arxiv.org/abs/0912.1624} {\path{arXiv:0912.1624}},
  \href {http://dx.doi.org/10.1016/j.nuclphysb.2010.01.008}
  {\path{doi:10.1016/j.nuclphysb.2010.01.008}}.

\bibitem{Beccaria:2009eq}
M.~Beccaria, V.~Forini, T.~Lukowski, S.~Zieme, {Twist-three at five loops,
  Bethe Ansatz and wrapping}, JHEP 03 (2009) 129.
\newblock \href {http://arxiv.org/abs/0901.4864} {\path{arXiv:0901.4864}},
  \href {http://dx.doi.org/10.1088/1126-6708/2009/03/129}
  {\path{doi:10.1088/1126-6708/2009/03/129}}.

\bibitem{Velizhanin:2010cm}
V.~N. Velizhanin, {Six-Loop Anomalous Dimension of Twist-Three Operators in N=4
  SYM}, JHEP 11 (2010) 129.
\newblock \href {http://arxiv.org/abs/1003.4717} {\path{arXiv:1003.4717}},
  \href {http://dx.doi.org/10.1007/JHEP11(2010)129}
  {\path{doi:10.1007/JHEP11(2010)129}}.

\bibitem{Fiamberti:2009jw}
F.~Fiamberti, A.~Santambrogio, C.~Sieg, {Five-loop anomalous dimension at
  critical wrapping order in N=4 SYM}, JHEP 1003 (2010) 103.
\newblock \href {http://arxiv.org/abs/0908.0234} {\path{arXiv:0908.0234}},
  \href {http://dx.doi.org/10.1007/JHEP03(2010)103}
  {\path{doi:10.1007/JHEP03(2010)103}}.

\bibitem{Gromov:2009zb}
N.~Gromov, V.~Kazakov, P.~Vieira, {Exact Spectrum of Planar ${\cal N}=4$
  Supersymmetric Yang-Mills Theory: Konishi Dimension at Any Coupling}, Phys.
  Rev. Lett. 104 (2010) 211601.
\newblock \href {http://arxiv.org/abs/0906.4240} {\path{arXiv:0906.4240}},
  \href {http://dx.doi.org/10.1103/PhysRevLett.104.211601}
  {\path{doi:10.1103/PhysRevLett.104.211601}}.

\bibitem{Roiban:2009aa}
R.~Roiban, A.~A. Tseytlin, {Quantum strings in AdS$_5$ x S$^5$: strong-coupling
  corrections to dimension of Konishi operator}, JHEP 11 (2009) 013.
\newblock \href {http://arxiv.org/abs/0906.4294} {\path{arXiv:0906.4294}},
  \href {http://dx.doi.org/10.1088/1126-6708/2009/11/013}
  {\path{doi:10.1088/1126-6708/2009/11/013}}.

\bibitem{Gromov:2011de}
N.~Gromov, D.~Serban, I.~Shenderovich, D.~Volin, {Quantum folded string and
  integrability: From finite size effects to Konishi dimension}, Journal of
  High Energy Physics 2011 (2011) 1--23.
\newblock \href {http://arxiv.org/abs/1102.1040} {\path{arXiv:1102.1040}}.

\bibitem{Roiban:2011fe}
R.~Roiban, A.~Tseytlin, {Semiclassical string computation of strong-coupling
  corrections to dimensions of operators in Konishi multiplet}, Nucl.Phys. B848
  (2011) 251--267.
\newblock \href {http://arxiv.org/abs/1102.1209} {\path{arXiv:1102.1209}},
  \href {http://dx.doi.org/10.1016/j.nuclphysb.2011.02.016}
  {\path{doi:10.1016/j.nuclphysb.2011.02.016}}.

\bibitem{Vallilo:2011fj}
B.~C. Vallilo, L.~Mazzucato, {The Konishi multiplet at strong coupling},
  Journal of High Energy Physics 2011 (2011) 1--9.
\newblock \href {http://arxiv.org/abs/1102.1219} {\path{arXiv:1102.1219}},
  \href {http://dx.doi.org/10.1007/JHEP12(2011)029}
  {\path{doi:10.1007/JHEP12(2011)029}}.

\bibitem{Gromov:2011bz}
N.~Gromov, S.~Valatka, {Deeper Look into Short Strings}, JHEP 1203 (2012) 058.
\newblock \href {http://arxiv.org/abs/1109.6305} {\path{arXiv:1109.6305}}.

\bibitem{Gromov:2014bva}
N.~Gromov, F.~Levkovich-Maslyuk, G.~Sizov, S.~Valatka, {Quantum Spectral Curve
  at Work: From Small Spin to Strong Coupling in $\mathcal N$=4 SYM}\href
  {http://arxiv.org/abs/1402.0871} {\path{arXiv:1402.0871}}.

\bibitem{Bajnok:2012bz}
Z.~Bajnok, R.~A. Janik, {Six and seven loop Konishi from Luscher corrections},
  JHEP 1211 (2012) 002.
\newblock \href {http://arxiv.org/abs/1209.0791} {\path{arXiv:1209.0791}},
  \href {http://dx.doi.org/10.1007/JHEP11(2012)002}
  {\path{doi:10.1007/JHEP11(2012)002}}.

\bibitem{Bombardelli:2013yka}
D.~Bombardelli, {A next-to-leading Luscher formula}, JHEP 1401 (2014) 037.
\newblock \href {http://arxiv.org/abs/1309.4083} {\path{arXiv:1309.4083}},
  \href {http://dx.doi.org/10.1007/JHEP01(2014)037}
  {\path{doi:10.1007/JHEP01(2014)037}}.

\bibitem{Zamolodchikov:1989cf}
A.~B. Zamolodchikov, {Thermodynamic Bethe Ansatz in relativistic models.
  Scaling three state Potts and Lee-Yang models}, Nucl. Phys. B342 (1990)
  695--720.
\newblock \href {http://dx.doi.org/10.1016/0550-3213(90)90333-9}
  {\path{doi:10.1016/0550-3213(90)90333-9}}.

\bibitem{Gromov:2009tv}
N.~Gromov, V.~Kazakov, P.~Vieira, {Exact Spectrum of Anomalous Dimensions of
  Planar N=4 Supersymmetric Yang-Mills Theory}, Phys. Rev. Lett. 103 (2009)
  131601.
\newblock \href {http://arxiv.org/abs/0901.3753} {\path{arXiv:0901.3753}},
  \href {http://dx.doi.org/10.1103/PhysRevLett.103.131601}
  {\path{doi:10.1103/PhysRevLett.103.131601}}.

\bibitem{Bombardelli:2009ns}
D.~Bombardelli, D.~Fioravanti, R.~Tateo, {Thermodynamic Bethe Ansatz for planar
  AdS/CFT: A Proposal}, J.Phys. A42 (2009) 375401.
\newblock \href {http://arxiv.org/abs/0902.3930} {\path{arXiv:0902.3930}},
  \href {http://dx.doi.org/10.1088/1751-8113/42/37/375401}
  {\path{doi:10.1088/1751-8113/42/37/375401}}.

\bibitem{Gromov:2009bc}
N.~Gromov, V.~Kazakov, A.~Kozak, P.~Vieira, {Exact Spectrum of Anomalous
  Dimensions of Planar N = 4 Supersymmetric Yang-Mills Theory: TBA and excited
  states}, Lett.Math.Phys. 91 (2010) 265--287.
\newblock \href {http://arxiv.org/abs/0902.4458} {\path{arXiv:0902.4458}},
  \href {http://dx.doi.org/10.1007/s11005-010-0374-8}
  {\path{doi:10.1007/s11005-010-0374-8}}.

\bibitem{Arutyunov:2009ur}
G.~Arutyunov, S.~Frolov, {Thermodynamic Bethe Ansatz for the AdS$_5$xS$^5$
  Mirror Model}, JHEP 05 (2009) 068.
\newblock \href {http://arxiv.org/abs/0903.0141} {\path{arXiv:0903.0141}},
  \href {http://dx.doi.org/10.1088/1126-6708/2009/05/068}
  {\path{doi:10.1088/1126-6708/2009/05/068}}.

\bibitem{Frolov:2010wt}
S.~Frolov, {Konishi operator at intermediate coupling}, J. Phys. A44 (2011)
  065401.
\newblock \href {http://arxiv.org/abs/1006.5032} {\path{arXiv:1006.5032}},
  \href {http://dx.doi.org/10.1088/1751-8113/44/6/065401}
  {\path{doi:10.1088/1751-8113/44/6/065401}}.

\bibitem{Frolov:2012zv}
S.~Frolov, {Scaling dimensions from the mirror TBA}, J.Phys. A45 (2012) 305402.
\newblock \href {http://arxiv.org/abs/1201.2317} {\path{arXiv:1201.2317}},
  \href {http://dx.doi.org/10.1088/1751-8113/45/30/305402}
  {\path{doi:10.1088/1751-8113/45/30/305402}}.

\bibitem{Arutyunov:2010gb}
G.~Arutyunov, S.~Frolov, R.~Suzuki, {Five-loop Konishi from the Mirror TBA},
  JHEP 04 (2010) 069.
\newblock \href {http://arxiv.org/abs/1002.1711} {\path{arXiv:1002.1711}},
  \href {http://dx.doi.org/10.1007/JHEP04(2010)069}
  {\path{doi:10.1007/JHEP04(2010)069}}.

\bibitem{Balog:2010xa}
J.~Balog, A.~Hegedus, {5-loop Konishi from linearized TBA and the XXX magnet},
  JHEP 06 (2010) 080.
\newblock \href {http://arxiv.org/abs/1002.4142} {\path{arXiv:1002.4142}},
  \href {http://dx.doi.org/10.1007/JHEP06(2010)080}
  {\path{doi:10.1007/JHEP06(2010)080}}.

\bibitem{Eden:2012fe}
B.~Eden, P.~Heslop, G.~P. Korchemsky, V.~A. Smirnov, E.~Sokatchev, {Five-loop
  Konishi in N=4 SYM}, Nucl.Phys. B862 (2012) 123--166.
\newblock \href {http://arxiv.org/abs/1202.5733} {\path{arXiv:1202.5733}},
  \href {http://dx.doi.org/10.1016/j.nuclphysb.2012.04.015}
  {\path{doi:10.1016/j.nuclphysb.2012.04.015}}.

\bibitem{Gromov:2010km}
N.~Gromov, V.~Kazakov, S.~Leurent, Z.~Tsuboi, {Wronskian Solution for AdS/CFT
  Y-system}, JHEP 1101 (2011) 155.
\newblock \href {http://arxiv.org/abs/1010.2720} {\path{arXiv:1010.2720}},
  \href {http://dx.doi.org/10.1007/JHEP01(2011)155}
  {\path{doi:10.1007/JHEP01(2011)155}}.

\bibitem{Gromov:2011cx}
N.~Gromov, V.~Kazakov, S.~Leurent, D.~Volin, {Solving the AdS/CFT Y-system},
  JHEP 1207 (2012) 023.
\newblock \href {http://arxiv.org/abs/1110.0562} {\path{arXiv:1110.0562}},
  \href {http://dx.doi.org/10.1007/JHEP07(2012)023}
  {\path{doi:10.1007/JHEP07(2012)023}}.

\bibitem{Balog:2012zt}
J.~Balog, A.~Hegedus, {Hybrid-NLIE for the AdS/CFT spectral problem}, JHEP 1208
  (2012) 022.
\newblock \href {http://arxiv.org/abs/1202.3244} {\path{arXiv:1202.3244}},
  \href {http://dx.doi.org/10.1007/JHEP08(2012)022}
  {\path{doi:10.1007/JHEP08(2012)022}}.

\bibitem{Leurent:2012ab}
S.~Leurent, D.~Serban, D.~Volin, {Six-loop Konishi anomalous dimension from the
  Y-system}, Phys.Rev.Lett. 109 (2012) 241601.
\newblock \href {http://arxiv.org/abs/1209.0749} {\path{arXiv:1209.0749}},
  \href {http://dx.doi.org/10.1103/PhysRevLett.109.241601}
  {\path{doi:10.1103/PhysRevLett.109.241601}}.

\bibitem{Leurent:2013mr}
S.~Leurent, D.~Volin, {Multiple zeta functions and double wrapping in planar
  $N=4$ SYM}, Nucl.Phys. B875 (2013) 757--789.
\newblock \href {http://arxiv.org/abs/1302.1135} {\path{arXiv:1302.1135}},
  \href {http://dx.doi.org/10.1016/j.nuclphysb.2013.07.020}
  {\path{doi:10.1016/j.nuclphysb.2013.07.020}}.

\bibitem{Broadhurst:1995km}
D.~J. Broadhurst, D.~Kreimer, {Knots and numbers in Phi**4 theory to 7 loops
  and beyond}, Int.J.Mod.Phys. C6 (1995) 519--524.
\newblock \href {http://arxiv.org/abs/hep-ph/9504352}
  {\path{arXiv:hep-ph/9504352}}, \href
  {http://dx.doi.org/10.1142/S012918319500037X}
  {\path{doi:10.1142/S012918319500037X}}.

\bibitem{Gromov:2013pga}
N.~Gromov, V.~Kazakov, S.~Leurent, D.~Volin, {Quantum spectral curve for
  $AdS_5/CFT_4$}, Phys.Rev.Lett. 112 (2014) 011602.
\newblock \href {http://arxiv.org/abs/1305.1939} {\path{arXiv:1305.1939}},
  \href {http://dx.doi.org/10.1103/PhysRevLett.112.011602}
  {\path{doi:10.1103/PhysRevLett.112.011602}}.

\bibitem{Gromov:2014caa}
N.~Gromov, V.~Kazakov, S.~Leurent, D.~Volin, {Quantum spectral curve for
  arbitrary state/operator in AdS$_5$/CFT$_4$}\href
  {http://arxiv.org/abs/1405.4857} {\path{arXiv:1405.4857}}.

\bibitem{Gromov:2013qga}
N.~Gromov, F.~Levkovich-Maslyuk, G.~Sizov, {Analytic solution of Bremsstrahlung
  TBA II: turning on the sphere angle}, JHEP 1310 (2013) 036.
\newblock \href {http://arxiv.org/abs/1305.1944} {\path{arXiv:1305.1944}},
  \href {http://dx.doi.org/10.1007/JHEP10(2013)036}
  {\path{doi:10.1007/JHEP10(2013)036}}.

\bibitem{Alfimov:2014bwa}
M.~Alfimov, N.~Gromov, V.~Kazakov, {QCD Pomeron from AdS/CFT Quantum Spectral
  Curve}\href {http://arxiv.org/abs/1408.2530} {\path{arXiv:1408.2530}}.

\bibitem{Blumlein:2009cf}
J.~Blumlein, D.~Broadhurst, J.~Vermaseren, {The Multiple Zeta Value Data Mine},
  Comput.Phys.Commun. 181 (2010) 582--625.
\newblock \href {http://arxiv.org/abs/0907.2557} {\path{arXiv:0907.2557}},
  \href {http://dx.doi.org/10.1016/j.cpc.2009.11.007}
  {\path{doi:10.1016/j.cpc.2009.11.007}}.

\bibitem{Broadhurst:2014fga}
D.~Broadhurst, O.~Schnetz, {Algebraic geometry informs perturbative quantum
  field theory}, PoS 211 (2014) 078.
\newblock \href {http://arxiv.org/abs/1409.5570} {\path{arXiv:1409.5570}}.

\bibitem{Belitsky:2007kf}
A.~V. Belitsky, {Strong coupling expansion of Baxter equation in N=4 SYM},
  Phys.Lett. B659 (2008) 732--740.
\newblock \href {http://arxiv.org/abs/0710.2294} {\path{arXiv:0710.2294}},
  \href {http://dx.doi.org/10.1016/j.physletb.2007.11.023}
  {\path{doi:10.1016/j.physletb.2007.11.023}}.

\bibitem{Volin:2008kd}
D.~Volin, {The 2-loop generalized scaling function from the BES/FRS
  equation}\href {http://arxiv.org/abs/0812.4407} {\path{arXiv:0812.4407}}.

\bibitem{Belitsky:2009mu}
A.~Belitsky, {Baxter equation beyond wrapping}, Phys.Lett. B677 (2009) 93--99.
\newblock \href {http://arxiv.org/abs/0902.3198} {\path{arXiv:0902.3198}},
  \href {http://dx.doi.org/10.1016/j.physletb.2009.05.004}
  {\path{doi:10.1016/j.physletb.2009.05.004}}.

\bibitem{Velizhanin:2013vla}
V.~Velizhanin, {Twist-2 at five loops: Wrapping corrections without wrapping
  computations}, JHEP 1406 (2014) 108.
\newblock \href {http://arxiv.org/abs/1311.6953} {\path{arXiv:1311.6953}},
  \href {http://dx.doi.org/10.1007/JHEP06(2014)108}
  {\path{doi:10.1007/JHEP06(2014)108}}.

\bibitem{Schnetz}
O.~Schnetz,
  \href{"http://www.raumzeitmaterie.de/veranstaltungen.php?evt=select\&sqn=8970\&lang=en"}{{Single-valued
  multiple-zeta-values}}, SFB Colloquium Novel Methods for Perturbative QFT,
  Humboldt-Universit\"at zu Berlin, May 2013.
\newline\urlprefix\url{"http://www.raumzeitmaterie.de/veranstaltungen.php?evt=select\&sqn=8970\&lang=en"}

\bibitem{Brown:2013gia}
F.~Brown, {Single-valued periods and multiple zeta values}\href
  {http://arxiv.org/abs/1309.5309} {\path{arXiv:1309.5309}}.

\bibitem{Schnetz:2013hqa}
O.~Schnetz, {Graphical functions and single-valued multiple
  polylogarithms}\href {http://arxiv.org/abs/1302.6445}
  {\path{arXiv:1302.6445}}.

\bibitem{SchnetzPrivate}
O.~Schnetz, {Private communication}.

\bibitem{Santambrogio:2002sb}
A.~Santambrogio, D.~Zanon, {Exact anomalous dimensions of N = 4 Yang-Mills
  operators with large R charge}, Phys. Lett. B545 (2002) 425--429.
\newblock \href {http://arxiv.org/abs/hep-th/0206079}
  {\path{arXiv:hep-th/0206079}}, \href
  {http://dx.doi.org/10.1016/S0370-2693(02)02627-8}
  {\path{doi:10.1016/S0370-2693(02)02627-8}}.

\bibitem{Arutyunov:2012tx}
G.~Arutyunov, S.~Frolov, A.~Sfondrini, {Exceptional Operators in N=4 super
  Yang-Mills}, JHEP 1209 (2012) 006.
\newblock \href {http://arxiv.org/abs/1205.6660} {\path{arXiv:1205.6660}},
  \href {http://dx.doi.org/10.1007/JHEP09(2012)006}
  {\path{doi:10.1007/JHEP09(2012)006}}.

\bibitem{Faddeev:1996iy}
L.~D. Faddeev, {How Algebraic Bethe Ansatz works for integrable model}\href
  {http://arxiv.org/abs/hep-th/9605187} {\path{arXiv:hep-th/9605187}}.

\bibitem{Freyhult:2010kc}
L.~Freyhult, {Review of AdS/CFT Integrability, Chapter III.4: Twist states and
  the cusp anomalous dimension}, Lett.Math.Phys. 99 (2012) 255--276.
\newblock \href {http://arxiv.org/abs/1012.3993} {\path{arXiv:1012.3993}},
  \href {http://dx.doi.org/10.1007/s11005-011-0483-z}
  {\path{doi:10.1007/s11005-011-0483-z}}.

\bibitem{Faddeev:1994zg}
L.~Faddeev, G.~Korchemsky, {High-energy QCD as a completely integrable model},
  Phys.Lett. B342 (1995) 311--322.
\newblock \href {http://arxiv.org/abs/hep-th/9404173}
  {\path{arXiv:hep-th/9404173}}, \href
  {http://dx.doi.org/10.1016/0370-2693(94)01363-H}
  {\path{doi:10.1016/0370-2693(94)01363-H}}.

\bibitem{Korchemsky:1994um}
G.~Korchemsky, {Bethe ansatz for QCD pomeron}, Nucl.Phys. B443 (1995) 255--304.
\newblock \href {http://arxiv.org/abs/hep-ph/9501232}
  {\path{arXiv:hep-ph/9501232}}, \href
  {http://dx.doi.org/10.1016/0550-3213(95)00099-E}
  {\path{doi:10.1016/0550-3213(95)00099-E}}.

\bibitem{Eden:2006rx}
B.~Eden, M.~Staudacher, Integrability and transcendentality, J. Stat. Mech.
  0611 (2006) P014.
\newblock \href {http://arxiv.org/abs/hep-th/0603157}
  {\path{arXiv:hep-th/0603157}}.

\bibitem{Janik:2013nqa}
R.~A. Janik, {Twist-two operators and the BFKL regime - nonstandard solutions
  of the Baxter equation}, JHEP 1311 (2013) 153.
\newblock \href {http://arxiv.org/abs/1309.2844} {\path{arXiv:1309.2844}},
  \href {http://dx.doi.org/10.1007/JHEP11(2013)153}
  {\path{doi:10.1007/JHEP11(2013)153}}.

\bibitem{Beccaria:2007cn}
M.~Beccaria, {Anomalous dimensions at twist-3 in the sl(2) sector of N=4 SYM},
  JHEP 0706 (2007) 044.
\newblock \href {http://arxiv.org/abs/0704.3570} {\path{arXiv:0704.3570}},
  \href {http://dx.doi.org/10.1088/1126-6708/2007/06/044}
  {\path{doi:10.1088/1126-6708/2007/06/044}}.

\bibitem{Yang:1968rm}
C.-N. Yang, C.~P. Yang, {Thermodynamics of a one-dimensional system of bosons
  with repulsive delta-function interaction}, J. Math. Phys. 10 (1969)
  1115--1122.

\bibitem{MathematicaNotebook}
C.~Marboe, D.~Volin,
  \href{http://www.maths.tcd.ie/~dvolin/QSC/loop10sl2.zip}{{{\it Mathematica}
  notebook for the paper: "Quantum spectral curve as a tool for a perturbative
  quantum field theory"}}.
\newline\urlprefix\url{http://www.maths.tcd.ie/~dvolin/QSC/loop10sl2.zip}

\end{thebibliography}

\end{document}